\documentstyle[12pt,aaspp4,amssym]{article}
\lefthead{The Sunyaev-Zel'dovich Effect}
\righthead{Mark Birkinshaw}
\slugcomment{To appear in Physics Reports}
\def \arcmin   {{\rm arcmin}}
\def \cm       {{\rm cm}}
\def \GHz      {{\rm GHz}}
\def \K        {{\rm K}}
\def \keV      {{\rm keV}}
\def \kms      {{\rm km \, s^{-1}}}
\def \kmsMpc   {{\rm km \, s^{-1} \, Mpc^{-1}}}
\def \kpc      {{\rm kpc}}
\def \mJy      {{\rm mJy}}
\def \mK       {{\rm mK}}
\def \Mpc      {{\rm Mpc}}
\def \Msol     {{M_\odot}}
\def \muJy     {{\rm \mu Jy}}
\def \muK      {{\rm \mu K}}
\def \W        {{\rm W}}
\def \etal     {{\it et~al.\/}}
\def \SZ       {{Sunyaev-Zel'dovich}}
\def\boltz     {{k_{\rm B}}}
\def\DT        {{\Delta T}}
\def\DTRJ      {{\Delta T_{\rm RJ}}}
\def\DTRJzero  {{\Delta T_{\rm RJ0}}}
\def\me        {{m_{\rm e}}}
\def\mp        {{m_{\rm p}}}
\def\rc        {{r_{\rm c}}}
\def\te        {{T_{\rm e}}}
\def\thc       {{\theta_{\rm c}}}
\def\trad      {{T_{\rm rad}}}
\newbox\grsign    \setbox\grsign=\hbox{$>$} 
\newdimen\grdimen \grdimen=\ht\grsign
\newbox\simlessbox \newbox\simgreatbox
\setbox\simgreatbox=\hbox{\raise.5ex\hbox{$>$}\llap
     {\lower.5ex\hbox{$\sim$}}}\ht1=\grdimen\dp1=0pt
\setbox\simlessbox=\hbox{\raise.5ex\hbox{$<$}\llap
     {\lower.5ex\hbox{$\sim$}}}\ht2=\grdimen\dp2=0pt
\def\simgreat{\mathrel{\copy\simgreatbox}}
\def\simless{\mathrel{\copy\simlessbox}}
\begin{document}
\title{THE SUNYAEV-ZEL'DOVICH EFFECT}
\author{Mark Birkinshaw\altaffilmark{1}}
\affil{Department of Physics, University of Bristol,
       Tyndall Avenue, Bristol BS8 1TL, UK}
\authoraddr{Department of Physics, University of Bristol,
            Tyndall Avenue, Bristol BS8 1TL, UK}
\altaffiltext{1}{Also Smithsonian Institution Astrophysical Observatory, 
 60 Garden Street, Cambridge, MA 02138, USA}
%
%
\begin{abstract}
The \SZ\ effect causes a change in the apparent brightness of the
Cosmic Microwave Background Radiation towards a cluster of galaxies or
any other reservoir of hot plasma. Measurements of the effect provide
distinctly different information about cluster properties than X-ray
imaging data, while combining X-ray and \SZ\ effect data leads to new
insights into the structures of cluster atmospheres. The effect is
redshift-independent, and so provides a unique probe of the 
structure of the Universe on the largest scales. The
present review discusses the theory of the \SZ\ effect
and collects published results for many clusters, presents the
overall conclusions that may be drawn from the detections so
far, and discusses the prospects for future research on the
\SZ\ effects.
\end{abstract}
%
%
%
%
\section{Astrophysical Context}
Compton scattering is one of the major physical processes that couples
matter and radiation. Its importance is often stressed in highly
relativistic environments where large energy transfers occur: for
example, in the synchrotron self-Compton process that may be
responsible for much of the X-radiation from active galactic nuclei
(e.g., Fabian \etal\markcite{fab86} 1986). However, the Compton
process also has observable 
consequences in low-energy environments, where small energy transfers
occur. The Sunyaev-Zel'dovich effect, which arises from the scattering
of electrons in clusters of galaxies on the cosmic microwave
background radiation field, is perhaps the most important
astrophysical example. The effect provides a cosmological
probe, it has been used to measure the properties of gas in clusters of
galaxies, and it has been discussed as a means of measuring the
motions of clusters of galaxies and hence studying the evolution of 
structure in the Universe.
\par
The purpose of this review is to provide a comprehensive introduction
to the Sunyaev-Zel'dovich effect. I aim to provide both a
theoretical treatment that can be followed by non-specialists, and an
introduction to the observation of the effect with a critical review
of data in the literature. The latter is more difficult today than it
would have been five years ago because of the rapid increase in the
number of papers on the \SZ\ effect, and the improvement in the
quality of the results that are being gained.
\subsection{The Cosmic Microwave Background Radiation}
The cosmic microwave background radiation (CMBR) is the dominant
radiation field in the Universe, and one of the most
powerful cosmological tools that has yet been found. 
25~years after its discovery by Penzias \& Wilson\markcite{pw65}
(1965)\footnote{Low-significance indications of excess
microwave radiation had been reported earlier (e.g.,
Shmaonov\markcite{shm57} 1957; Ohm\markcite{ohm61} 1961), but not
attributed to cosmic processes or lost in the 
error estimates. In hindsight a universal radiation field could 
have been deduced from the excitation of some interstellar molecules
(Thaddeus\markcite{thad72} 1972).}
much is now known about the properties of the
radiation (see the recent review by Partridge\markcite{part95} 1995),
and a vigorous community studies the CMBR to extract all the
cosmological and astrophysical data that it carries.
\par
Within a few years of the discovery of the CMBR, it was established
the radiation field is close to isotropic, with a spectrum
characterized by a single temperature, $\trad\approx 2.7 \ \rm K$.
The specific intensity of the radiation is therefore close to 
\begin{equation}
 \label{eq-planckspec}
 I_\nu = {2 \, h \, \nu^3 \over c^2} \,
           \left( e^{h\nu/k_{\rm B}\trad} - 1 \right)^{-1}
\end{equation}
which corresponds to a peak brightness
$I_{\rm max} \sim 3.7 \times 10^{-18} \ \rm W \, m^{-2} \, Hz^{-1} \,
sr^{-1}$ at $\nu_{\rm max} \sim 160 \ \rm GHz$, a photon
density $n_\gamma \sim 4 \times 10^8 \ \rm photons \, m^{-3}$, and an
energy density $u_\gamma \sim 4 \times 10^{-14} \ \rm J \,
m^{-3}$, which can also be expressed as a mass density 
$\rho_\gamma \sim 5 \times 10^{-31} \ \rm kg \, m^{-3}$, 
much less than the critical density
\begin{eqnarray}
  \rho_{\rm crit} & = & {3 H_0^2 \over {8 \pi G}} 
                        \nonumber \\
                  & = & 1.88 \times 10^{-26} \, h_{100}^2 \quad 
                        \rm kg \, m^{-3} 
  \label{eq-rhocrit}
\end{eqnarray}
required to close the Universe. In these equations, $h$ is Planck's
constant, $c$ is the speed of light, $\nu$ is the frequency, $k_{\rm
B}$ is the Boltzmann constant, $G$ is the gravitational constant, and
$h_{100} = H_0/100 \ \rm km \, s^{-1} \, Mpc^{-1}$ is a dimensionless
measure of the value of the Hubble constant, $H_0$. Recent estimates
give $0.5 \lesssim h_{100} \lesssim 0.8$ (e.g., Sandage
\etal\markcite{san96} 1996; Falco \etal\markcite{fal97} 1997; Sandage
\& Tammann\markcite{st97} 1997; Freedman, Madore \&
Kennicutt\markcite{fmk97} 1997).
\par
\begin{figure}[p]
\epsscale{1.0}
\plotone{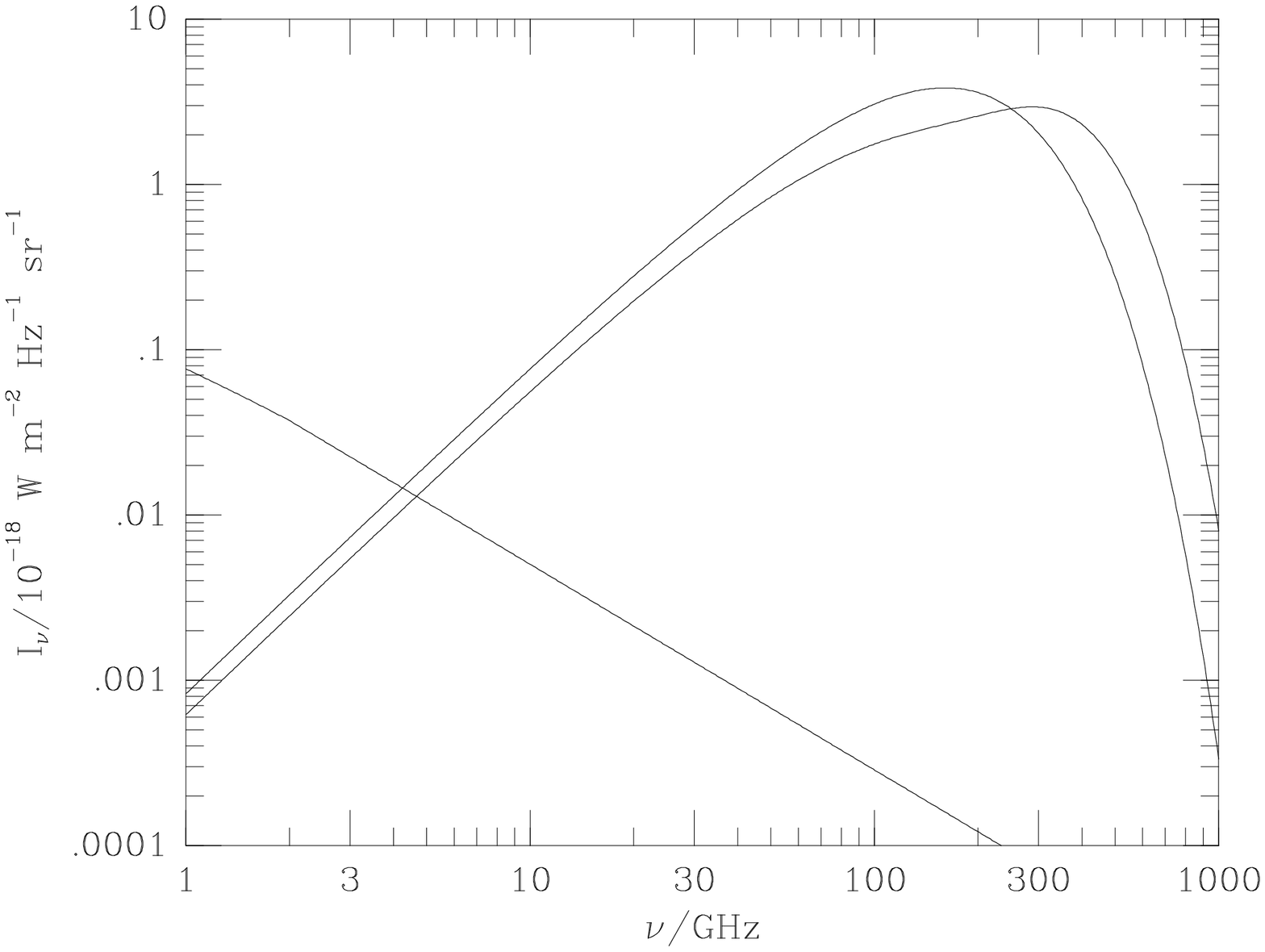}
\caption{\label{fig-cmbspec}
The spectrum of the microwave background
radiation, and the microwave background radiation after passage
through an (exaggerated) scattering atmosphere with $y = 0.1$ and
$\tau\beta = 0.05$ (as defined in Sections~\ref{sec-scat}
and~\ref{sec-kinematic}), compared with the integrated emission from the
bright radio source Cygnus~A as observed by a telescope with solid angle 
$\Omega_{\rm beam} = 1$~square deg. Note that the microwave background
radiation dominates at high frequencies. Scattering (the
Sunyaev-Zel'dovich effect) causes a fractional decrease in the 
low-frequency intensity of the CMBR that is proportional
to $y$. The location of the cross-over point, where the scattered CMBR
and the unscattered CMBR have equal brightness, 
is a measure of $\tau\beta$. This scattered spectrum was calculated
using the Kompaneets formula (\ref{eq-kompaneetsdeltan}), rather than
the relativistic results (eq.~\ref{eq-deltaIintegral}) of Rephaeli
(1995a), and hence is only accurate for 
low cluster gas temperatures (see Sec.~\ref{sec-effectspectrum}),
although the difference is imperceptible in this figure.}
\end{figure}

Although specific small parts of the sky (stars, radio sources, and
so on) are brighter than the CMBR, overall the CMBR constitutes the
major electromagnetic radiation field in the Universe and contributes
about 60~per cent of the relativistic energy density (the other 40~per
cent being provided by the neutrinos, assumed to be massless
here). The integrated brightness of the sky in the CMBR is not small,
and a comparison with a bright radio source may be useful. Cygnus~A is
one of the brightest extragalactic radio sources at low
frequencies. A comparison of the relative brightness of Cygnus~A and
the CMBR, as observed by a telescope with a 1~square degree beam, is
shown in Figure~\ref{fig-cmbspec}. It can be seen that the CMBR easily
dominates over a wide range of frequencies above 10~GHz. It is not signal
strength that makes measuring the intensity of the CMBR difficult, but
rather the problem of making absolute measurements, 
since the CMBR is present in all directions with almost
equal intensity.
\subsection{Thermal history of the Universe and the CMBR}
The origins of the CMBR lie in an early hot phase of the expansion of
the Universe, where the details of its generation are erased by the
close coupling of radiation and matter. Later energy
releases, interactions with matter at different temperatures, 
and other effects can modify the spectrum and brightness distribution
of the CMBR. Cosmological data on the gross properties of the
Universe are contained in the integrated properties of the CMBR, such
as the spectrum and the large-scale brightness structure. Detailed
information about the properties and formation of present-day objects,
such as clusters of galaxies, is encoded in the small-scale
structures in the brightness.
\par
A critical stage in the development of the CMBR occurs when the
expansion of the Universe causes the temperature to drop to about
$3000 \ \rm K$. At earlier times (higher redshifts), matter and
radiation were in good thermal contact because of the abundance of 
free electrons. But at this stage the number of free electrons drops
rapidly as matter becomes neutral, and the radiation and matter become
thermally decoupled, so that the temperatures of the photon and matter
fluids evolve almost independently. We can distinguish three events
that occur at almost the same time: the non-relativistic and
relativistic (photon plus neutrino) mass densities
are equal at redshift
\begin{equation}
   z_{\rm eq} = 2.5 \times 10^4 \, \Omega_0 h_{100}^2 ,
\end{equation}
most electrons have become bound to ions at the redshift of
recombination, 
\begin{equation}
   z_{\rm rec} = 1.4 \times 10^3 \, (\Omega_{\rm B} h_{100}^2)^{0.02}, 
\end{equation}
and the interaction length of photons and electrons exceeds the scale
of the Universe at the redshift of decoupling
\begin{equation}
   z_{\rm dec} = 1.1 \times 10^3 \, (\Omega_0/\Omega_{\rm B})^{0.02}
\end{equation}
(approximate forms taken from Kolb \& Turner\markcite{kt90} 1990). In
these relations, $\Omega_0$ is the present-day mass density of the Universe,
and $\Omega_{\rm B}$ is the present-day baryon density, both in units
of the critical density, $\rho_{\rm crit}$ (equation~\ref{eq-rhocrit}). The
redshifts of recombination and decoupling are similar, and neither
phenomenon is sharply-defined, so that there was
a moderately broad redshift range from $1500$ to $1000$ 
(about $1.6 \times 10^5 (\Omega_0 h_{100}^2)^{-1/2}$
years after the Big Bang) when the Universe was becoming
neutral, matter-dominated, and transparent to radiation. At some time
about then, most of the photons that are now in the
cosmic background radiation were scattered by electrons for the last
time, and we often refer to a sphere of last scattering or 
redshift of last scattering at this epoch.
\par
One of the important changes that occurred during this period, because
of the change in the interactions of photons and electrons, was that
the length scale on which gravitational collapse can occur dropped
dramatically, so that fluctuations in the mass density that were
stabilized by the radiation field before recombination became unstable
after recombination, and were able to collapse (slowly --- the
expansion of the Universe causes the collapse of gravitationally bound
objects to be power-law rather than exponential in time: Landau \&
Lifshitz\markcite{ll62} 1962; see descriptions in Kolb \&
Turner\markcite{kt90} 1990). Matter 
over-densities and under-densities present at recombination, and which
later became the large-scale objects that we see in the present-day
Universe, such as clusters of galaxies, 
caused fluctuations in the intensity of the radiation field through
their gravitational perturbations (the Sachs-Wolfe effect; Sachs \&
Wolfe\markcite{sw67} 1967), through thermodynamic fluctuations in the
density of 
radiation coupled to the matter, and through Doppler shifts due to
motions of the surface of last scattering. Recent reviews of the
introduction of primordial structure in the CMBR by objects near
recombination are given by Bond\markcite{bond95} (1995) and White,
Scott \& Silk\markcite{wss94} (1994). 
\subsection{COBE and the CMBR\label{sec-cobe}}
Much of the best data on the large-scale structural and spectral
properties of the CMBR was gathered by the Cosmic Background Explorer
(COBE) satellite (Boggess \etal\markcite{bog92} 1992). The accuracy
with which the 
spectrum of the radiation matches a black body with temperature $\trad
= 2.728 \pm 0.002 \ \K$ (Fixsen \etal\markcite{fix96}
1996)\footnote{This, and all 
later, limits have been converted to $1\sigma$ from the 95~per cent
confidence limits quoted in the Fixsen \etal\ paper.} demonstrates that
the Universe has been through a dense, hot, phase and provides strong
limits on non-thermalized cosmological energy transfers to the
radiation field (Wright \etal\markcite{wr94} 1994). The
previously-known dipolar 
term in the CMBR anisotropy was better measured --- Fixsen {\it et
al.} find an amplitude $3.372 \pm 0.004$~mK. This dipole is
interpreted as arising mostly from our peculiar
motion relative to the sphere of last scattering, and this motion was
presumably induced by local masses (within 100~Mpc or so). Our implied
velocity is $371 \pm 1 \ \kms$ towards galactic coordinates
$l = 264^\circ\llap{.}14 \pm 0^\circ\llap{.}15$,
$b =  48^\circ\llap{.}26 \pm 0^\circ\llap{.}15$ .
It is interesting to note that this dipolar anisotropy
shows an annual modulation from the motion of the
Earth around the Sun (Kogut \etal\markcite{kog94} 1994) and a spectral shape
consistent with the first derivative of a black body spectrum (Fixsen
\etal\markcite{fix94} 1994), as expected. This modulation was used to
check the calibration of the COBE~data.
\par
After the uniform (monopole) and dipolar
parts of the structure of the CMBR are removed, there remain
significant significant correlated signals in the
angular power spectrum. These signals correspond to an
an rms scatter of $35 \pm 2 \ \muK$ on the $7^\circ$ scale of the COBE
DMR~beam (Banday \etal\markcite{ban97} 1997), much larger than any likely
residual systematic errors (Bennett \etal\markcite{ben96} 1996), and
hold information about the radiation fluctuations at the sphere of last
scattering which are caused by density and temperature fluctuations
associated with the formation of massive structures (such as clusters
of galaxies). Their amplitude can be 
described by a multipole expansion of the brightness temperature
variations
\begin{equation}
   \Delta T(\theta,\phi) = \sum_{lm} \, a_{lm} \, Y_{lm}(\theta,\phi)
\end{equation}
with power spectrum 
\begin{equation}
   P_l = | a_l |^2 
             = {1 \over 2 l+1} \, \sum_m \, | a_{lm} |^2.
\end{equation}
It is usually assumed that the $a_{lm}$ obey Gaussian statistics, as
measured by a set of observers distributed over the Universe. The
ensemble of values of $a_{lm}$ for each $(l,m)$ then has a zero mean with a
standard deviation dependent on $l$ only and a phase that is uniformly
distributed over $0$ to $2\pi$. In that case, the temperature field is 
completely specified by the two-point correlation function 
\begin{equation}
   C(\theta) = \langle T({\bf n}_{\rm a}) \, T({\bf n}_{\rm b}) \rangle
 \label{eq-correlation}
\end{equation}
where the average is over all observers, and 
$\cos\theta = {\bf n}_{\rm a} {\bf .} {\bf n}_{\rm b}$
is the angle between the directions ${\bf n}_{\rm a}$
and ${\bf n}_{\rm b}$. For a Gaussian random field, 
\begin{equation}
   C(\theta) = {1 \over 4\pi} \sum_l \, (2 l + 1 ) \, C_l \,
               P_l(\cos\theta)
\end{equation}
where $C_l = \langle | a_{lm} |^2 \rangle$. For such a
spectrum and correlation function, it can be shown that a power-law
initial density fluctuation spectrum, $P(k) \propto k^n$ will produce
a spectrum with 
\begin{equation} 
   C_l = C_2 \,
    {\Gamma\left(l + {1 \over 2}(n-1)\right) \,
     \Gamma\left({1 \over 2}(9-n)\right)     \over
     \Gamma\left(l + {1 \over 2}(5-n)\right) \,
     \Gamma\left({1 \over 2}(3+n)\right) } 
\end{equation}
if $n < 3$, $l \ge 2$, and the Sachs-Wolfe effect
dominates the primordial fluctuations
(Bond \& Efstathiou\markcite{be87} 1987). In this case, the
character of the fluctuations is usually described by the
best-fitting index $n$ and 
\begin{equation} 
   Q_{\rm rms-PS} = \left( {5 \over 4 \pi} C_2 \right)^{1/2} 
\end{equation}
which is the mean rms temperature fluctuation expected in the
quadrupole component of the anisotropy averaged over all cosmic
observers and obtained by fitting the correlation function by a flat
spectrum of fluctuations.
\par
For the 4-year COBE DMR data, the best-fitting power spectrum of the
fluctuations has 
$n = 1.2 \pm 0.3$ and $Q_{\rm rms-PS} = 15^{+4}_{-3} \ \muK$
(Gorski \etal\markcite{gor96} 1996), although different analyses of
the data by the 
COBE team give slightly different errors and central values (Wright
\etal\markcite{wr96} 1996; Hinshaw \etal\markcite{hin96} 1996). These
values are consistent with 
the scale-invariant Harrison-Zel'dovich spectrum
(Harrison\markcite{har70} 1970; Zel'dovich\markcite{zel72} 1972;
Peebles \& Yu\markcite{py70} 1970), with $n=1$ ($P(k) \propto k$), and 
hence with the usual picture of random fluctuations growing to form
galaxies and clusters of galaxies following a phase of inflation 
(Starobinsky\markcite{sta80} 1980; Guth\markcite{guth81} 1981;
Bardeen, Steinhardt \& Turner\markcite{bst83} 1983). 
\subsection{Clusters of galaxies and the CMBR}
If the CMBR were undisturbed from the epoch of decoupling, where it
picks up these ``primordial'' anisotropies from structure formation,
to the present, then all perturbations in the background could be
interpreted in terms of early processes in the Universe. If there are
strong interactions between the epoch of decoupling and the present,
then all the perturbations associated with the formation of structure
might have been overwritten by later effects (e.g., from a smoothly
re-ionized and dense intergalactic medium; Tegmark, Silk \&
Blanchard\markcite{tsb94} 1994).
\par
The true appearance of the CMBR lies between these two extremes.
Even away from obvious local structures (such as stars and radio
sources) there are a number of structures in the Universe that can
affect the propagation of radiation. For example, gravitational lenses
redistribute radiation from the epoch of 
recombination. Were this radiation to be isotropic, then there would
be no effect from a static lens. However, a lens would affect the
detailed pattern of anisotropies that are imposed on the CMBR at
recombination, and detailed studies of these anisotropies should
take that effect into account, especially on the smallest angular
scales (e.g., Blanchard \& Schneider\markcite{bs87} 1987;
Sasaki\markcite{sas89} 1989; Watanabe \& Tomita\markcite{wt91}
1991). Even an isotropic radiation field may pick up anisotropies from
lenses, if those lenses are not static. Examples of such effects have
been discussed by Rees \& Sciama\markcite{rs68} (1968),
Dyer\markcite{dyer76} (1976), Nottale\markcite{not84} (1984),
Gott\markcite{gott85} (1985), Gurvits \& Mitrofanov\markcite{gm86}
(1986), and Birkinshaw\markcite{b89} (1989).
\par
These {\it metric} (Rees-Sciama) perturbations of the isotropy of the
background radiation tend to be small, of order the gravitational
lensing angle implied by 
the mass ($\Delta\theta \sim {4 G M / R c^2}$, where $M$ is the object's
mass and $R$ its size or the impact parameter) multiplied by a
dimensionless measure of the extent to which the lens is non-static. 
For example, the fractional intensity change is of order
$\Delta\theta \, (v/c)$ for a lens moving across the line of sight
with velocity $v$. For even the largest masses (of clusters of
galaxies), for which $\Delta\theta \sim 1$~arcmin),
and the largest likely velocities ($\sim 10^3 \ \kms$), the
fractional intensity change $\Delta I_\nu / I_\nu \lesssim 10^{-6}$. 
It is interesting that redshift and angular effects introduced by
{\it spatial} and {\it temporal} metric variations of a perturbing mass are
closely related (Pyne \& Birkinshaw\markcite{pb93} 1993), and can be
fitted into the same formalism as the Sachs-Wolfe effect (Sachs \&
Wolfe\markcite{sw67} 1967), which is the dominant source of
anisotropy in the microwave background radiation on the angular scale
of the COBE experiments.
\par
The most likely sources for metric perturbations of the CMBR are 
clusters of galaxies, which are the most massive well-differentiated
structures in the Universe. However the
structures introduced by metric effects associated with
clusters of galaxies will be very difficult to see because of the
presence of the Sunyaev-Zel'dovich effects, which are also introduced
by clusters, but which are far more intense. 
\par
The basic physics of the Sunyaev-Zel'dovich effect is simple. Clusters
of galaxies have masses that often exceed $3 \times 10^{14} \ \Msol$,
with effective gravitational radii, $R_{\rm eff}$, of order Mpc. Any gas in
hydrostatic equilibrium within a cluster's
gravitational potential well must have electron temperature $\te$ given by
\begin{eqnarray}
   \boltz \te &\approx & {G M \mp \over 2 R_{\rm eff}} \nonumber \\
              &\approx & 7 \, (M/3 \times 10^{14} \, \Msol) \,
               (R_{\rm eff} /\Mpc)^{-1} \qquad \keV .
\end{eqnarray}
At this temperature, thermal emission from the gas appears in the X-ray part
of the spectrum, and is composed of thermal bremsstrahlung and line
radiation. 
\par
About a quarter of the mass of
clusters of galaxies is in the form of distributed gas (e.g., White \&
Fabian\markcite{wf95} 1995; Elbaz, Arnaud \&
B\"ohringer\markcite{eab95} 1995; David \etal\markcite{dav95} 1995;
Dell'Antonio, Geller \& Fabricant\markcite{dgf95} 1995). The density
of the gas is sufficiently high that clusters of galaxies are luminous
X-ray sources (e.g., Figure~\ref{fig-0016pspc};
see the reviews of Forman \& Jones\markcite{fj82} 1982;
Sarazin\markcite{sar88} 1988), with the bulk 
of the X-rays being produced as bremsstrahlung rather than line
radiation. Electrons in the intracluster gas are not only scattered by
ions, but can themselves scatter photons of the CMBR: for these
low-energy scatterings the cross-section is the Thomson scattering
cross-section,
$\sigma_{\rm T}$, so that the scattering optical depth 
$\tau_{\rm e} \approx n_{\rm e} \, \sigma_{\rm T} \, R_{\rm eff} \sim
10^{-2}$. In any
one scattering the frequency of the photon will be shifted slightly,
and up-scattering is more likely. On average a scattering
produces a slight mean change of photon energy 
$(\Delta \nu / \nu) \approx (\boltz\te/m_{\rm e} c^2) \sim 10^{-2}$.
The overall change in brightness of the microwave background radiation
from inverse Compton (Thomson) scattering is therefore about 1~part
in $10^4$, a signal which is about ten times larger than the
cosmological signal in the microwave background radiation detected by
COBE. 
\par
The primordial and Sunyaev-Zel'dovich effects are both detectable, and
can be distinguished by their different spatial distributions.
Sunyaev-Zel'dovich effects are {\it localized}: they are seen towards
clusters of galaxies, which are large-scale structures visible to
redshifts $> 0.5$ in the optical and X-ray bands. Furthermore, the
amplitude of the signal should be related to other observable
properties of the clusters. Primordial structures in the CMBR are {\it
non-localized}: they are not associated with structures seen at other
wavebands, and are distributed at random over the entire sky, with
almost constant correlation amplitude in different patches of sky.
\par
It is on the Sunyaev-Zel'dovich effects that the present review
concentrates. Although the original discussion and detection of the
effects were driven by the question of whether cluster X-ray emission
arose from the hot gas in cluster potential wells or from non-thermal
electrons interacting with magnetic fields or the cosmic background
radiation (Sunyaev \& Zel'dovich\markcite{sz72} 1972),
more recently the effects have been studied for the
information that they can provide on cluster structures, on the
motions of clusters of galaxies relative to the Hubble flow, and on
the Hubble flow itself (and the cosmological constants that
characterize it). The last few years have seen many new detections of
\SZ\ effects from clusters with strong X-ray emission --- and the special
peculiarity of the \SZ\ effects, that they are redshift-independent,
and therefore almost as easy to observe at high as at low redshift,
has been illustrated by detecting clusters as distant as CL~0016+16,
at $z = 0.5455$, or at even higher redshift.
%
%
\section{Radiation basics}
Although the CMBR is close to being an isotropic and thermal radiation
background with simple spectral and angular distributions, it is
useful to recall the formalism needed to deal with a 
general radiation field, since the details of the small perturbations
have great physical significance. The notation used here is similar to
that of Shu\markcite{shu91} (1991), which may be consulted for more detailed
descriptions of the quantities employed.
\par
The state of a radiation field can be described by
distribution functions $f_\alpha({\bf r},{\bf p},t)$, such that
the number of photons in real space volume $d^3r$ about $\bf r$ and
momentum space volume $d^3p$ about $\bf p$ at time $t$ with
polarization $\alpha$ ($= 1$ or $2$) is $f_\alpha \, d^3r \, d^3p$. This
distribution function is related to the photon occupation number in
polarization state $\alpha$, $n_\alpha({\bf r},{\bf p},t)$, by
\begin{equation}
   f_\alpha({\bf r},{\bf p},t) = h^{-3} \, n_\alpha({\bf r},{\bf p},t)
\end{equation}
and to the specific intensity in the radiation,
$I_\nu({\bf \hat k},{\bf r},t)$, by
\begin{equation}
   I_\nu({\bf \hat k},{\bf r},t) = \sum_{\alpha=1}^2 \,
      \left( { h^4 \, \nu^3 \over c^2 } \right) \,
      f_\alpha({\bf r},{\bf p},t)
\end{equation}
where ${\bf \hat k}$ is a unit vector in the direction of the radiation
wavevector, $\nu$ is the photon frequency, and $h$ and $c$
are Planck's constant and the speed of light. The meaning of the
specific intensity is that the energy crossing area element $\bf dS$
in time $dt$ from within solid angle $d\Omega$ about $\bf \hat k$ and
with frequency in the range $\nu$ to $\nu + d\nu$ is
$I_\nu \, ({\bf \hat k . dS}) \, d\Omega \, d\nu \, dt$. 
\par
If the occupation number is of Planck form
\begin{equation}
   n_\alpha = \left( e^{h \nu / \boltz \trad} -1 \right)^{-1}
   \quad \rm for \ \alpha=1,2
   \label{eq-occnum}
\end{equation}
then the radiation field has the form of (\ref{eq-planckspec}).
The number density of photons in the Universe is then
\begin{eqnarray}
  n_\gamma & = & \sum_\alpha \int f_\alpha({\bf p}) \, d^3p 
                 \nonumber \\
           & = & 16 \, \pi \, \zeta(3) \, \left(
                 { \boltz \trad \over h c } \right)^3 \nonumber \\
           & = & (4.12 \pm 0.01) \times 10^8 \quad
                 \rm photons \, m^{-3}
                 \qquad if \ \trad = 2.728 \pm 0.002 \ \rm K
  \label{eq-numphot}
\end{eqnarray}
from which can be calculated the baryon to photon number ratio,
$\eta = {n_{\rm B}/n_{\gamma}} = 2.7 \times 10^{-8} \Omega_{\rm B}
h_{100}^2$. In (\ref{eq-numphot}) $\zeta(x)$ is the Riemann zeta
function ($\zeta(3) \approx 1.202$) and the value of $\trad$ is taken
from a recent analysis of COBE data on the CMBR spectrum (Fixsen
\etal\markcite{fix96} 1996).
\par
Similarly, the energy density of the radiation field is
\begin{eqnarray}
  u_\gamma & = & \sum_\alpha \int h \nu \, 
                 f_\alpha({\bf p} ) \, d^3 p \nonumber \\
           & = & { 8 \, \pi^5 \, h \, c \over 15} \, 
                 \left( {\boltz \trad \over h \, c} \right)^4
                 \nonumber \\
           & = & (4.19 \pm 0.01) \times 10^{-14} \quad
                 \rm J \, m^{-3}
                 \qquad if \ \trad = 2.728 \pm 0.002 \ \rm K .
  \label{eq-enphot}
\end{eqnarray}
It is apparent that the errors on $u_\gamma$ and $n_\gamma$ in
(\ref{eq-numphot}) and (\ref{eq-enphot}) are so small as to have no
significant astrophysical impact, and may safely be dropped.
\par
It is common for the specific intensity of a radiation field to be
described by radio-astronomers in units of brightness temperature,
$T_{\rm RJ}$. This is defined as the temperature of a thermal
radiation field which in
the Rayleigh-Jeans limit of low frequency would have the same
brightness as the radiation that is being described. In the limit of
low frequency (\ref{eq-planckspec}) reduces to 
$I_\nu = 2 \boltz \trad \nu^2 / c^2$, so that
\begin{equation}
   T_{\rm RJ}(\nu) = {c^2 I_\nu \over 2 \boltz \nu^2} \quad . 
\end{equation}
Thus the brightness temperature of a thermal spectrum as described by
(\ref{eq-planckspec}) is frequency-dependent, with a peak value equal
to the radiation temperature at low frequencies, and tending to zero
in the Wien tail.
\par
In the presence of absorption, emission and scattering processes, 
and in a flat spacetime, $I_\nu$ obeys a transport equation 
\begin{equation}
   {1 \over c} \, {\partial I_\nu \over \partial t}
  + {\bf \hat k}.{\bf \nabla} I_\nu 
  =
   j_\nu - \alpha_{\nu,\rm abs} I_\nu
         - \alpha_{\nu,\rm sca} I_\nu
         + \alpha_{\nu,\rm sca} \int \phi_\nu({\bf \hat k},{\bf \hat
                           k'}) \, I_\nu({\bf k'}) d\Omega'
  \label{eq-transeqn}
\end{equation}
where $j_\nu$ is the emissivity along the path (the energy emitted per
unit time per unit frequency per unit volume per unit solid angle),
$\alpha_{\nu,\rm abs}$ is the absorption coefficient (the fractional
loss of intensity of the radiation per unit length of propagation
because of absorption by material in the beam), $\alpha_{\nu,\rm sca}$
is the scattering coefficient (the fractional loss of intensity of the
radiation per unit length of propagation because of scattering by
material in the beam), and $\phi_{\nu}({\bf \hat k},{\bf \hat k'})$ is
the scattering redistribution function --- the probability of a
scattering from direction $\bf \hat k'$ to $\bf \hat k$. The
absorption coefficient is regarded as containing both true absorption
and simulated emission. While this is important in astrophysical
masers, where $\alpha_{\nu,\rm abs}$ is negative, this subtlety will not
affect the discussions in the present review. An important
property of $I_\nu$ that follows from its definition (or
equation~\ref{eq-transeqn}) is that it is conserved in flat spacetimes
in the absence of radiation sources or absorbers. 
\par
The specific intensity of a radiation field may be changed in several
ways. One is to make the photon distribution function anisotropic,
for example by the Doppler effect due to the peculiar motion of the
Earth relative to the sphere of last scattering, which causes the
radiation temperature becomes angle-dependent
\begin{equation}
  T^\prime_{\rm rad}(\theta) 
  = {\trad \over \gamma \left( 1 - {v \over c}
                         \cos\theta \right) },
\end{equation}
but otherwise leaves the form of (\ref{eq-occnum}) unchanged. 
$\gamma = \left( 1 - {v^2 \over c^2} \right)^{-1/2}$ and
$\theta$ is the angle between the line of sight and the
observer's velocity vector (Peebles \& Wilkinson\markcite{pw68} 1968).
The specific intensity may also be changed by redistributing photons
to different directions and frequencies (e.g., by scattering
processes), or by absorbing or emitting radiation (e.g., by thermal
bremsstrahlung). The choice of whether to describe these effects in
the photon distribution function, or in the specific intensity, is
made for reasons of convenience. Although the statistical
mechanics of photon scattering is often related to the
occupation numbers, $n_\alpha$, most astrophysical work is done in
the context of the specific intensity, $I_\nu$.
%
%
\section{Inverse-Compton Scattering\label{sec-scat}}
The theoretical foundation of the Sunyaev-Zel'dovich effect was laid
in the early 1970s (Sunyaev \& Zel'dovich\markcite{sz70} 1970), but is
based on earlier work on the interactions of photons and free electrons
(Kompaneets\markcite{kom56} 1956; Dreicer\markcite{dr64} 1964;
Weymann\markcite{wey65} 1965). Excellent recent reviews of the physics
of the Sunyaev-Zel'dovich effect have been given by Bernstein \&
Dodelson\markcite{bd90} (1990) and Rephaeli\markcite{rep95b} (1995b),
while discussions of the more general problem of comptonization of a
radiation field by passage through an ionized gas have been given by
Blumenthal \& Gould\markcite{bg70} (1970), Sunyaev \&
Zel'dovich\markcite{sz80a} (1980a), Pozdnyakov, Sobol' \&
Sunyaev\markcite{pss83} (1983), and Nagirner \&
Poutanen\markcite{np94} (1994). Comptonization is also an essential
ingredient in the discussion of the X-ray and gamma-ray 
emission of active galactic nuclei (see, for example, Zbyszewska
\& Zdziarski\markcite{zz91} 1991; Zdziarski \etal\markcite{zzsb93}
1993; Skibo \etal\markcite{skib95} 1995). The present section relies
heavily on this work, and on the material on inverse-Compton
scatterings in Rybicki \& Lightman\markcite{rl80} (1980), and the 
papers by Wright\markcite{wr79} (1979) and Taylor \&
Wright\markcite{tw89} (1989).
\subsection{Single photon-electron scattering}
\begin{figure}[p]
\epsscale{0.8}
\plotone{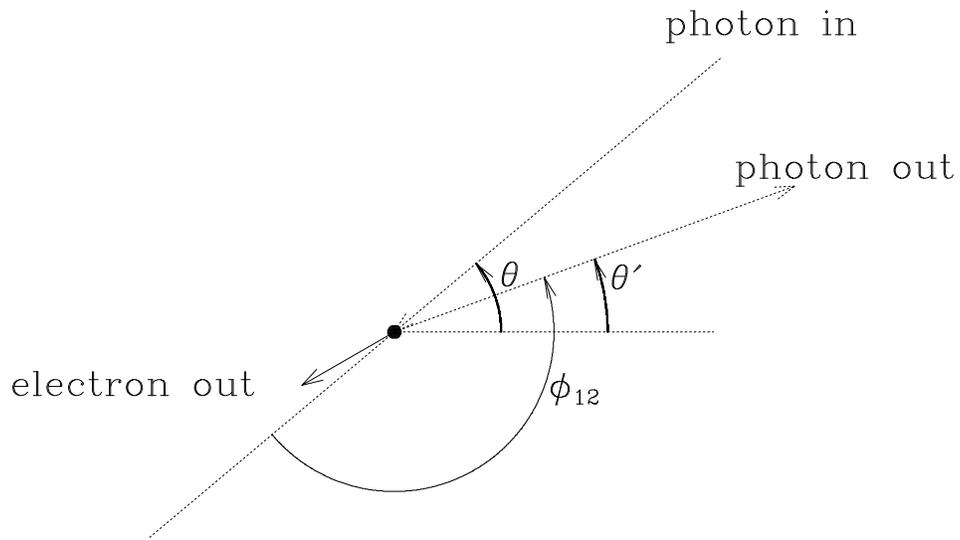}
\caption{\label{fig-scatgeom}
The scattering geometry, in the frame of
rest of the electron before the interaction. An incoming photon, at
angle $\theta$ relative to the $x_{\rm e}$ axis, is deflected by angle
$\phi_{12}$, and emerges after the scattering at
angle $\theta^\prime$ with almost unchanged energy
(equation~\ref{eq-comscat}). In the observer's frame,
where the electron is moving with velocity $\beta c$ along the 
$x_{\rm e}$ axis, the photon changes energy by an amount depending on
$\beta$ and the angles $\theta$ and $\theta^\prime$
(equation~\ref{eq-frqshift}).
}
\end{figure}

When a photon is scattered by an electron, the energy and direction of
motion of both the photon and the electron are usually altered. The
change in properties of the photon is described by the usual Compton
scattering formula
\begin{equation}
   \epsilon^\prime = { \epsilon \over
                 1 + {\epsilon \over m_{\rm e} c^2} \,
                 (1 - \cos\phi_{12}) }
                 \label{eq-comscat}
\end{equation}
where the electron is assumed to be at rest before the interaction,
$\epsilon$ and $\epsilon^\prime$ are the photon energies before and
after the interaction, and $\phi_{12}$ is the angle by which the
photon in deflected in the encounter (see Fig.~\ref{fig-scatgeom}).
\par
For low-energy photons and mildly relativistic or non-relativistic
electrons, $\epsilon \ll m_{\rm e} c^2$ and the scattering is
almost elastic ($\epsilon^\prime = \epsilon$). This limit
is appropriate for the scatterings in clusters of galaxies that
cause the Sunyaev-Zel'dovich effect, and causes a considerable
simplification in the physics. Although the scatterings are usually
still referred to as inverse-Compton processes, they might better be
described as Thomson scatterings in this limit.
\par
Scatterings of this type will also cause Sunyaev-Zel'dovich effects
from the relativistic plasma of radio galaxies. The lobes of radio
galaxies emit strong synchrotron radiation, and must contain
electrons with Lorentz factors $\gamma_{\rm e} \gtrsim 10^8$. In the
rest frames of such electrons the microwave background
radiation appears to have a peak at photon energies 
$\sim 0.1 m_{\rm e} c^2$, and the assumption of elastic scattering
will be inappropriate. Little theoretical work has been done on the
spectrum of the scattered radiation in this limit, but see
Section~\ref{sec-nonthermal}.
\par
In this thermal scattering limit, 
the interaction cross-section for a microwave background photon
with an electron need not be described using
the Klein-Nishina formula, 
\begin{equation}
  {d\sigma \over d\Omega} = {r_{\rm e}^2 \over 2} \,
  \left( {\epsilon^\prime \over \epsilon} \right)^2
  \left( {\epsilon \over \epsilon^\prime} +
         {\epsilon^\prime \over \epsilon} -
         \sin^2 \phi_{12} \right)
\end{equation}
but rather the classical Thomson cross-section formula which results
in the limit $\epsilon^\prime \rightarrow \epsilon$. Then if the
geometry of the collision process in the electron rest frame is as
shown in Fig.~\ref{fig-scatgeom}, the probability of a scattering with
angle $\theta$ is 
\begin{eqnarray}
   p(\theta) \, d\theta & = & p(\mu) \, d\mu \nonumber \\
                        & = & \left( 2 \gamma^4 ( 1 - \beta \mu )^3
                           \right)^{-1} \, d\mu
   \label{eq-pthetadef}
\end{eqnarray}
where the electron velocity $v_{\rm e} = \beta c$, and
$\mu = \cos\theta$. The probability of a scattering to angle
$\theta^\prime$ is
\begin{equation}
   \phi(\mu^\prime; \mu) d\mu^\prime = {3 \over 8} \left(
   1 + \mu^2 {\mu^\prime}^2 + {1 \over 2}(1 - \mu^2)(1 - {\mu^\prime}^2 ) 
   \right) \, d\mu^\prime
   \label{eq-scatprob}
\end{equation}
(Chandrasekhar\markcite{ch50} 1950; Wright\markcite{wr79} 1979), and
the change of photon direction causes the scattered photon to appear
at frequency 
\begin{equation}
   \nu^{\prime\prime} = \nu \, (1 + \beta\mu^\prime ) \,
                            (1 - \beta \mu)^{-1}
   \label{eq-frqshift}
\end{equation}
with $\mu^\prime = \cos\theta^\prime$. 
\par
\begin{figure}[p]
\epsscale{0.8}
\plotone{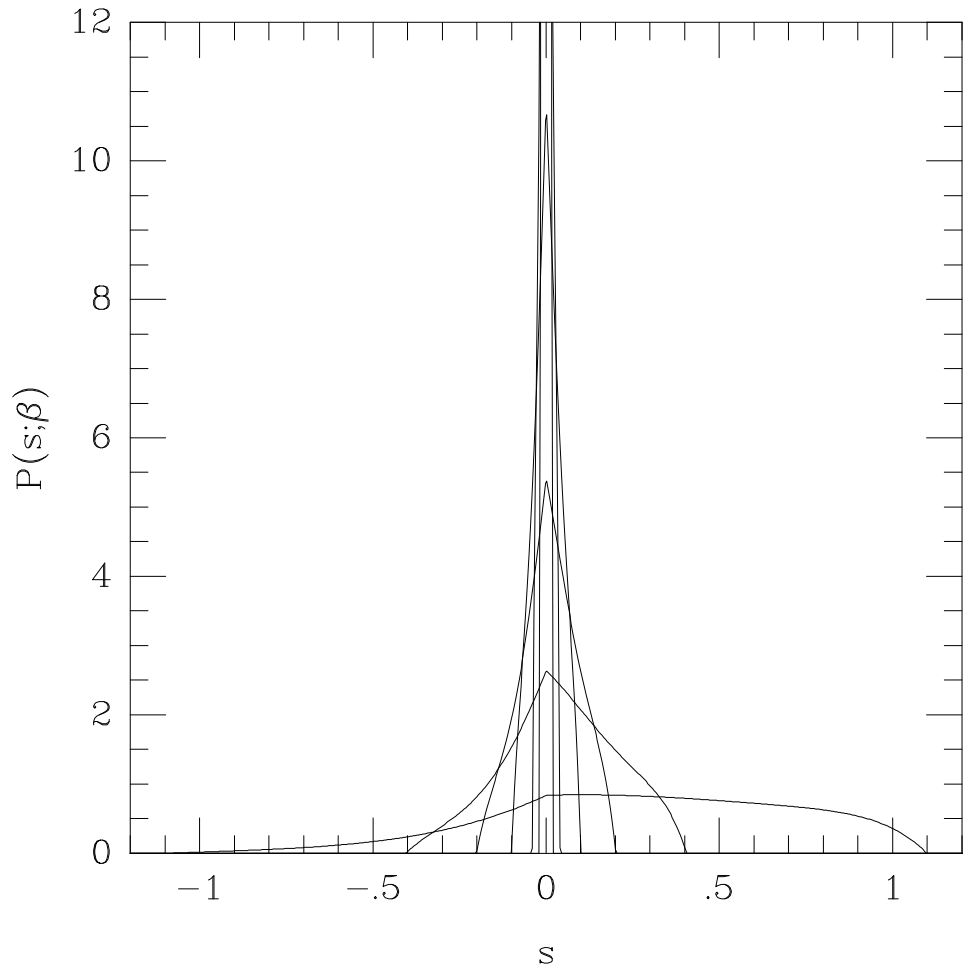}
\caption{\label{fig-scatprob}
The scattering probability function
$P(s;\beta)$, for $\beta = 0.01$, $0.02$, $0.05$, $0.10$, $0.20$, and
$0.50$. The function becomes increasingly asymmetric and broader as
$\beta$ increases.}
\end{figure}

It is conventional (Wright\markcite{wr79} 1979;
Sunyaev\markcite{sun80} 1980; Rephaeli\markcite{rep95b} 1995b) to 
express the resulting scattering in terms of the logarithmic frequency
shift caused by a scattering, $s$ (Sunyaev uses $u$ for a related
quantity), 
\begin{equation}
   s = \log(\nu^{\prime\prime} / \nu)
   \label{eq-sdef}
\end{equation}
when the probability that a single scattering of the photon causes a
frequency shift $s$ from an electron with speed $\beta c$ is
\begin{equation}
   P(s;\beta) \, ds = \int p(\mu) \, d\mu \,
                      \phi(\mu^\prime;\mu) \, 
                      \left({ d\mu^\prime \over ds} \right) \, ds. 
\end{equation}
Using (\ref{eq-pthetadef}--\ref{eq-frqshift}), this becomes
\begin{equation}
   P(s;\beta) = {3 \over 16 \gamma^4 \beta} \,
   \int_{\mu_1}^{\mu_2} \, (1 + \beta\mu^\prime) \, 
   \left( 1 + \mu^2 {\mu^\prime}^2 + {1 \over 2}(1 - \mu^2)
   (1 - {\mu^\prime}^2) 
   \right) (1 - \beta \mu)^{-3} \, d\mu
   \label{eq-psbetadef}
\end{equation}
where $\mu^\prime$ can be expressed in terms of $\mu$ and $s$ as
\begin{equation}
   \mu^\prime = {e^s(1 - \beta\mu)  - 1 \over \beta}
\end{equation}
(from equations~\ref{eq-frqshift} and~\ref{eq-sdef}), and the integral is
performed only over real angles, so that 
\begin{eqnarray}
 \mu_1 & = & \cases{
                 -1                                & $s \le 0$ \cr
                 {1 - e^{-s}(1+\beta) \over \beta} & $s \ge 0$ \cr
          } \\
 \mu_2 & = & \cases{
                 {1 - e^{-s}(1-\beta) \over \beta} & $s \le 0$ \cr
                  1                                & $s \ge 0$ \cr
          }
\end{eqnarray}
in (\ref{eq-psbetadef}). The integration can be done easily, 
and Fig.~\ref{fig-scatprob} shows the resulting function for several
values of $\beta$. The increasing asymmetry of $P(s;\beta)$ as $\beta$
increases is caused by relativistic beaming, and the width of the
function to zero intensity in $s$, 
\begin{equation}
   \Delta s_0 = 2 \log\left({1+\beta \over 1-\beta}\right),
\end{equation}
increases because increasing $\beta$ causes the
frequency shift related to a given photon angular deflection to
increase.
\subsection{Scattering of photons by an electron
 population\label{sec-elpopscat}} 
The distribution of photon frequency shifts caused by scattering by a
population of electrons is calculated from $P(s;\beta)$ by
averaging over the electron $\beta$ distribution. Thus for photons
that have been scattered once, the probability distribution of $s$,
$P_1(s)$, is given by
\begin{equation}
   P_1(s) = \int_{\beta_{\rm lim}}^1 \, 
            p_{\rm e}(\beta) \, d\beta \, P(s;\beta) 
   \label{eq-p1sdef}
\end{equation}
where $\beta_{\rm lim}$ is the minimum value of $\beta$ capable of
causing a frequency shift $s$, 
\begin{equation}
   \beta_{\rm lim} = {e^{|s|} - 1 \over e^{|s|} +1}.
\end{equation}
\par
The limitations of equation~(\ref{eq-p1sdef}) are evident from the
assumptions made to derive equation~(\ref{eq-psbetadef}). That is, the
electron distribution
$p_{\rm e}(\beta)$ must not extend to sufficiently large Lorentz factors,
$\gamma$, that the assumptions of elastic scattering with the Thomson
scattering cross-section are violated. For photons of the microwave
background radiation these assumptions are amply satisfied provided
that $\gamma \lesssim 2 \times 10^9$. In clusters of galaxies the
typical electron temperatures may be as much as 15~keV ($1.8 \times
10^8 \ \K$), but the corresponding Lorentz factors are still small, so
that we may ignore relativistic corrections to the scattering
cross-section. 
\par
If the electron velocities are assumed to follow a relativistic
Maxwellian distribution, 
\begin{equation}
   p_{\rm e}(\beta) \, d\beta = {\gamma^5 \, \beta^2 \, 
   \exp \left( - {\gamma \over \Theta}
   \right) \, d\beta \over \Theta \, K_2({1 \over \Theta})}, 
   \label{eq-thermale}
\end{equation}
where $\Theta$ is the dimensionless electron temperature
\begin{equation}
   \Theta =
        \left( {k_{\rm B} \, T_{\rm e} \over m_{\rm e} \, c^2} \right)
\end{equation}
and $K_2(z)$ is a modified Bessel function of the second kind
and second order, then the resulting distribution of photon frequency
shift factors can be calculated by a numerical integration of 
equation~(\ref{eq-p1sdef}).
\par
\begin{figure}[p]
\epsscale{0.8}
\plottwo{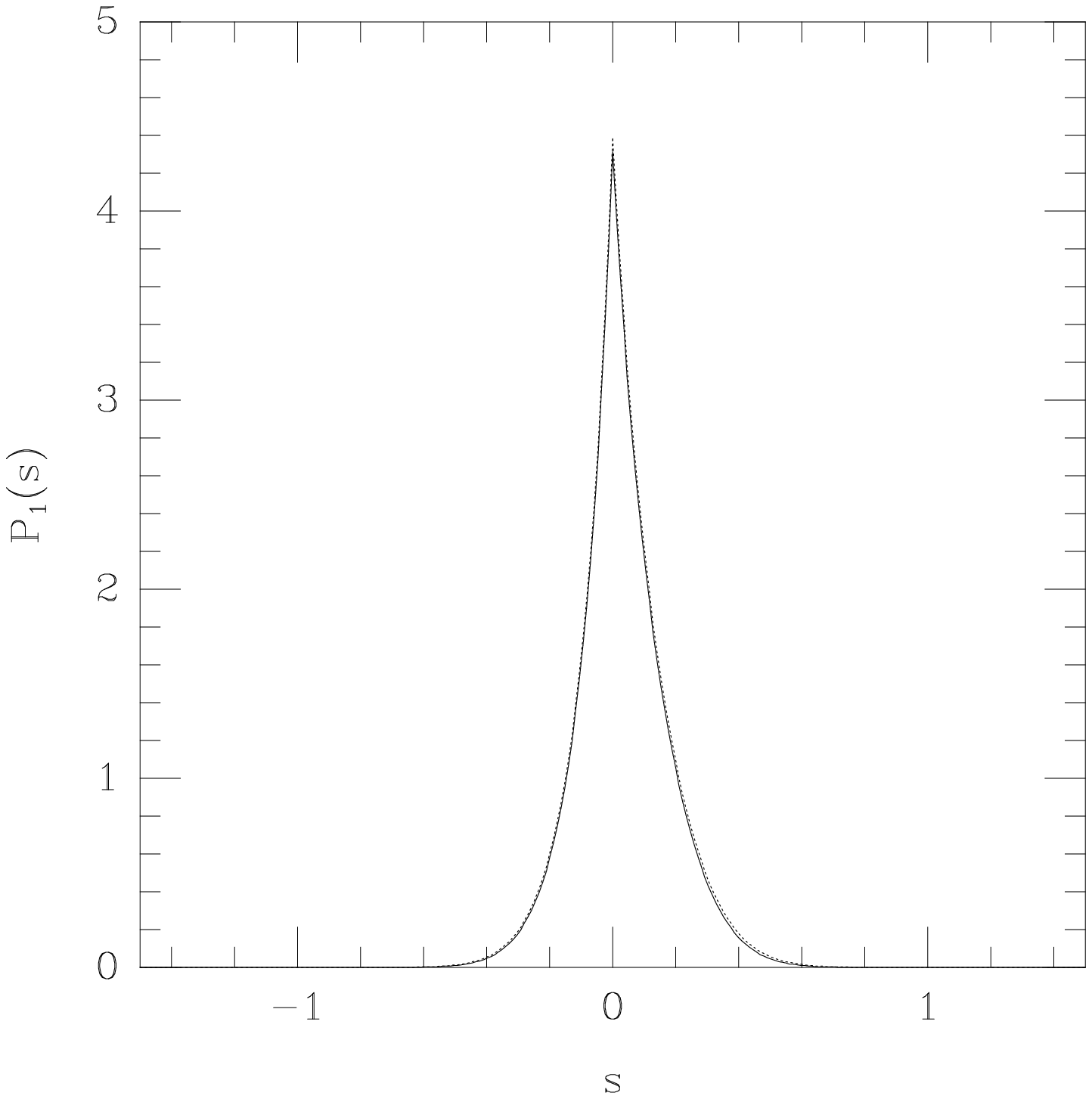}{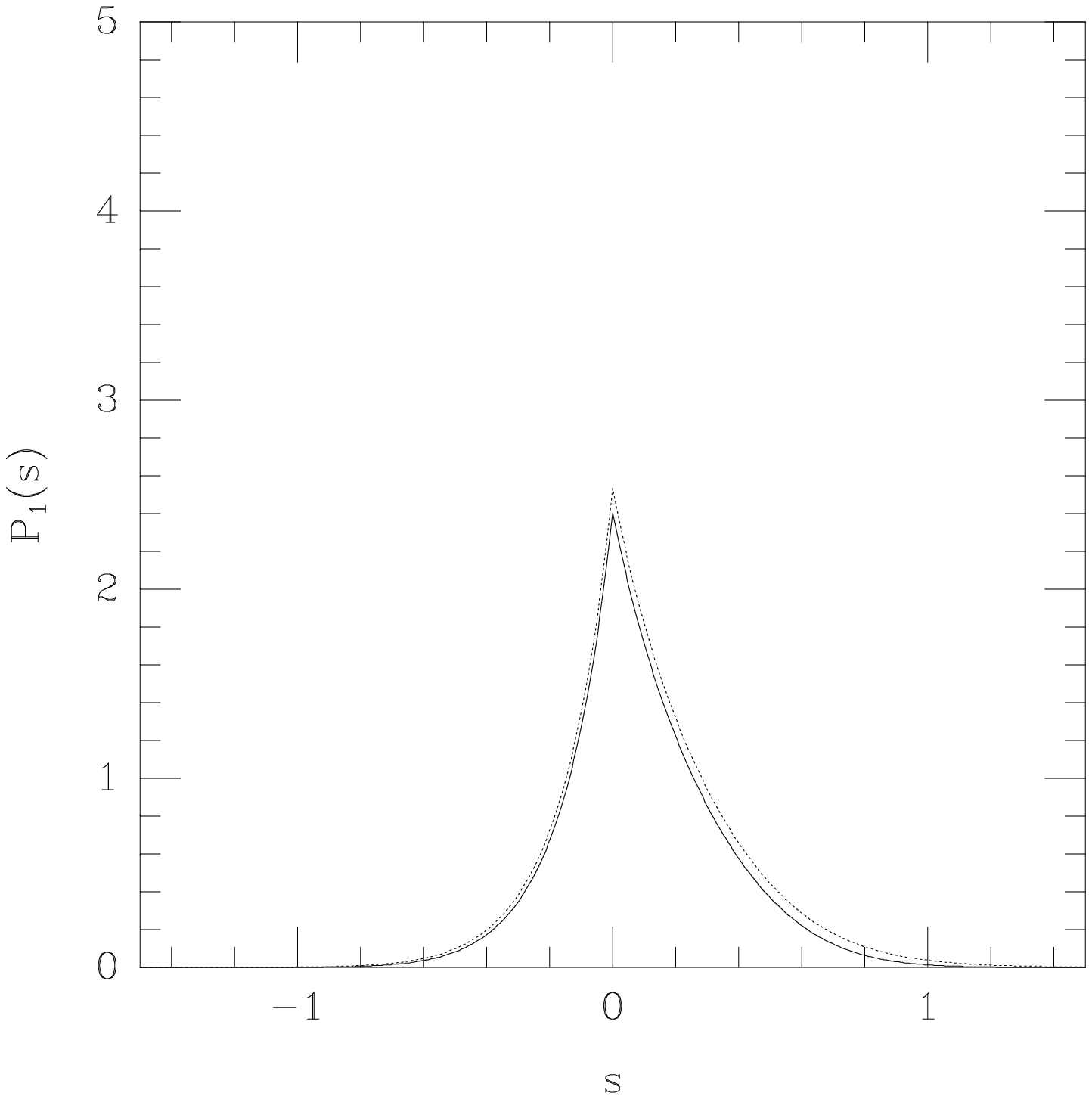}
\caption{\label{fig-p1func}
The scattering kernel, $P_1(s)$, for
gases at 5.1~keV and 15.3~keV. The solid line shows the scattering
kernels calculated according to (\ref{eq-p1sdef}), as derived by
Rephaeli (1995a). The dotted line indicates the scattering kernels as
calculated by Sunyaev (1980), based on the results of Babuel-Peyrissac
\& Rouvillois (1969).}
\end{figure}

The result of performing this calculation
for $k_{\rm B} T_{\rm e} = 5.1$ and 15.3~keV
is shown in Fig.~\ref{fig-p1func}, where it is compared with the result
given by Sunyaev\markcite{sun80} (1980). It can be seen that
the distribution of scattered photon frequencies is significantly
asymmetric, with a stronger upscattering ($s > 1$) tail than a
downscattering tail. This is the origin of the mean frequency increase
caused by scatterings. As the temperature of the
electron distribution increases, this upscattering tail increases in
strength and extent. Sunyaev's\markcite{sun80} (1980) distribution
function tends to have a stronger tail at large values of $s$ and a
larger amplitude near $s=0$ than does the form derived using
(\ref{eq-p1sdef}).
\par
\begin{figure}[p]
\epsscale{0.8}
\plotone{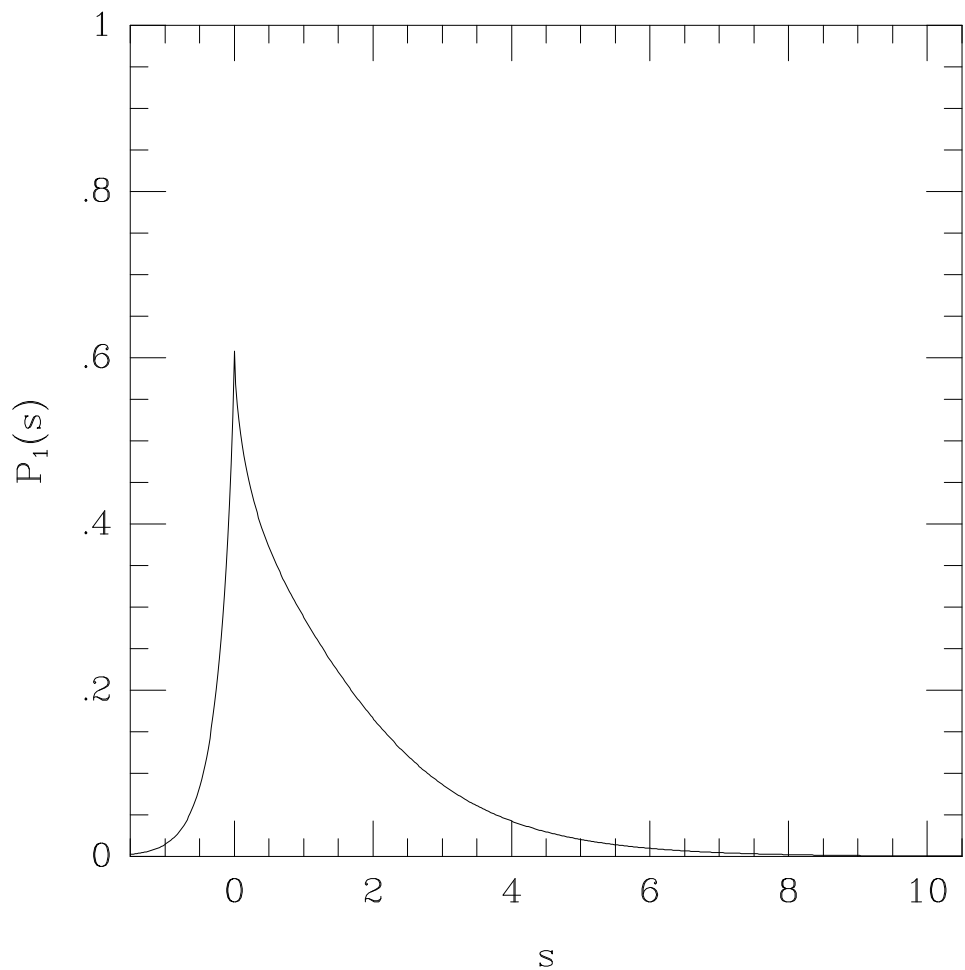}
\caption{\label{fig-p1power}
The scattering kernel, $P_1(s)$, for
a power-law electron distribution with energy index $\alpha=2.5$ (see
equation~\ref{eq-powere}). The stronger upscattering tail here, relative to
Fig.~\ref{fig-p1func}, is caused by the higher proportion of fast electrons in
distribution~(\ref{eq-powere}) than~(\ref{eq-thermale}).}
\end{figure}

It is also of interest to calculate the form of $P_1(s)$ for a
power-law distribution of electron energies in some range of Lorentz
factors $\gamma_1$ to $\gamma_2$
\begin{equation}
   p_{\rm e}(\gamma) d\gamma = \cases{
          A \, \gamma^{-\alpha} \, d\gamma
          & $\gamma_1 \le \gamma \le \gamma_2$ \cr
          0
          & otherwise \cr }
          \label{eq-powere}
\end{equation}
with normalizing constant
\begin{equation}
   A = \cases {
           \log \gamma_2 - \log \gamma_1 
           & $\alpha = 1$ \cr
           (1 - \alpha) \, \left(
           \gamma_2^{1-\alpha} - \gamma_1^{1-\alpha} \right)^{-1}
           & $\alpha \ne 1$ \cr
       }
\end{equation}
since such a population, which might be found in a radio galaxy lobe,
can also produce a Sunyaev-Zel'dovich effect. Synchrotron emission
from radio galaxies has a range of spectral indices, but values of
$\alpha \approx 2.5$ are common. Thus Fig.~\ref{fig-p1power}
shows the result of a calculation for an electron population with
$\alpha = 2.5$. As might be expected, the upscattering
tail is much more prominent in Fig.~\ref{fig-p1power} than in 
Fig.~\ref{fig-p1func}, since there are more electrons with 
$\gamma \gg 1$ in distribution~(\ref{eq-powere}) than in
distribution~(\ref{eq-thermale}) for the values of $\Theta$ and
$\alpha$ chosen.
\subsection{Effect on spectrum of radiation\label{sec-effectspectrum}}
Finally it is necessary to use the result for the frequency shift in
a single scattering to calculate the form of the scattered spectrum of
the CMBR. If every photon in the incident spectrum, 
\begin{equation}
   I_0(\nu) = {2 \, h \, \nu^3 \over c^2} \,
           \left( e^{h\nu/k_{\rm B}\trad} - 1 \right)^{-1},
\end{equation}
is scattered once, then the resulting spectrum is given by
\begin{equation}
   {I(\nu) \over \nu} = \int_0^{\infty} d\nu_0 \, P_1(\nu,\nu_0) \,
                       {I_0(\nu_0) \over \nu_0}
\end{equation}
where $P_1(\nu,\nu_0)$ is the probability that a scattering occurs from
frequency $\nu_0$ to $\nu$, and $I(\nu)/h\nu$ is the spectrum in photon
number terms. Since $P_1(\nu,\nu_0) = P_1(s)/\nu$, where $P_1(s)$ is
the frequency shift function in (\ref{eq-p1sdef}), this can be
rewritten as a convolution in $s = \ln (\nu/\nu_0)$, 
\begin{equation}
   I(\nu) = \int\limits_{-\infty}^{\infty} P_1(s) \, I_0(\nu_0) \, ds
            \quad .
   \label{eq-scatconvol}
\end{equation}
The change in the radiation spectrum at frequency $\nu$ is then
\begin{eqnarray}
 \Delta I (\nu) & \equiv & I(\nu) - I_0(\nu) \nonumber \\
                & =      & {2 h \over c^2} \,
           \int\limits_{-\infty}^{\infty} P_1(s) \, ds \,
           \left( {\nu_0^3 \over e^{h\nu_0/\boltz\trad} - 1}
           -
                  {\nu^3   \over e^{h\nu  /\boltz\trad} - 1}\right)
 \label{eq-diint}
\end{eqnarray}
where the normalization of $P_1(s)$ has allowed the 
(trivial) integral over $I_0(\nu)$ to be included in (\ref{eq-diint})
to give a form that is convenient for numerical calculation.
\par
\begin{figure}[p]
\epsscale{0.8}
\plottwo{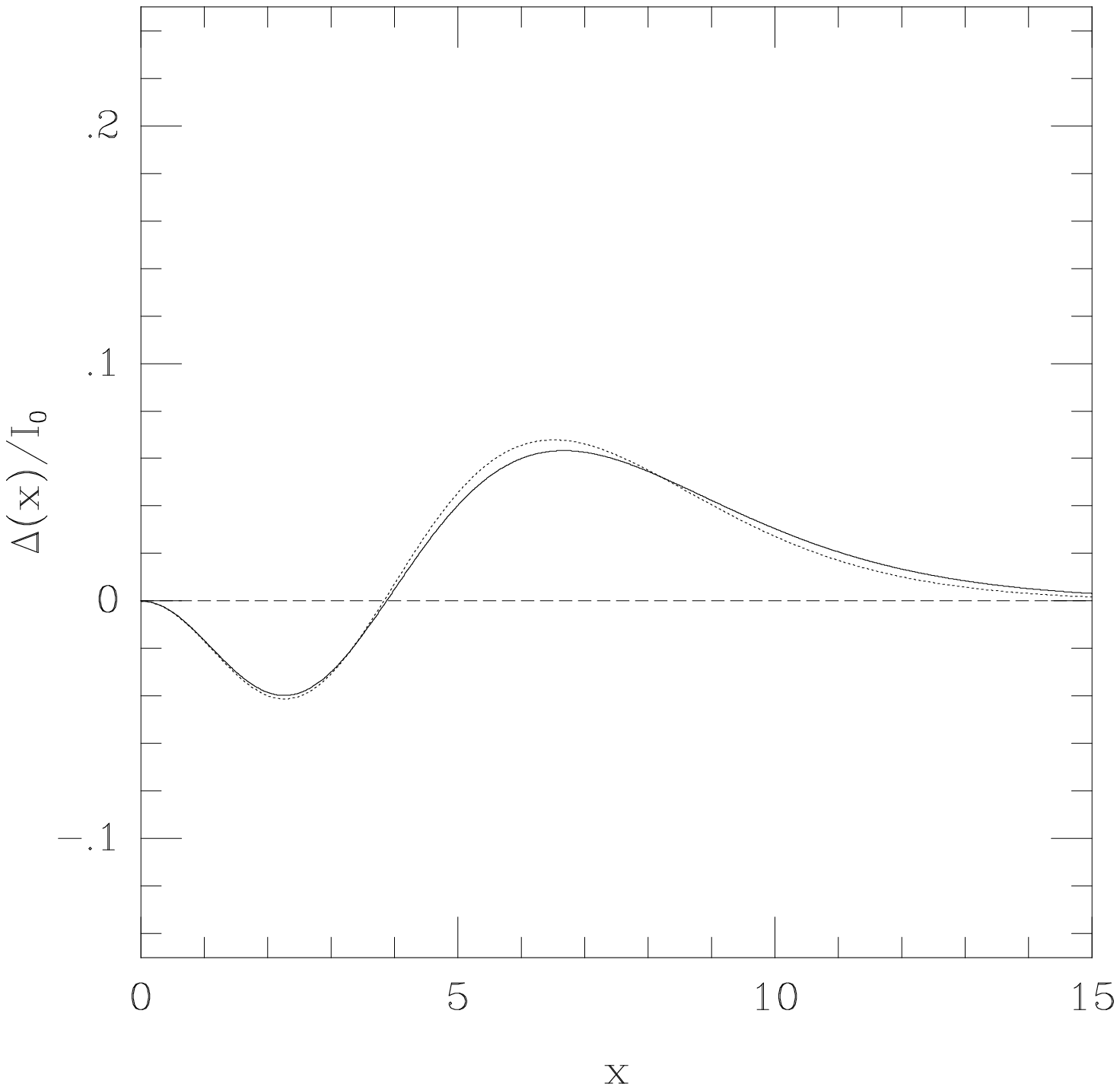}{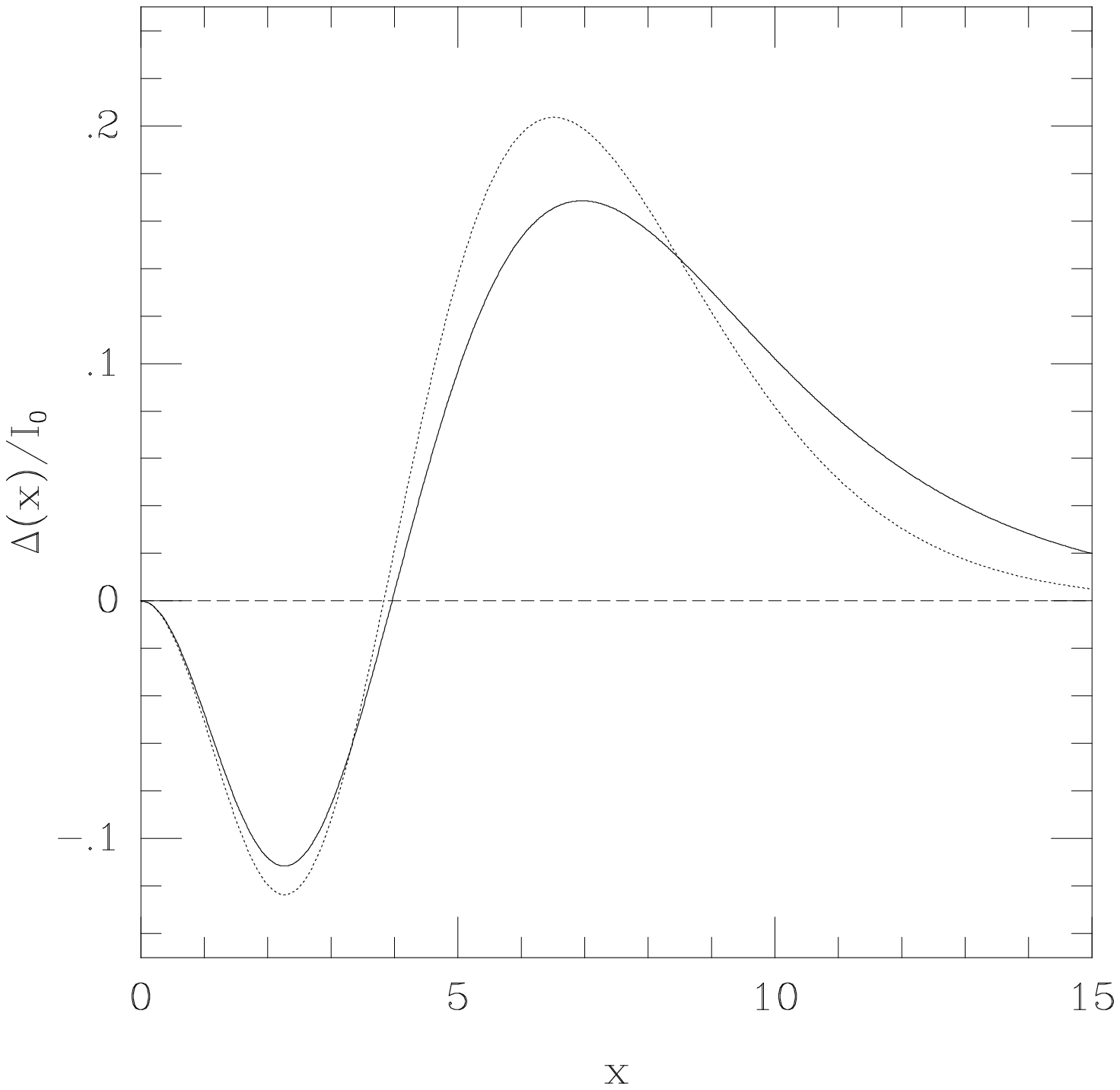}
\caption{\label{fig-dithermal}
The spectral deformation
caused by inverse-Compton scattering of an incident Planck spectrum
after a single scattering from a thermal population of
electrons as a function of dimensionless frequency
$x = h\nu/\boltz \trad = 0.0176(\nu/{\rm GHz})$, with scaling
$I_0 = {2 h \over c^2} \left( {\boltz \trad \over h} \right)^3$.
Left, for electrons at $\boltz\te = 5.1$~keV;
right for electrons at $\boltz\te = 15.3$~keV. 
The result obtained from the Kompaneets kernel is shown as a dotted
line. The shape of the distortion is independent of $\te$ (and the
amplitude is proportional to $\te$) for the Kompaneets kernel, but the
relativistic expression leads to a more complicated form.}
\end{figure}
\begin{figure}[p]
\epsscale{0.8}
\plotone{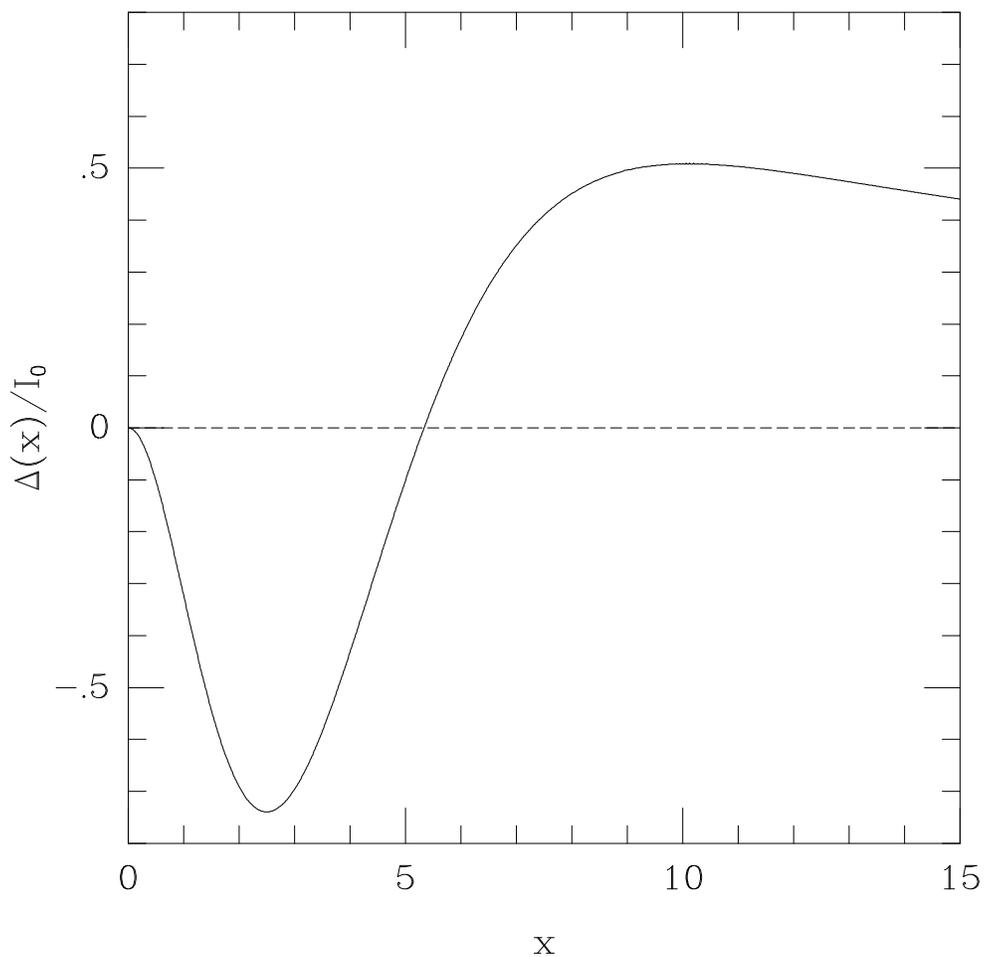}
\caption{\label{fig-dipower}
The fractional spectral deformation
caused by inverse-Compton scattering of an incident Planck spectrum by
a single scattering from a power-law population of electrons 
with $\alpha = 2.5$ (equation~\ref{eq-powere}). The spectral deformation has a
similar shape to that seen in Fig.~\ref{fig-dithermal}, but with a deeper
minimum and more extended tail. This arises from the larger frequency
shifts caused by the higher Lorentz factors of the electrons (see
Fig.~\ref{fig-p1power}).}
\end{figure}

The integrations in (\ref{eq-scatconvol}) or (\ref{eq-diint}) are
performed using 
the $P_1(s)$ function appropriate for the spectrum of the scattering
electrons. The results are shown in Fig.~\ref{fig-dithermal}
and~\ref{fig-dipower} for two
temperatures of the electron gas and for the power-law electron
distribution. In these figures, $x = h\nu/\boltz\trad$ is a
dimensionless frequency.
The functions $\Delta I(x)$ show broadly similar features for thermal
or non-thermal electron distributions: a decrease in
intensity at low frequency (where the mean upward shift of the photon
frequencies caused by scattering causes the Rayleigh-Jeans part of the
spectrum to shift to higher frequency, and hence to show an intensity
decrease: see Fig.~\ref{fig-cmbspec}) 
and an increase in intensity in the Wien part
of the spectrum. The detailed shapes of the spectra differ
because of the different shapes of the scattering functions
$P_1(s)$ (Figs.~\ref{fig-p1func} and~\ref{fig-p1power}). 
\par
More generally, a photon entering the electron distribution may be
scattered 0, 1, 2, or more times by encounters with the electrons. If the
optical depth to scattering through the electron cloud is 
$\tau_{\rm e}$, then the probability that a photon penetrates the cloud
unscattered is $e^{-\tau_{\rm e}}$, the probability that it is once
scattered is $\tau_{\rm e} e^{-\tau_{\rm e}}$, and in general the
probability of N scatterings is 
\begin{equation}
   p_{\rm N} = {\tau_{\rm e}^N e^{-\tau_{\rm e}} \over N!}
\end{equation}
and the full frequency redistribution function from scattering is 
\begin{equation}
   P(s) = e^{-\tau_{\rm e}} \left( \delta(s)
                                  + \tau_{\rm e} \, P_1(s)
                                  + {1 \over 2!} \tau_{\rm e}^2 P_2(s)
                                  + \ldots \right). 
   \label{eq-generalps}
\end{equation}
The redistribution function $P_{\rm n}(s)$ after $n$ scatterings is
given by a repeated convolution
\begin{eqnarray}
 P_{\rm 2}(s) & = & \int dt_1 P_1(t_1) P_1(s-t_1) \nonumber \\
 P_{\rm 3}(s) & = & \int dt_1 dt_2 P_1(t_1) P_1(t_2) P_1(s-t_1-t_2) \\
 \vdots       &   & \nonumber
\end{eqnarray}
but as pointed out by Taylor \& Wright\markcite{tw89} (1989), the
expression for $P(s)$ can be written in much simpler form using
Fourier transforms, with $P(s)$ obtained by the back transform
\begin{equation}
   P(s) = {1 \over \sqrt{2\pi}} \, \int_{-\infty}^{\infty} 
            \tilde P(k) \, e^{iks} \, ds 
\end{equation}
of
\begin{equation}
   \tilde P(k) = e^{-\tau_{\rm e} (\tilde P_1(k) -1 ) }
\end{equation}
where the Fourier transform of $P_1(s)$ is
\begin{equation}
   \tilde P_1(k) = {1 \over \sqrt{2\pi}} \, \int_{-\infty}^{\infty} 
            P_1(s) \, e^{-iks} \, ds \quad . 
\end{equation}
The generalization of (\ref{eq-scatconvol}) for an arbitrary
optical depth is then 
\begin{equation}
   I(\nu) = \int\limits_{-\infty}^{\infty} P(s) \, I_0(\nu_0) \, ds 
   \label{eq-generalkernel}
\end{equation}
but this full formalism will rarely be of interest, since in most
situations the electron scattering medium is optically thin, with
$\tau_{\rm e} \ll 1$, so that the approximation
\begin{equation}
   P(s) = (1 - \tau_{\rm e})\, \delta(s) + \tau_{\rm e} \, P_1(s) 
\end{equation}
will be sufficient (but see Molnar \& Birkinshaw\markcite{mb98b}
1998b). The resulting intensity change has the form shown in
Fig.~\ref{fig-dithermal} or~\ref{fig-dipower}, but with an amplitude
reduced by a factor $\tau_{\rm e}$. This is given explicitly as 
\begin{equation}
   \Delta I(\nu) = {2 h \over c^2} \, \tau_{\rm e} \,
                   \int\limits_{-\infty}^{\infty} P_1(s) \, ds \,
                   \left( {\nu_0^3 \over e^{h\nu_0/\boltz\trad} - 1}
                   -
                          {\nu^3   \over e^{h\nu  /\boltz\trad} - 1}
                   \right).
   \label{eq-deltaIintegral}
\end{equation}
and this form of $\Delta I(\nu)$ will be used extensively later. One
important result is already clear from (\ref{eq-deltaIintegral}): the
intensity change caused by the \SZ\ effect is redshift-independent,
depending only on intrinsic properties of the scattering medium
(through the $\tau_{\rm e}$ factor and $P_1(s)$), and the \SZ\ effect
is therefore a remarkably robust indicator of gas properties at a
wide range of redshifts.
\subsection{The Kompaneets approximation}
The calculations that led to equation (\ref{eq-deltaIintegral}) are
accurate to the appropriate order in photon frequency and
electron energy for our purposes, and take account of the relativistic
kinematics and statistics of the scattering process. In the
non-relativistic limit the scattering process simplifies
substantially, and may be described by the Kompaneets\markcite{kom56}
(1956) equation 
\begin{equation}
   {\partial n \over \partial y } =
   {1 \over x_{\rm e}^2} \, {\partial \over \partial x_{\rm e}} \,
   x_{\rm e}^4 \, \left( {\partial n \over \partial x_{\rm e}}
          + n + n^2 \right)
   \label{eq-kompaneets}
\end{equation}
which describes the change in the occupation number, $n({\nu})$ by a
diffusion process. In (\ref{eq-kompaneets}), $x_{\rm e} = {h \nu /
\boltz \te}$, which should not be confused with $x = {h \nu / \boltz
\trad}$ used previously, and 
\begin{equation}
   y = {\boltz \te \over m_{\rm e} c^2} \, 
       {c \, t \over \lambda_{\rm e}}
   \label{eq-ydef}
\end{equation}
is a dimensionless measure of time spent in the electron
distribution.\footnote{(\ref{eq-ydef}) corrects equation (11) in
the Kompaneets paper for an obvious typographical error.}
$\lambda_{\rm e}$ is the ``Compton range'', or the
scattering mean free path, $(n_{\rm e} \sigma_{\rm T})^{-1}$.
For a radiation field passing though an electron cloud, $y$, which is
usually known as the Comptonization parameter, can be
rewritten in the more usual form 
\begin{equation}
   y = \int \, n_{\rm e} \, \sigma_{\rm T}\, dl \,
       {\boltz \te \over m_{\rm e} c^2} \quad .
\end{equation}
\par
Note that a time- (or $y$-) independent solution of
(\ref{eq-kompaneets}) is given when the 
electrons and photons are in thermal equilibrium, so that
$n = (e^{x_{\rm e}} -1 )^{-1}$, as expected, and that the more general
Bose-Einstein distributions $n = (e^{x_{\rm e}+\alpha}-1)^{-1}$ are also
solutions. A derivation of equation (\ref{eq-kompaneets}) from the
Boltzmann equation is given by Bernstein \& Dodelson\markcite{bd90} (1990). 
\par
\begin{figure}[p]
\epsscale{0.8}
\plotone{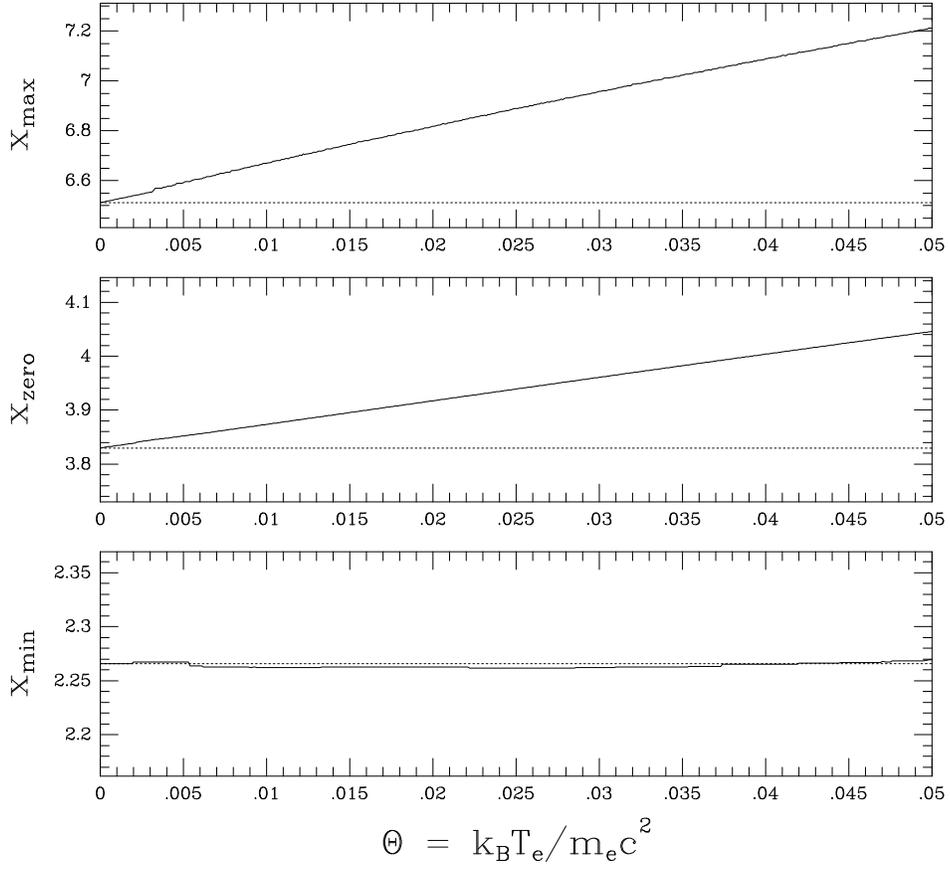}
\caption{\label{fig-zerothermal}
The variation of the positions of the
minimum, zero, and maximum of the spectrum of the thermal
Sunyaev-Zel'dovich effect, $\Delta I(x)$, as the electron temperature
varies. The positions of the spectral features are described by the
dimensionless frequency $x = h\nu/\boltz\trad$, and the electron
temperature is characterized by $\Theta = \boltz\te/m_{\rm e}c^2$.}
\end{figure}

In the limit
of small $x_{\rm e}$, which is certainly appropriate for the CMBR and hot
electrons, ${\partial n \over \partial x_{\rm e}} \gg n$, $n^2$, and
(\ref{eq-kompaneets}) becomes
\begin{equation}
   {\partial n \over \partial y } =
   {1 \over x_{\rm e}^2} \, {\partial \over \partial x_{\rm e}} \,
   x_{\rm e}^4 \, {\partial n \over \partial x_{\rm e}}. 
   \label{eq-kompaneetslimit}
\end{equation}
The homogeneity of the right hand side of this equation allows us to
replace $x_{\rm e}$ by $x$, and a change of variables from 
$x, y$ to $\xi,y$, where $\xi = 3y + \ln x_{\rm e}$, reduces 
(\ref{eq-kompaneetslimit}) to
the canonical form of the diffusion equation,
\begin{equation}
   \left( {\partial n \over \partial y } \right) =
          {\partial^2 n \over \partial \xi^2}.
\end{equation}
In the format of equation (\ref{eq-generalkernel}), this indicates
that the solution of the Kompaneets equation can be written
\begin{equation}
   I(\nu) = \int\limits_{-\infty}^{\infty} P_{\rm K}(s) \, I_0(\nu_0)
            \, ds 
   \label{eq-ikompaneets}
\end{equation}
where the Kompaneets scattering kernel is of Gaussian form
\begin{equation}
   P_{\rm K}(s) = {1 \over \sqrt{4 \pi y}} \,
                  \exp \left( - {(s + 3 y)^2 \over 4 y} \right)
   \label{eq-kompaneetskernel}
\end{equation}
(Sunyaev\markcite{sun80} 1980; Bernstein \& Dodelson\markcite{bd90} 1990). 
The difference between the
scattering kernels~(\ref{eq-p1sdef}) and (\ref{eq-kompaneetskernel}) is
small but significant for the mildly 
relativistic electrons that are important in most cases, and can
become large where the electron distribution becomes more relativistic
(compare the solid and dashed lines in Fig.~\ref{fig-p1func}).
These shape changes lead to spectral differences
in the predicted $\Delta I(\nu)$, 
which for thermal electrons can be characterized by the changing
positions of the minimum, zero, and maximum of the spectrum of the
\SZ\ effect with changing electron temperature. This is illustrated in
Fig.~\ref{fig-zerothermal} for a range of temperatures that is of most
interest for the clusters of galaxies (Section~\ref{sec-thermal}).
\par
Most work on the \SZ\ effect has used the Kompaneets equation, and
implicitly the Kompaneets scattering kernel, rather than the more
precise relativistic kernel (\ref{eq-generalps})
advocated by Rephaeli\markcite{rep95a} (1995a). The
two kernels are the same at small $\te$, as expected since the
Kompaneets equation is correct in the low-energy limit. At
low optical depth and low temperatures, where the
Comptonization parameter $y$ is small, and for
an incident photon spectrum of the form of equation~(\ref{eq-occnum}),
the approximation  
${\partial n \over \partial y} = {\Delta n \over y}$ may be used in
(\ref{eq-kompaneetslimit}) to obtain a simple form for the spectral
change caused by scattering 
\begin{equation}
   \Delta n = x \, y \, {e^x \over (e^x - 1)^2} \,
              \left( x \, \coth (x/2) - 4 \right)
   \label{eq-kompaneetsdeltan}
\end{equation}
with a corresponding $\Delta I(x) = x^3 \Delta n(x) I_0$, where 
$I_0 = {2 h \over c^2} \left( {\boltz \trad \over h} \right)^3$ again. 
This result can also be obtained directly from the integral 
(\ref{eq-ikompaneets}) for $I(x)$ in the limit of small
$y$. Fig.~\ref{fig-dithermal} 
compares the Kompaneets approximation for
$\Delta I(x)$ with the full relativistic results.
\par
There are three principal simplifications gained by using 
(\ref{eq-kompaneetsdeltan}), rather than the relativistic results.
\begin{enumerate}
\item The spectrum of the effect is given by a simple analytical
function (\ref{eq-kompaneetsdeltan}).
\item The location of the spectral maxima, minima, and zeros are
independent of $\te$ in the Kompaneets approximation, but vary with
$\te$ in the relativistic expressions. To first order in $\Theta$,
which is adequate for
temperatures $\boltz \te < 20$~keV ($\Theta < 0.04$),
\begin{eqnarray}
 x_{\rm min}  & = & 2.26                        \nonumber \\
 x_{\rm zero} & = & 3.83 \, ( 1 + 1.13 \Theta ) \label{eq-approxroots} \\
 x_{\rm max}  & = & 6.51 \, ( 1 + 2.15 \Theta ) \nonumber 
\end{eqnarray}
as seen in Fig.~\ref{fig-zerothermal}. Earlier approximations for $x_{\rm
zero}(\Theta)$ are given by Fabbri\markcite{fabb81} (1981) and
Rephaeli\markcite{rep95a} (1995a).
\item The amplitude of the intensity (or brightness temperature)
change depends only on $y$ ($\propto \te \tau_{\rm e}$)
in the Kompaneets approximation, but in the
relativistic expression the amplitude is proportional to 
$\tau_{\rm e}$ (for small $\tau_{\rm e}$) and also depends on a
complicated function of $\te$.
\end{enumerate}
\par
It is possible to improve on the Kompaneets result
(\ref{eq-kompaneetsdeltan}) by working to higher order in $\Theta$
from (\ref{eq-deltaIintegral}) or the Boltzmann equation. The
resulting expressions for $\Delta n$ or $\Delta I$ are usually written
as a series in increasing powers of $\Theta$ (Challinor \&
Lasenby\markcite{cl98} 1998; Itoh, Kohyama \& Nozawa\markcite{ikn98}
1998; Stebbins\markcite{ste98} 1998). Taken to four or 
five terms these series provide a useful analytical expression for the
Sunyaev-Zel'dovich effect for hot clusters for a wide range of
$x$. However, the expressions cannot be used blindly since they are
asymptotic approximations, and are still of poor accuracy for some
$(x,\Theta)$ values. The approximations also rely on the assumptions
that the cluster is optically thin and that the electron distribution
function is that of a single-temperature gas (\ref{eq-thermale}). Both
assumptions are questionable when precise results (to better than
1~per cent) are needed, and so the utility of the idealized multi-term
expressions is limited. For the most precise work, especially work on
the kinematic Sunyaev-Zel'dovich effect (Section~\ref{sec-kinematic}),
the relativistic expressions with suitable forms for the distribution
functions, and proper treatment of the cluster optical depth, must be
used if accurate spectra for the effect, and hence estimates for the
cluster velocities, are to be obtained (Molnar \&
Birkinshaw\markcite{mb98b} 1998b).
%
%
\section{The thermal \SZ\ effect\label{sec-thermal}}
The results in Section~\ref{sec-scat} indicate that passage of
radiation through an 
electron population with significant energy content will produce a
distortion of the radiation's spectrum. In the present section the
question of the effect of thermal electrons on the CMBR is addressed
in terms of the three likely sites for such a distortion to occur:
\begin{enumerate}
\item the atmospheres of clusters of galaxies
\item the ionized content of the Universe as a whole, and
\item ionized gas close to us.
\end{enumerate}
\subsection{The \SZ\ effect from clusters of galaxies\label{sec-szcluster}}
\begin{figure}[p]
\epsscale{0.8}
\plotone{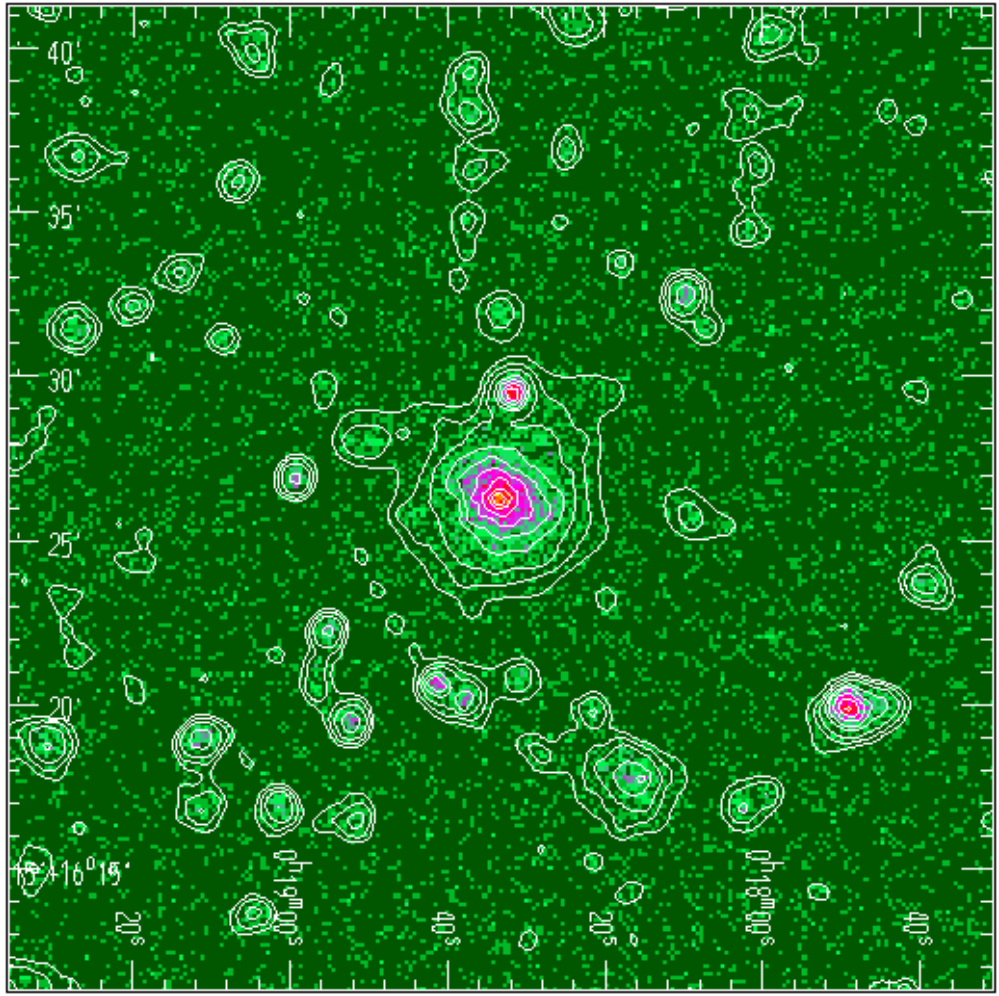}
\caption{\label{fig-0016pspc}
The central region of the {\it ROSAT}
PSPC X-ray image containing the distant cluster CL~0016+16
($z=0.5455$), showing the extended X-rays produced by the
thermal emission of gas in approximate hydrostatic equilibrium in the
cluster's potential well. The coordinates are in epoch J2000. The
data, extracted from PI bins 0.4--2.4 keV, have been background
subtracted, exposure corrected, and adaptively smoothed. The effective
spatial resolution of this image is $\sim$30$^{\prime\prime}$
(half-power diameter). Contour levels start at a value of $1.8 \times
10^{-4}$ counts s$^{-1}$ arcmin$^{-2}$ (75~per cent of the average background
level) and increase by multiplicative factors of 1.94.  The bright
X-ray source immediately to the north of the cluster is an AGN,
QSO~0015+162, at a redshift $z = 0.554$.  Note the extended source to
the southwest which is a poor cluster, RX~J0018.3+1618, at a redshift
$z = 0.5506$ (Hughes, Birkinshaw \& Huchra 1995).}
\end{figure}

By far the commonest references to the \SZ\ effect in the literature
are to the effect that the atmosphere of a cluster of galaxies has
on the CMBR. Cluster atmospheres are usually detected through their
X-ray emission, as in the example shown in Fig.~\ref{fig-0016pspc},
although the existence of such gas can also be inferred from its
effects on radio source morphologies (e.g., Burns \& Balonek\markcite{bb82}
1982) --- `disturbed' lobe shapes and head-tail sources being typical
indicators of the presence of cluster gas.
\par
If a cluster atmosphere contains gas with electron concentration
$n_{\rm e}({\bf r})$, then the scattering optical depth, 
Comptonization parameter, and X-ray spectral surface brightness along a
particular line of sight are
\begin{eqnarray}
   \tau_{\rm e} &=& \int n_{\rm e}({\bf r}) \, \sigma_{\rm T} \, dl 
                    \label{eq-taueclus} \\
   y            &=& \int n_{\rm e}({\bf r}) \, \sigma_{\rm T} 
                    \, {\boltz \te({\bf r}) \over m_{\rm e} c^2} \, dl
                    \label{eq-yclus} \\
   b_{\rm X}(E) &=& {1 \over 4\pi (1 + z)^3} \,
                    \int n_{\rm e}({\bf r})^2 \, \Lambda(E,T_{\rm e})
                    \, dl \label{eq-bxclus}
\end{eqnarray}
where $z$ is the redshift of the cluster, and $\Lambda$ is the
spectral emissivity of the gas at observed X-ray energy $E$ or into some
bandpass centered on energy $E$ (including both line and continuum
processes). The factor of $4\pi$ in the expression for $b_{\rm X}$
arises from the assumption that this emissivity is
isotropic, while the $(1+z)^3$ factor takes account of the
cosmological transformations of spectral surface brightness and
energy. 
\par
By far the most detailed information on the structures of cluster
atmospheres is obtained from X-ray astronomy satellites, such as 
ROSAT and ASCA. Even though these satellites also provide some
information about the spectrum of 
$b_{\rm X}$ (and hence an emission-weighted measure of the average gas
temperature along the line of sight) there is no unique inversion of
(\ref{eq-bxclus}) to $n_{\rm e}({\bf r})$ and $T_{\rm e}({\bf r})$.
Thus it is not possible to predict accurately the distribution of $y$
on the sky, and hence the shape of the Sunyaev-Zel'dovich effect
(which we will, in the current section, take to be close to the shape
of $y$, although Sec.~\ref{sec-effectspectrum} indicates that an
accurate prediction of the Sunyaev-Zel'dovich effect requires a
more complicated calculation which includes both the electron
scattering optical depth and the gas temperature).  
\par
In many cases it is then convenient to introduce a parameterized model
for the properties of the scattering gas in the cluster, and to fit the
the values of these parameters to the X-ray data. The integral
(\ref{eq-yclus})
can then be performed to predict the appearance of the cluster in the
Sunyaev-Zel'dovich effect. A form that is convenient, simple, and
popular is the isothermal $\beta$ model, where it is assumed that the
electron temperature $\te$ is constant and that the electron number
density follows the spherical distribution 
\begin{equation}
   n_{\rm e}({\bf r}) = n_{\rm e0} \, \left( 1 + {r^2 \over
       \rc^2} \right)^{-{3 \over 2}\beta} 
 \label{eq-nebeta}
\end{equation}
(Cavaliere \& Fusco-Femiano\markcite{cff76} 1976,\markcite{cff78} 1978:
the so-called `isothermal beta model'). This has been much used to fit
the X-ray structures of clusters of galaxies and individual galaxies
(see the review of Sarazin\markcite{sar88} 1988). Under these
assumptions the cluster will produce circularly-symmetrical patterns
of scattering optical depth, Comptonization parameter and X-ray
emission, with
\begin{eqnarray}
  \tau_{\rm e}(\theta) &=& \tau_{\rm e0}
         \left( 1 + {\theta^2 \over \thc^2} 
         \right)^{{1 \over 2} - {3 \over 2}\beta} \\
  y(\theta)            &=& y_0
         \left( 1 + {\theta^2 \over \thc^2} 
         \right)^{{1 \over 2} - {3 \over 2}\beta} 
         \label{eq-ycirc} \\
  b_{\rm X}(\theta)    &=& b_{\rm X0}
         \left( 1 + {\theta^2 \over \thc^2} 
         \right)^{{1 \over 2} - 3 \beta}
         \label{eq-bxisob}
\end{eqnarray}
where the central values are
\begin{eqnarray}
   \tau_{\rm e0} &=& n_{\rm e0} \, \sigma_{\rm T} \, r_{\rm c}
                    \, \sqrt\pi \, 
                    {\Gamma({3 \over 2}\beta - {1 \over 2}) \over
                     \Gamma({3 \over 2}\beta) } \quad , 
                    \label{eq-taue0circ} \\
   y_0           &=& \tau_{\rm e0} \, 
                    {\boltz \te \over m_{\rm e} c^2} \quad , \\
   b_{\rm X0}    &=& {1 \over 4 \pi (1+z)^3} \,
                    n_{\rm e0}^2 \, \Lambda(E,\te) \, r_{\rm c}
                    \, \sqrt\pi \, 
                    {\Gamma( 3\beta - {1 \over 2}) \over
                     \Gamma( 3\beta              ) } \quad .
                    \label{eq-bx0circ}
\end{eqnarray}
$\theta$ is the angle between the center of the cluster and the
direction of interest and $\thc = \rc / D_{\rm A}$ 
is the angular core radius of the cluster as deduced from
the X-ray data. $D_{\rm A}$ is the angular diameter distance of the
cluster, given in terms of redshift, deceleration parameter $q_0$, and
Hubble constant by
\begin{equation}
   D_{\rm A} = {c \over H_0 q_0^2} \, 
               {\left( q_0 z + (q_0 - 1) (\sqrt{1 + 2 q_0 z} - 1) \right)
               \over
               (1 + z)^2 }
   \label{eq-angdia}
\end{equation}
if the cosmological constant is taken to be zero (as it is throughout
this review).
\par
A useful variation on this model was introduced by Hughes
\etal\markcite{hugh88} (1988) on the basis of observations of the Coma
cluster. Here the divergence in gas mass which arises for typical
values of $\beta$ that fit X-ray images is eliminated by truncating
the electron density distribution. A model structure of similar form
describes the decrease of gas temperature at large radius. The 
density and temperature functions used are
\begin{eqnarray}
   n_{\rm e}(r) &=& \cases{n_{\rm e0} \left( 
                           1 + {r^2 \over r_{\rm c}^2} 
                           \right)^{\!-{3 \over 2}\beta} 
                           & $r \le r_{\rm lim}$ \cr
                           \cr
                           0
                           & $r > r_{\rm lim}$ \cr }
                    \label{eq-hughesne} \\
   T_{\rm e}(r) &=& \cases{T_{\rm e0} 
                           & $r \le r_{\rm iso}$ \cr
                           \cr
                           T_{\rm e0} \left(
                           1 + {r^2 \over r_{\rm iso}^2} 
                           \right)^{-\gamma}
                           & $r > r_{\rm iso}$ \cr }
                    \label{eq-hugheste}
\end{eqnarray}
where $r_{\rm lim}$ is the limiting gas radius, $r_{\rm iso}$ is the
isothermal radius, and $\gamma$ is some index. Not all choices of
these parameters are physically reasonable, but the forms above
provide an adequate description of at least some cluster structures.
Further modifications of models (\ref{eq-hughesne} -- \ref{eq-hugheste})
are required in cases where the cluster displays a cooling flow
(Fabian, Nulsen \& Canizares\markcite{fnc84} 1984), but this will be
important only in Sec.~\ref{sec-cosmolpar} in the present review.
\par
For cluster CL~0016+16 shown in Fig.~\ref{fig-0016pspc}, the redshift
of $0.5455$ implies an angular diameter distance $D_{\rm A} = 760
h_{100}^{-1}$~Mpc (for $q_0=0.5$). The X-ray emission mapped with the
ROSAT PSPC best matches a circular distribution of form
(\ref{eq-bxisob}) with structural parameters $\beta = 0.73 \pm 0.02$
and $\thc = 0.69 \pm 0.04$~arcmin, so that $r_{\rm c} = (150 \pm
10)h_{100}^{-1}$~kpc. The corresponding cluster central X-ray
brightness, $b_0 = 0.047 \pm 0.002 \ \rm counts \, s^{-1} \,
arcmin^{-2}$ (Hughes \& Birkinshaw\markcite{hub98} 1998). X-ray
spectroscopy of the cluster using data from the GIS on the ASCA
satellite and the ROSAT PSPC led to a gas temperature $\boltz\te = 7.6
\pm 0.6$~keV and a metal abundance in the cluster that is only $0.07$
solar. The X-ray spectrum is absorbed by a line-of-sight column with
equivalent neutral hydrogen column density $N_{\rm H} = (5.6 \pm 0.4)
\times 10^{20} \ \rm cm^{-2}$. These spectral parameters are
consistent with the results obtained by Yamashita\markcite{yam94}
(1994) using the ASCA data alone.\footnote{In this section all errors
given in Hughes \& Birkinshaw\markcite{hub98} (1998) have been
converted to symmetrical $\pm 1\sigma$ errors, for simplicity. Better
treatments of the errors are used in more critical calculations, for
example in Sec.~\ref{sec-cosmo}.}
\par
Using the known response of the ROSAT PSPC and the spectral
parameters of the X-ray emission found from the ASCA data, 
the emissivity of the intracluster gas in CL~0016+16 is 
$\Lambda_{\rm e0} = (2.70 \pm 0.06) \times 10^{-13} \ \rm counts \,
s^{-1} \, cm^{-5}$. The central electron number density
can then be found from (\ref{eq-bx0circ}) to be about 
$1.2 \times 10^{-2} h_{100}^{1/2} \ \rm electrons \, cm^{-3}$.
The corresponding central optical depth through the cluster is
$\tau_{\rm e0} = 0.01 h_{100}^{-1/2}$, which
corresponds to a central Comptonization parameter $y_0 = 1.5 \times
10^{-4}h_{100}^{-1/2}$. At such a small optical depth, the Sunyaev-Zel'dovich
effect through the cluster should be well described by
(\ref{eq-deltaIintegral}), so that the brightness change through the
cluster center should be 
$\DT_0 = -2y_0\trad \approx -0.82 h_{100}^{-1/2} \ \mK$ at low frequency.
\par
\begin{figure}[p]
\epsscale{0.8}
\plottwo{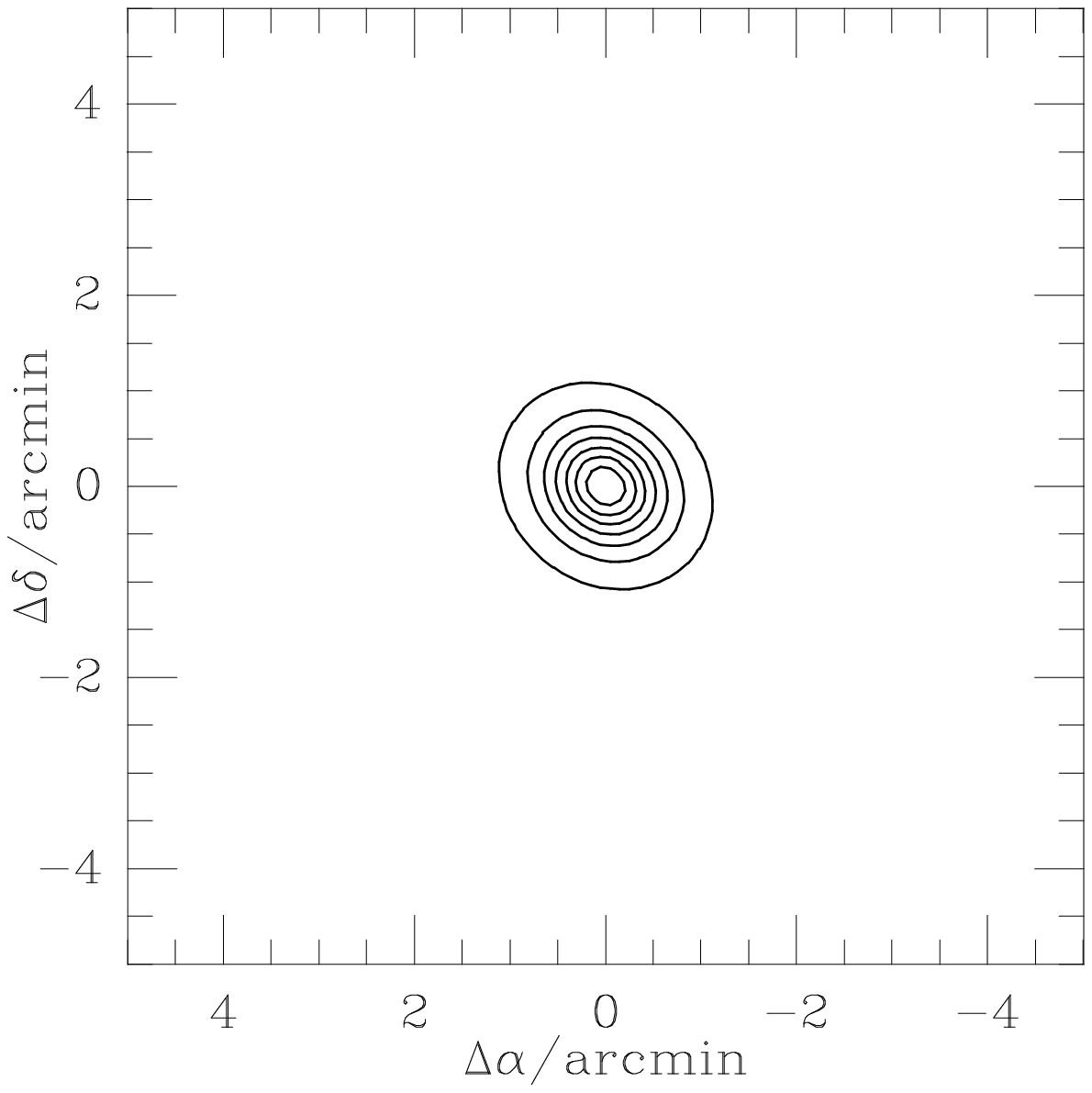}{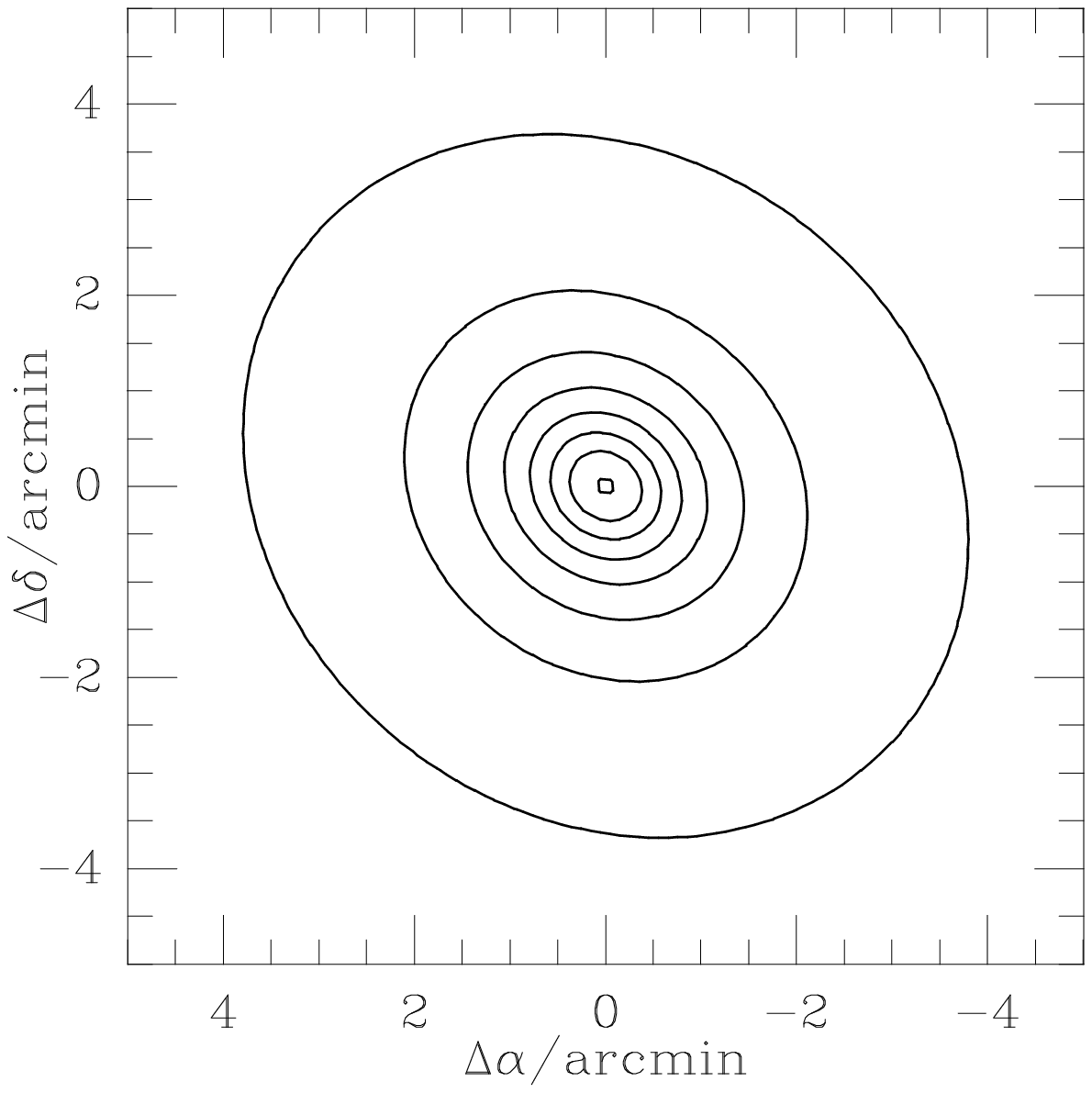}
\caption{\label{fig-0016model}
A model for CL~0016+16 
assuming that the cluster is oblate with the symmetry axis
in the plane of the sky, and with the structural parameters 
fixed by fits to the ROSAT PSPC image (Fig.~\ref{fig-0016pspc}).
Left: the model X-ray surface brightness. 
Right: the model \SZ\ effect. 
The contours in both plots are spaced at
intervals of 12.5~per cent of the peak effect --- note that the 
\SZ\ effect shows a much greater angular extent than the X-ray
emission (compare the angular dependences in equations~\ref{eq-ycirc}
and \ref{eq-bxisob}). 
The central \SZ\ effect predicted on the basis of the X-ray data 
is $-0.84h_{100}^{-1/2}$~mK.}
\end{figure}

More complicated models of the cluster density and temperature can be
handled either analytically, or by numerical integrations. For
example, Fig.~\ref{fig-0016pspc} clearly shows a non-circular structure
for CL~0016+16: a better representation of the structure of the
atmosphere may then be to replace (\ref{eq-nebeta}) by an ellipsoidal
model, with  
\begin{equation}
   n_{\rm e}({\bf r}) = n_{\rm e0} \, \left( 1 +
        {{\bf r}^{\rm T}.{\bf M}.{\bf r} \over r_{\rm c}^2}
        \right)^{-{3 \over 2}\beta} . 
  \label{eq-neellipse}
\end{equation}
where the matrix ${\bf M}$ encodes the orientation and relative sizes
of the semi-major axes of the cluster. If CL~0016+16 is assumed to be
intrinsically oblate, with its symmetry axis in the plane of the sky
(at position angle $\approx -40^\circ$) and with structural parameters
that match the X-ray image, then the intrinsic axial ratio is $1.17
\pm 0.03$, with best-fitting values for $\beta$ and the major axis
core radius of $0.751 \pm 0.025$ and $0.763 \pm 0.045$~arcmin (Hughes
\& Birkinshaw\markcite{hub98} 1998). The central X-ray surface
brightness is almost unchanged, reflecting the good degree of
resolution of the cluster structure effected by the ROSAT PSPC. With
these structural parameters, the predicted \SZ\ effect map of the
cluster is as shown in Fig.~\ref{fig-0016model}. The predicted central
\SZ\ effect is now $\DT_0 \approx -0.84h_{100}^{-1/2} \ \mK$ at low
frequency, very little changed from the prediction of the circular
model.
\par
Whether (\ref{eq-nebeta}) or (\ref{eq-neellipse}) is used to describe
the density structure of a cluster, it is important to be aware that
these representations are not directly tied to physical descriptions
of the gas physics and mass distribution in real clusters, but simply
choices of convenience. In principle a comparison of the \SZ\ effect
map and X-ray image of a cluster could be used to derive interesting
information about the structure of the gas, particularly when combined
with other information on cluster structure, such as weak lensing maps
of the cluster mass distribution, velocity dispersion measurements,
and data on the locations of cluster galaxies. Even based on the X-ray
and \SZ\ effect data alone there are several possibilities for finding
out more about cluster structures.
\begin{enumerate}
\item A comparison of the \SZ\ and X-ray images might be used to
determine the intrinsic three-dimensional shape of the
cluster. However, the leverage that the data have 
on the three-dimensional projection is poor. Changing the
model for CL~0016+16 from oblate to prolate only results in a change
of about $9$~per cent in the central predicted \SZ\ effect. Thus this
is unlikely to be a useful tool, at least for simple X-ray structures.
\item Since the X-ray emission depends on some average of 
$n_{\rm e}^2$ along the line of sight, while the 
\SZ\ effect depends on an average of $n_{\rm e}$, 
the shape of the \SZ\ effect image that is predicted
is sensitive to variations of the clumping factor 
$C_{\rm n} = \langle n_{\rm e}^2 \rangle 
           / \langle n_{\rm e}   \rangle ^2$
on the different lines of
sight through the cluster if clumping occurs at constant gas
temperature. The amplitude of the \SZ\ effect, indeed,
scales as $\sqrt C_{\rm n}$. Thus it might be possible to measure the
sub-beam scale clumping in the cluster gas. However, if 
the clumping occurs with a compensating temperature change then the
effect may be reduced. For example, if the X-ray emissivity is
proportional to $\te^{-1/2}$ and the clumping is adiabatic, then
changes in the X-ray emissivity are matched by equal changes in the 
Comptonization parameter and no difference will be seen in the \SZ\
effect image predicted based on the X-ray data. 
\item Probably the most useful astrophysical result that can be
extracted from the comparison is information on 
thermal structure in the intracluster gas. The X-ray and
\SZ\ effect images depend on $T_{\rm e}$ in different ways, with the
X-ray image from a particular satellite being a complicated function
of temperature, while the \SZ\ effect image is close to being an image
of the electron pressure. A comparison of the two images therefore
gives information about thermal structure --- particularly the thermal
structure of the outer part of the cluster, which has a greater
fractional contribution to the \SZ\ effect than to the X-ray emission
(since the X-ray emission depends on $n_{\rm e}^2$ while the \SZ\
effect depends on $n_{\rm e}$). However, it is likely that this
information will be more easily gained using spatially-resolved
spectroscopy on the next generation of X-ray satellites. 
\item Finally, as has been emphasized by Myers \etal\markcite{myers97}
(1997), the \SZ\ effect is a direct measure of the 
projected mass of gas in the cluster on the line of sight if the
temperature structure of the cluster is simple. This implies that the
baryonic surface mass density in the cluster can be measured directly,
and compared with other measurements of the mass density, for example
from gravitational lensing. A discussion of this in relation to the
cluster baryon problem appears in Section~\ref{sec-cluster}.
\end{enumerate}
\par 
However, most of the recent interest in the \SZ\ effects of clusters
has not been because of their use as diagnostics of the cluster
atmospheres, but rather because the effects can be used as 
cosmological probes. A detailed explanation of the method and its
limitations is given in Section~\ref{sec-cosmo}, but the essence of
the method is a comparison of the \SZ\ effect predicted from the X-ray
data with the measured effect. Since the predicted effect is
proportional to $h_{100}^{-1/2}$ via the dependence on the angular
diameter distance (equation~\ref{eq-taue0circ} with $\rc = D_{\rm A}
\thc$; see also the discussion of CL~0016+16 above), this comparison
measures the value of the Hubble constant, and potentially other
cosmological parameters.
\par
It should be emphasized that the 
\SZ\ effect has the unusual property of being redshift independent:
the effect of a cluster is to cause some 
fractional change in the brightness of the CMBR, and this fractional
change is then seen at all positions on the line of sight through the
cluster, at whatever redshift. Thus
the central \SZ\ effect through a cluster
with the properties of CL~0016+16 will have the same value whether the
cluster is at redshift $0.55$, $0.055$, or $5.5$. This makes the 
\SZ\ effect exceptionally valuable as a cosmological probe of hot
electrons, since it should be detectable at any redshift for which
regions with large electron pressures exist. 
\par
There have now been a number of detections of the \SZ\ effects of
clusters, and recent improvements in the sensitivities of
interferometers with modest baselines have led to many maps of the
effects. A discussion of the detection strategies, and the
difficulties involved in comparing the results from different
instruments, is given in Section~\ref{sec-meas}. 
\subsection{Superclusters of galaxies\label{sec-supercl}}
While the discussion in Sec.~\ref{sec-szcluster} has concentrated on
clusters of galaxies, other objects also contain extended atmospheres
of hot gas and may be sources of detectable \SZ\ effects. One
possibility is in superclusters, large-scale groups of clusters, where small
enhancements of the baryon density over the mean cosmological baryon
density (which is well constrained by nucleosynthesis arguments;
Walker \etal\markcite{walk91} 1991; Smith, Kawano \&
Malaney\markcite{skm93} 1993) are expected, but where the path lengths
may be long so that a significant \SZ\ effect builds up. Supercluster
atmospheres may originate in left-over baryonic matter that did not
collapse into clusters of galaxies after a phase of inefficient
cluster formation, and could be partially enriched with heavy metals
through mass loss associated with early massive star formation or
stripping from merging clusters and protoclusters. A measurement of
the mass and extent of supercluster gas would be a useful indication
of the processes involved in structure formation. 
\par
Most work on supercluster gas has been conducted through X-ray
searches. Persic \etal\markcite{per88} (1988,\markcite{per90} 1990)
searched for X-ray emission from superclusters in the HEAO-1 A2 data,
finding no evidence for emission from the gas. Day
\etal\markcite{day91} (1991) searched for intra-supercluster gas in 
the Shapley supercluster using GINGA scans, and were able to set strong
limits on the X-ray emission. More recently, Bardelli
\etal\markcite{bar96} (1996) have used ROSAT PSPC data to claim that
there is some diffuse X-ray emission in the Shapley supercluster
between two of its component clusters.
\par
The thermal \SZ\ effect provides another potential probe for
intrasupercluster gas. Since this effect is proportional to the line
of sight integral of $n_{\rm e}$, it should be a more sensitive probe
than the X-ray emission for studying the diffuse gas expected in
superclusters. The angular scales of the well-known superclusters are
large (degrees), so that the COBE DMR database is the best source of
information on their \SZ\ effects: ground-based work is always on too
small an angular scale, and the balloon searches do not cover such a
large fraction of the sky at present.
\par
Indeed, Hogan\markcite{hogan92} (1992) suggested that much of the
anisotropy in the CMBR detected by the COBE DMR might be produced by local
superclusters. This has been tested by Boughn \& Jahoda\markcite{bj93}
(1993), who found no sign of the anticorrelation of the HEAO-1 A2 and
COBE DMR sky maps that would be expected from such a mechanism and
concluded that the COBE DMR signal was not produced by supercluster
\SZ\ effects.
\par
Limits to the {\it average} \SZ\ effects from clusters of galaxies
(and their associated superclusters) were derived by Banday
\etal\markcite{ban96} (1996) through a cross-correlation analysis of
the COBE DMR 4-year data with catalogues of clusters of galaxies. The
result, that the average \SZ\ effect is less than $8 \ \muK$ (95~per
cent confidence limit at $7^\circ$ angular scale), suggests that these
\SZ\ effects are not strong. However, although population studies of
this type show that average superclusters do not contain atmospheres
with  significant gas pressures, the COBE DMR database can also be
searched for indications of a non-cosmological signal towards
particular superclusters of galaxies.
\par
Banday \etal\markcite{ban96} (1996) were able to set limits
$\DTRJ(7^\circ) \lesssim 50 \ \muK$ for the \SZ\ effects
towards the well-known Virgo, Coma, Hercules, and Hydra clusters. 
Since the largest (in angular-size) of these clusters have low
X-ray luminosity, and the highest X-ray luminosity object (Coma) is
strongly beam diluted, it is not surprising that no signals were
found. However, equivalent results for superclusters should set
interesting new limits on their gas contents. 
\par
The most prominent supercluster near us (at $z < 0.1$) is the Shapley
supercluster, which consists of many Abell and other clusters centered
on Abell~3558, lies at a distance $140 h_{100}^{-1}$~Mpc, and has a
core radius of about $20 h_{100}^{-1}$~Mpc. The estimated overdensity
of the Shapley supercluster is the largest known on such a scale, and
this supercluster may be the largest gravitationally-bound structure
in the observable Universe (Raychaudhury \etal\markcite{ray91} 1991;
Fabian\markcite{fab91} 1991). Searches for gas in the supercluster,
conducted by Day \etal\markcite{day91} (1991) and others, have not led
to any convincing detection of such gas. A rough scaling argument
suggests that the peak \SZ\ effect to be expected from a supercluster
of scale $R$ is about
\begin{equation}
   \DT_{\rm th} \approx -6 \, (L_{\rm X}  /10^{37} \, \W)^{1/2}
                           \, (T_{\rm e}  /\keV)^{3/4}
                           \, (R/10 \, \Mpc)^{-1/2} \quad \muK .  
 \label{eq-superclszest}
\end{equation}
For the Shapley supercluster, and a gas temperature of a few keV, the
Day \etal\markcite{day91} (1991) limit on the X-ray surface brightness
corresponds to an \SZ\ effect of about $-20 \ \muK$. Molnar \&
Birkinshaw\markcite{mb98a} (1998a) have used the COBE DMR 4-year
database to set a limit of about $-100 \ \muK$ on the thermal \SZ\
effect. This result is an improvement on the constraint of Day
\etal\markcite{day91} only if the atmosphere is hot, with $\boltz \te
\gtrsim 15 \ \keV$. However, improvements in the microwave background
data from the next generation of satellites will achieve a factor $10$
or more improvement in sensitivity to the \SZ\ effect, and will
strengthen the limits on the mass of gas in the supercluster at all
likely gas temperatures. 
\par
Superclusters are sufficiently massive objects that they also produce
CMBR anisotropies through their distortion of the Hubble flow (Rees \&
Sciama\markcite{rs68} 1968; Dyer\markcite{dyer76} 1976;
Nottale\markcite{not84} 1984) as well as any \SZ\ effects that 
they produce. A supercluster of mass $M$ and radius $R$ will cause a
Rees-Sciama effect of order
\begin{equation}
 \DT_{\rm RS} \approx -2 \, (M/10^{16} \, \Msol)^{3/2}
                         \, (R/10 \, \Mpc)^{-3/2} \quad \muK 
\end{equation}
which is of the same order as the \SZ\ effect (\ref{eq-superclszest}),
but with a different spectrum (that of primordial anisotropies) and
angular structure. Since the intrinsic anisotropies in the CMBR are
expected to be larger than these supercluster-generated effects, it is
unlikely that even statistical information about $\DT_{\rm RS}$ 
can be obtained, but the next
generation of microwave background satellites should be able to use
\SZ\ effect data to constrain supercluster 
properties. No useful limits on the mass of superclusters (or the
Shapley supercluster in particular) are obtained using the COBE DMR
4-year data to search for Rees-Sciama effects (Molnar \&
Birkinshaw\markcite{mb98a} 1998a).
\subsection{Local \SZ\ effects \label{sec-local}}
While the above discussions have concentrated on distant clusters of
galaxies and on the integrated \SZ\ effects of clusters and the
diffuse intergalactic medium, it is also interesting to consider the
possibility of distortions of the microwave background
radiation induced by gas in the local group. 
\par
Suto \etal\markcite{suto96} (1996) have proposed that gas in the local group 
may contribute to the apparent large-scale anisotropy of the CMBR
(specifically, the quadrupolar anisotropy) through the \SZ\ effect. If the
local group contains a spherical gas halo, described by an isothermal
$\beta$ model (equation~\ref{eq-nebeta}) and with the Galaxy offset a
distance $x_0$ from its center, then the limit on the value of $y$
from the COBE FIRAS data implies that
\begin{equation}
   (n_{\rm e0}/\cm^{-3}) \, (\boltz T_{\rm e}/\keV) \, 
   (r_{\rm c}/100 \ \kpc)
   \lesssim 0.03
\end{equation}
if $x_0 \ll r_{\rm c}$. Electron concentrations this small
cause a dipole anisotropy of the CMBR that is much
smaller than the observed dipole anisotropy, but may produce 
a significant quadrupole. Suto \etal\markcite{suto96} suggest that
this quadrupole may be as large as $40 \ \muK$ without violating
either the X-ray background 
limits or the COBE FIRAS limits. Since the observed COBE quadrupole is
only $Q_{\rm rms} = 6 \pm 3 \ \muK$ (Bennett \etal\markcite{ben94} 1994),
significantly less than $Q_{\rm rms-PS}$ derived from the overall
spectrum of fluctuations, a
local \SZ\ effect may help to explain why we observe an
anomalously small quadrupole moment in the CMBR. 
\par
This ingenious explanation of the COBE quadrupole in terms of a local
\SZ\ effect has been criticized by Pildis \& McGaugh\markcite{pm96}
(1996), who note that to produce a significant quadrupole the electron
density in the local group needs to exceed the value typical of 
distant groups of galaxies by a factor $\gtrsim 10$. Thus gas in the
local group is unlikely to produce a significant contribution to the
COBE quadrupole. Furthermore, Banday \& Gorski\markcite{bg96} (1996)
found that the full model \SZ\ effect predicted by Suto \etal, when
fitted to the COBE dataset, cannot produce a large enough quadrupolar
term to be interesting. Nevertheless, it is clear that local gas may cause
some small contributions to microwave background anisotropies on
angular scales normally thought to be ``cosmological'', and care will
be needed in interpreting signals at levels $\approx 0.1 \ \muK$.
%
%
\section{The non-thermal \SZ\ effect\label{sec-nonthermal}}
\begin{figure}[p]
\epsscale{0.8}
\plotone{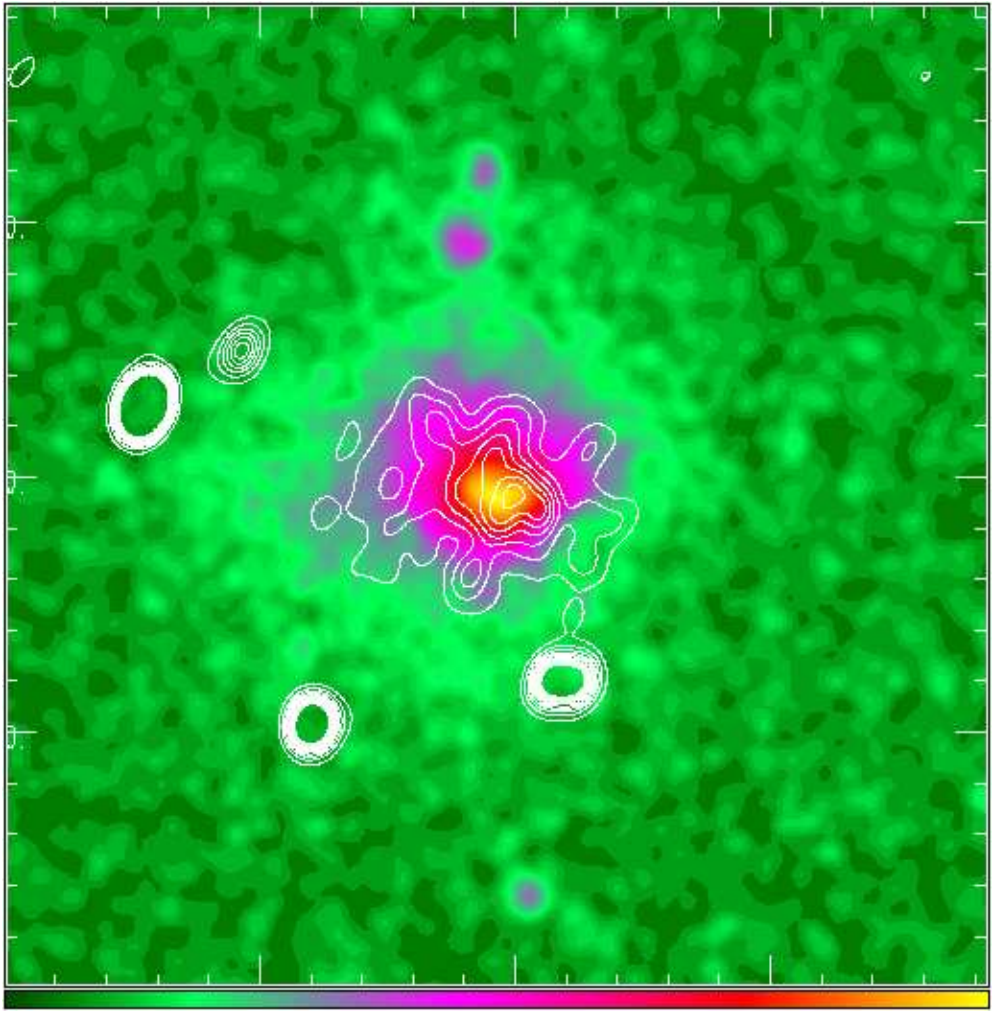}
\caption{\label{fig-a2163xr}
1400-MHz radio contours superimposed on
a soft X-ray image of Abell~2163 (Herbig \& Birkinshaw 1995). Note 
the close resemblance of the radio and X-ray structures, and the
diffuseness of the radio source. This is a particularly luminous
example of a cluster radio halo source, with a radio 
luminosity $L_{\rm radio} \approx 10^{35} \, h_{100}^{-2}
\ \rm W$.}
\end{figure}

As was noted in Sec.~\ref{sec-effectspectrum}, a non-thermal population
of electrons must also scatter microwave background photons, and it
might be expected that a sufficiently dense relativistic electron
cloud would also produce a \SZ\ effect. Fig.~\ref{fig-a2163xr}, which
shows a radio map of Abell~2163 superimposed on a soft X-ray image,
indicates that in some clusters there are populations of highly
relativistic electrons (in cluster radio halo sources) that have
similar angular distributions to the populations of thermal electrons
which are more conventionally thought of as producing \SZ\
effects. Indeed, in many of the clusters in which \SZ\ effects have
been detected there is also evidence for radio halo sources, so it is
of interest to assess whether the detected effects are in fact from
the thermal or the non-thermal electron populations.
\par
Quick calculations based on the Kompaneets approximation (for example,
equation \ref{eq-kompaneetsdeltan}), suggest that at low frequencies
the amplitude of the \SZ\ effect should be 
\begin{eqnarray}
   \DTRJ &=& -2 \, y \, \trad \nonumber \\
         &=& -2 \, \trad \,
                \int {\boltz \te \over m_{\rm e} c^2} \, 
                n_{\rm e} \, \sigma_{\rm T} \, dl \nonumber \\
         &=& -2 \, {\sigma_{\rm T} \over m_{\rm e} c^2}
                \, \int p_{\rm e} dl
\end{eqnarray}
so that the effect depends on the line-of-sight integral of the
electron pressure alone. If a radio halo source, such as is seen in
Fig.~\ref{fig-a2163xr}, and the cluster gas which (presumably)
confines it are in approximate pressure balance, then this argument
would suggest that the thermal and non-thermal contributions to the
overall \SZ\ effect should be of similar amplitude if the angular
sizes of the radio source and the cluster gas are similar. Since the
spectra of the thermal and non-thermal effects are distinctly
different (compare Fig.~\ref{fig-dithermal} and~\ref{fig-dipower}),
the spectrum of the overall \SZ\ effect 
measures the energy densities in the thermal gas and in the radio halo
source separately. This would remove the need to use the
minimum energy argument (Burbidge\markcite{bur56} 1956) to deduce the
energetics of the source. 
\par
Matters are significantly more complicated if the full relativistic
formalism of Sec.~\ref{sec-scat} is used. But this is necessary, since the
electrons which emit radio radiation by the synchrotron process are
certainly highly relativistic and the use of the Kompaneets
approximation is invalid. Thus we must distinguish between the
effects of the electron spectrum and those of the electron scattering
optical depth, but the results of Sec.~\ref{sec-effectspectrum} can be
used to predict the expected \SZ\ effect intensity and spectrum from any
particular radio source. 
\par
Consider, for example the Abell~2163 radio
halo, for which we assume a spectral index $\alpha=-1.5$ (there is no
information on the spectral index, since the
halo has been detected only at 1400~MHz: $\alpha = -1.5$ is typical
of radio halo sources). The diameter of the halo is about
$1.2 \, h_{100}^{-1}$~Mpc, and the radio luminosity (in an assumed
frequency range from 10~MHz to 10~GHz) is 
$10^{35} \, h_{100}^{-2} \ \rm W$. 
Using the minimum energy argument in its traditional form (see the
review by Leahy\markcite{leahy90} 1990), the equipartition magnetic
field is about $0.06 h_{100}^{2/7} \ \rm nT$ and the energy density in
relativistic electrons is about $10^{-15} h_{100}^{4/7} \ \rm W \,
m^{-3}$. This estimate assumes that all the particle energy resides in
the electrons, and that the source is completely filled by the
emitting plasma. The equivalent electron density in the source is
$n_{\rm e} = 2 \times 10^{-3} \, h_{\rm 100}^{6/7} \ \rm m^{-3}$,
which is a factor $\sim 10^6$ less than the electron density in the
embedding thermal medium and 
corresponds to a scattering optical depth of only 
$6 \times 10^{-9} \, h_{100}^{-1/7}$, which is certainly much less than
the optical depth of the thermal atmosphere in which the radio source
resides. Although the power-law electron distribution is more
effective at scattering the microwave background radiation than the
intracluster gas, at low frequencies it is found that the predicted
\SZ\ effect from the halo radio source electron distribution is
$\DTRJ = -5 \ \rm nK$. This is about $10^5$ times smaller than the
\SZ\ effect from the thermal gas.
\par
The dominance of the thermal over the non-thermal effect from the
cluster arises principally from the lower density of relativistic than
non-relativistic electrons. Only a low relativistic electron density 
is inferred because of the high efficiency of the synchrotron
process if only a small range of electron energies is present. If the
frequency range of the synchrotron radiation is extended beyond
the 10~MHz to 10~GHz range previously assumed, then the optical
depth to inverse-Compton scattering depends on the lower frequency
limit as $\tau_{\rm e} \approx 10^{-12} \, h^{-1/7} \, (\nu_{\rm
min}/\GHz)^{-13/7}$ (which would be $\gg 1$ if the electron spectrum
extends down to thermal energies). A
reduction in the lower cutoff frequency of the
spectrum by a factor $\sim 10^3$ then increases the estimated relativistic
electron density to the point that the non-thermal \SZ\ effect makes a
significant contribution. Indeed, this strong dependence of 
$\tau_{\rm e}$ on $\nu_{\rm min}$ or, equivalently, on the minimum
electron energy, suggests that the non-thermal \SZ\ effect is a
potential test for the low-end cutoff energy of the relativistic
electron spectrum.
\par
Thus although the original purpose of searching for the non-thermal
\SZ\ effect (as was done by McKinnon, Owen \& Eilek\markcite{moe90}
1990) was to check on the applicability of the minimum energy formula,
it is more appropriate to think of it as a measurement of the minimum
energy of the electrons that produce the radio radiation. Although
limits on this minimum energy can be deduced from the polarization
properties of radio sources, these limits are model-dependent (e.g.,
Leahy\markcite{leahy90} 1990), and an independent check from the
non-thermal \SZ\ effect would be useful.
\par
The problem of detecting the \SZ\ effect from non-thermal electron
populations is likely to be severe because of the associated
synchrotron radio emission. At low radio frequencies, that synchrotron
emission will easily dominate over the small negative signal of the
\SZ\ effect. At high radio frequencies, or in the mm-wave bands, there
is more chance that the \SZ\ effect could be detectable, but even here
there are likely to be difficulties separating the \SZ\ effects from
the flattest-spectrum component of the synchrotron emission.
\par
Several inadvertent limits to the non-thermal \SZ\ effect are
available in the literature, from observations of clusters of galaxies
which contain powerful radio halo sources (such as Abell~2163) or
radio galaxies (such as Abell~426), but few detailed analyses of the
results in terms of the non-thermal effect have been possible, and a
treatment of the interpretation of the Abell~2163 data is deferred
until later (Sec.~9.2). 
\par
Only a single intentional search for the \SZ\ effect from a
relativistic electron population has been attempted to date (McKinnon
\etal\markcite{moe90} 1990), and that searched for the \SZ\ effect in
the lobes of several bright radio sources. No signals were seen, but a
detailed spectral fit of the data to separate residual synchrotron and
\SZ\ effect signals was not done, and the limits on the \SZ\ effects
(of $y \lesssim 2 \times 10^{-3}$ for the best two sources) do not
constrain the electron populations in the radio lobes strongly: the
lobes could be far from equipartition without violating the \SZ\
effect constraint.
\par
One difficulty with the analysis given above, and the discussion of
the testing of minimum electron energy or the minimum energy
formalism, is that radio sources are expected to be strongly
inhomogeneous, so single-dish \SZ\ effect observations are averaging 
over a wide variety of different radio source structures (such as
lobes and hot spots). This would mean, for example, that the spectral
curvature that might be predicted by a superposition of the source
spectrum and the \SZ\ effect might be also be produced by small
variations in the electron energy distribution function from place to
place within the radio source. If strong tests of the electron energy
distribution are to be made, the observations must be made with angular
resolution comparable with the scale of structures within the radio
sources. For all but the largest radio sources (such as the lobes of
Cen~A), this means that interferometers (or bolometer arrays on large
mm-wave telescopes) must be used. No work of such a type has yet been
attempted, and the sensitivity requirements for a successful detection
are formidable.
%
%
\section{The kinematic Sunyaev-Zel'dovich effect\label{sec-kinematic}}
Although early work on the \SZ\ effects concentrated on the
thermal effect, a second effect must also occur when
the thermal (or non-thermal) \SZ\ effect is present. This is the
velocity (or kinematic) \SZ\ effect, which
arises if the scattering medium causing the thermal (or non-thermal)
\SZ\ effect is moving relative to the Hubble
flow. In the reference frame of the scattering gas the microwave background
radiation appears anisotropic, and the effect of the inverse-Compton
scattering is to re-isotropize the radiation slightly. Back in the
rest frame of the observer the radiation field is no longer
isotropic, but shows a structure towards the scattering atmosphere
with amplitude proportional to $\tau_{\rm e} {v_{\rm z}\over c}$, where 
$v_{\rm z}$ is the component of peculiar velocity of the scattering
atmosphere along the line of sight (Sunyaev \&
Zel'dovich\markcite{sz72} 1972; Rephaeli \& Lahav\markcite{rl91} 1991).
\par
The most interesting aspect of the kinematic effect is that it 
provides a method for measuring one component of the peculiar velocity
of an object at large distance, provided that the velocity and thermal
effects can be separated, as they can using their different spectral
properties. Since there is evidence for large-scale motions of
clusters of galaxies in the local Universe, both from COBE (Fixsen
\etal\markcite{fix96} 1996) and from direct observations of galaxies
(e.g., Dressler \etal\markcite{dre87} 1987; Lynden-Bell
\etal\markcite{lynd88} 1988), and these motions place strong 
constraints on the dynamics of structure formation (see the review of
Davis \etal\markcite{defw92} 1992), other examples of large-scale
flows would be of considerable interest. This is particularly true if
those flows can be measured over a range of redshifts, so that the
development of the peculiar velocity field can be studied. At present
this is beyond the capability of the measurements
(Sec.~\ref{sec-meas}), but rapid progress is being made in this area.
\par
Even larger velocities are possible for scattering gas: radio source
lobes may be moving at speeds that approach the speed of light, and 
expanding at hundreds or thousands of $\kms$. Although this will boost
the kinematic effect, the optical depths of these lobes are probably
too small to make the effect observable at present
(Sec.~\ref{sec-nonthermal}; Molnar\markcite{mo98} 1998).
\par
Although Sunyaev \& Zel'dovich\markcite{sz72} (1972) quoted the result
that the radiation temperature decrease in the kinematic effect is
\begin{equation}
 {\Delta \trad \over \trad} \approx - \tau_{\rm e} {v_{\rm z} \over c}
 \label{eq-kinesz}
\end{equation}
and the spectrum of the kinematic effect has often been quoted (e.g.,
by Rephaeli\markcite{rep95b} 1995b), the first published derivation of
the size and spectrum of the kinematic effect was given by
Phillips\markcite{phil95} (1995). The version of the derivation given
here is similar to Phillips' argument, but uses different conventions
and the radiative transfer equation (\ref{eq-transeqn}) rather than
the Boltzmann equation.
\par
For the sake of simplicity, it is assumed that the kinematic and the
thermal effects are both small, and that only single scatterings are
important. Then the thermal effect, which depends on random motions of
the scattering electrons, and the kinematic effect, which depends on their
systematic motion, will decouple and we can derive the kinematic
effect by taking the electrons to be at rest in the frame of the
scattering medium. This approximation will ignore quantities which
involve cross-products 
$\left( {\boltz \te \over \me c^2}\right) \left({v_{\rm z}\over
 c}\right)$ 
relative to terms of order
$\left( {\boltz \te \over \me c^2}\right)$
and
$\left({v_{\rm z} \over c}\right)$.
Since the peculiar velocities and electron temperatures are
small for the thermal \SZ\ effect, this will not be a significant
limitation on the result for clusters of galaxies. However, this
approximation is not valid for the non-thermal effect, where electron
energies $\gg m_{\rm e} c^2$ are likely, and an alternative analysis
is necessary (Molnar\markcite{mo98} 1998; Nozawa \etal\markcite{nik98}
1998).
\par
\begin{figure}[p]
\epsscale{0.8}
\plotone{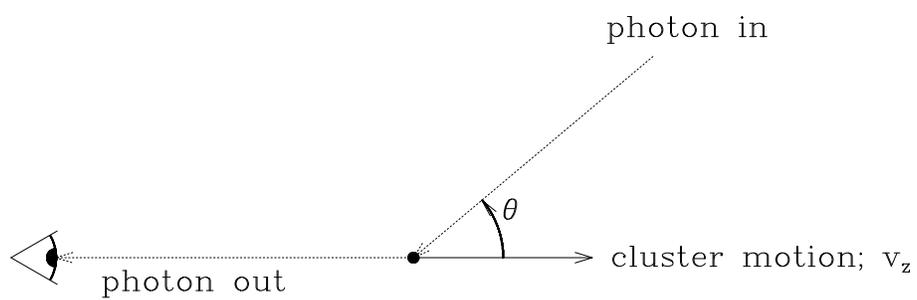}
\caption{\label{fig-kinegeom}
The geometry for the discussion of the
kinematic \SZ\ effect, as seen in the frame of an observer at rest in
the Hubble flow.}
\end{figure}

In the rest frame of the CMBR, the spectrum of the radiation follows
(\ref{eq-planckspec}), and the occupation number has the form
(\ref{eq-occnum}). The occupation number in mode $\alpha$
in a frame moving at speed $v_{\rm z}$ along the $z$ axis away
from the observer is then
\begin{equation}
   n_\alpha = \left( \exp{\left( x_1 \gamma_{\rm z}
                     (1 - \beta_{\rm z} \mu_1) \right)}
                     - 1 \right)^{-1}
   \label{eq-occnumboost}
\end{equation}
where $x_1 = {h \nu_1 \over \boltz T_1}$ 
is the dimensionless frequency of photons in the frame of
the scattering medium, the radiation temperature of the CMBR as seen
by an observer at rest in the Hubble flow near the scattering gas is
$T_1 = \trad (1 + z_{\rm H})$, $z_{\rm H}$ is the Hubble flow
redshift, and $\beta_{\rm z} = v_{\rm z}/c$ measures the peculiar
velocity. $\gamma_{\rm z}$ is the corresponding Lorentz factor. $\mu_1
= \cos\theta_1$ is the 
direction cosine of photons arriving at a scattering electron,
relative to the $z$ axis and measured in the frame of the moving
scattering medium (Fig.~\ref{fig-kinegeom}). The relativistic
transformation of frequency relates $\nu_1$ to $\nu$ in the frame at rest
relative to the CMBR by $\nu_1 = \gamma_{\rm z} ( 1 + \beta_{\rm z}) \nu$, 
since the observer at rest sees the scattered photons along the $z$
axis, where $\mu = \cos\theta = 1$.
\par
We may apply the Boltzmann equation (e.g., as in
Peebles\markcite{peeb93} 1993, Sec.~24 and Phillips\markcite{phil95}
1995), or the radiative transfer equation (\ref{eq-transeqn}), to
derive an equation for the scattered radiation intensity. Using
(\ref{eq-transeqn}), and the form (\ref{eq-scatprob}) for the
scattering redistribution function, the specific 
intensity is given by 
\begin{equation}
   {d I_{\nu_1}(\mu) \over d \tau_{\rm e}} =
   \int_{-1}^{+1} d\mu_1 \, \phi(\mu,\mu_1) \, 
   \biggl( I_{\nu_1}(\mu_1) - I_{\nu_1}(\mu) \biggr)  \quad .
\end{equation}
The optical depth, $\tau_{\rm e}$, enters from (\ref{eq-transeqn}) as
$\tau_{\rm e} = \int \alpha_{\nu,\rm sca} dz$. 
For small optical depth, so that photons are scattered only once, this
can be simplified to 
\begin{equation}
   {I_{\nu_1}(\tau_{\rm e};\mu) - I_{\nu_1}(0;\mu) \over \tau_{\rm e}}
   =
   \int_{-1}^{+1} d\mu_1 \, \phi(\mu,\mu_1) \,
   \biggl( I_{\nu_1}(0;\mu_1) - I_{\nu_1}(0;\mu) \biggr) .
\end{equation}
where the optical depth is now inserted as an explicit argument of
$I$. For $\mu = 1$, the
scattering redistribution function takes a particularly simple form,
and we can write the fractional 
change in the specific intensity in the frame of the scattering gas
as
\begin{equation}
   {\Delta I_{\nu_1} \over I_{\nu_1}}
   =
   \tau_{\rm e}  \,
   \int_{-1}^{+1} d\mu_1 
   {3 \over 8} ( 1 + \mu_1^2 )
   \left( {I_{\nu_1}(0;\mu_1) \over I_{\nu_1}(0;1)} - 1 \right) .
 \label{eq-kinint1}
\end{equation}
\par
The expression on the left-hand side of this equation is a
relativistic invariant: the same fractional
intensity change would seen by an observer in the rest frame of the
CMBR at frequency $\nu$, where $\nu$ is related to $\nu_1$ by a Lorentz
transform. Furthermore, this is also the fractional intensity change
seen by a distant observer, for whom the scattering medium lies at redshift
$z_{\rm H}$, after allowance is made for the redshifting of frequency
and radiation temperature. Using the expression for $n_\alpha$ in
(\ref{eq-occnumboost}), and working in terms of the frequency seen at
redshift zero, $\nu$, (\ref{eq-kinint1}) becomes 
\begin{equation}
   {\Delta I_{\nu} \over I_{\nu}}
   =
   \tau_{\rm e}  \,
   \int_{-1}^{+1} d\mu_1  \,
   {3 \over 8} ( 1 + \mu_1^2 ) \,
   \left( {e^x - 1 \over e^{x_2} - 1 } - 1 \right) .
\end{equation}
where 
$x_2 = x \gamma_{\rm z}^2 (1 + \beta_{\rm z}) ( 1 - \beta_{\rm z}
\mu_1)$ and $x = {h \nu \over \boltz \trad}$ as usual.
\par
For small $\beta_{\rm z}$, the integral can be expanded in powers of
$\beta_{\rm z}$, and the symmetry of the integrand ensures that only
terms in the expansion which are even powers of $\mu_1$ will appear in
the result. This enables the integral to be performed easily, giving
the result
\begin{equation}
   {\Delta I_{\nu} \over I_{\nu} }
   = 
   -\tau_{\rm e} \, \beta_{\rm z} \,
   {x \, e^x \over e^x - 1}
\end{equation}
so that the changes in specific intensity and brightness temperature
are given by
\begin{eqnarray}
  \Delta I &=& - \beta \, \tau_{\rm e} \, I_0 \,
                 {x^4 e^x \over \left( e^x - 1 \right)^2} 
                 \label{eq-deltaIkinematic} \\
  \DTRJ    &=& - \beta \, \tau_{\rm e} \, \trad \,
                 {x^2 e^x \over \left( e^x - 1 \right)^2} \quad .
\end{eqnarray}
This spectral form corresponds to a simple decrease in the radiation
temperature (\ref{eq-kinesz}), as stated by Sunyaev \&
Zel'dovich\markcite{sz72} (1972).
\par
For the cluster CL~0016+16 discussed in Sec.~\ref{sec-thermal}, the
X-ray data imply a central scattering optical depth 
$\tau_{e0} = 0.01 h_{100}^{-1/2}$. At low frequency the brightness
temperature change through the cluster center caused by the kinematic
effect is $- \tau_{\rm e0} \, (v_{\rm z} / c) \, \trad
= - 0.1 \, (v_{\rm z} / 1000 \ \kms) \, h_{100}^{-1/2} \ \mK$,
significantly less than the central thermal \SZ\ effect of
$-0.82 h_{100}^{-1/2} \ \mK$ for all likely $v_{\rm z}$. 
\par
It would be very difficult to locate the kinematic \SZ\ effect in the
presence of the thermal \SZ\ effect at low frequency. The ratio of the
brightness temperature changes caused by the effects is
\begin{eqnarray}
 {\Delta T_{\rm kinematic} \over \Delta T_{\rm thermal}}
     &=& {1 \over 2} \, {v_{\rm z} \over c} \,
        \left( {\boltz \te \over \me c^2} \right)^{-1} \nonumber \\
     &=& 0.085 \, (v_{\rm z} / 1000 \ \kms) \, (\boltz \te /10 \ \keV)^{-1}
\end{eqnarray}
which is small for the expected velocities of a few hundred $\kms$ or
less, and typical cluster temperatures of a few $\keV$. However, the
thermal and kinematic effects may be separated using their different
spectra: indeed, in the Kompaneets approximation it is easy to show
that the kinematic effect produces its maximum intensity
change at the frequency at which the thermal effect is zero.
\par
Thus observations near $x=3.83$ ($218 \ \GHz$) are sensitive mostly to the
kinematic effect, but in interpreting such observations it is necessary
to take careful account
of the temperature-dependence of the shape of the thermal \SZ\
effect's spectrum, and of the frequency of the 
null of the thermal effect (Equation~\ref{eq-approxroots};
Fig.~\ref{fig-zerothermal}), as emphasized by
Rephaeli\markcite{rep95a} (1995a). The first strong limits on the
peculiar velocities of clusters of galaxies derived using this
technique are now becoming available (e.g., Holzapfel
\etal\markcite{holz97b} 1997b; see Sec.~\ref{sec-clustervels}).  
\par
Although this technique measures only the peculiar radial velocity of
a cluster of galaxies, the other velocity components may be measured
using the specific intensity changes caused by gravitational lensing
(e.g., Birkinshaw \& Gull\markcite{bg83a} 1983a, corrected by Gurvits
\& Mitrofanov\markcite{gm86} 1986; Pyne \& Birkinshaw\markcite{pb93}
1993). These fractional intensity changes are 
small, of order $\theta_{\rm grav} \, (v_{\rm xy}/c)$, where
$\theta_{\rm grav}$ is the gravitational lensing angle, and 
$v_{\rm xy}$ is the velocity of the cluster across the observer's line
of sight.\footnote{This is a special case of a more general class of
intensity-changing effects, often referred to as Rees-Sciama effects
(after Rees \& Sciama\markcite{rs68} 1968), which arise when the
evolution of spacetime near a cluster (or other massive object)
differs from the evolution of the metric of the Universe as a whole
(Pyne \& Birkinshaw\markcite{pb93} 1993).} For typical cluster masses
and sizes, the gravitational 
lensing angle is less than about 1~arcmin, so that 
$ (\Delta I / I) \lesssim 10^{-6} \, (v_{\rm xy}/1000 \ \kms)$,
whereas the kinematic \SZ\ effect may be an order of magnitude
stronger. The possibility of measuring $v_{\rm z}$ and $v_{\rm xy}$
separately then depends on the different angular patterns of the
effects on the sky: the transverse motion produces a characteristic
dipole-like effect near the moving cluster, with an angular structure
which indicates the direction of motion on the plane of the
sky. Nevertheless, substantial improvements in techniques are going to
be required to measure these other velocity terms in this way.
\par
Clusters of galaxies produce further microwave background anisotropies
through the same spacetime effect, if they are expanding or contracting 
(Nottale\markcite{not84} 1984; Pyne \& Birkinshaw\markcite{pb93}
1993). A contaminating \SZ\ effect must also appear at the same time
if an expanding or collapsing cluster contains associated gas because
of the anisotropy of inverse-Compton scattering (Molnar \&
Birkinshaw\markcite{mo98b} 1998b), but the sizes of these effects are
too small to be detectable in the near future. 
\par
An interesting extension of this work would be to use the kinematic
\SZ\ effect from a radio source to measure the speed of the
radio-emitting plasma. Just as for a cluster of galaxies, the presence
of a scattering medium which is moving relative to the CMBR will
produce a kinematic \SZ\ effect which is proportional to $\tau_{\rm e}
\, (v_{\rm z}/c)$, but whereas $v_{\rm z}$ should be small for a cluster
relative to the Hubble flow, the velocity of the radio-emitting plasma
in a radio galaxy may be a substantial fraction of the speed of
light, and a large kinematic \SZ\ effect will be seen if the optical
depth of the radio-emitting plasma is sufficient. A complication here
is that the location of the null of the non-thermal \SZ\ effect 
depends on the spectrum of the electrons, and that the cross-terms
between the electron energy and the plasma velocity are no longer
small. Thus the approximation of a cold plasma that was made in the
derivation above, to calculate the spectrum of the kinematic \SZ\
effect, is no longer valid and a more complete calculation must be
performed using proper angular averages over the (anisotropic)
electron distribution function (see Molnar\markcite{mo98} 1998).
%
%
\section{Polarization and the \SZ\ effect\label{sec-polarization}}
In the previous sections of this review I have concentrated on the
\SZ\ effects in the specific intensity, the Stokes $I$
parameter. Any effects in the polarized intensity, the Stokes $Q$,
$U$, and $V$ terms, will be of smaller order by factors
$\tau_{\rm e}$, or $v/c$. 
An early reference was made to polarization terms in the paper by
Sunyaev \& Zel'dovich\markcite{sz80b} (1980b), with particular
reference to their use to measure the velocities of clusters of
galaxies across the line of sight. A more thorough discussion of
polarization effects in inverse-Compton scattering is given by
Nagirner \& Poutanen\markcite{np94} (1994). All the polarization terms
depend on higher powers of $\tau_{\rm e}$,  $v_{\rm xy}$, or $v_{\rm
z}$ than the thermal (or non-thermal) and kinematic effects discussed
earlier, and therefore are not detectable with the current generation
of experiments, although they may be measured in the future.
\par
The simplest polarization term arises from multiple scatterings of
photons within a scattering atmosphere. If a plasma distribution lies
in front of a radio source, then the scattering of unpolarized
radiation from that source by the surrounding atmosphere will produce
a polarized halo, with the fractional polarization proportional to
$\tau_{\rm e}$ but depending on the detailed geometry of the
scattering process. For a radio source located centrally behind a
spherical atmosphere, the pattern of polarization is
circumferential. Similarly, scattering of the thermal (or non-thermal)
\SZ\ effect by the same plasma producing the effect will produce a
polarization, which may be circumferential (at high frequencies, where
the \SZ\ effect appears as a source) or radial (at low frequencies,
where it appears as a ``hole''). The peak polarization in this case
will be less than a fraction $\tau_{\rm e}$ of the \SZ\ effect itself,
or less than $\tau_{\rm e} y$ relative to the overall
CMBR. For the particularly prominent cluster CL~0016+16
(Sec.~\ref{sec-thermal}), this factor is $\approx 2 \times 10^{-6}
h_{100}^{-1}$, so that polarized signals of order $1 \ \muK$
are the most that might be expected. 
\par
The motion of the plasma cloud across the line of sight also
introduces polarization effects, from the Thomson scattering of the
anisotropic radiation field in the frame of the moving cluster. The
two largest contributions to the 
polarized intensity in this case were identified by Sunyaev \&
Zel'dovich\markcite{sz80b} (1980b) as a component of 
about $0.1 \, \tau_{\rm e} \, (v_{\rm xy}/c)^2$ of the CMBR
intensity, due to single scatterings of the quadrupolar term in the 
anisotropic radiation field seen in the frame of the moving cluster, and a
component of about $0.025 \, \tau_{\rm e}^2 \, (v_{\rm xy}/c)$ from
repeated scatterings of the dipolar term in the radiation
field. Taking CL~0016+16 as an example again, the first of these
polarizations is roughly a fraction
$3 \times 10^{-9} \, h_{100}^{-1/2} \, (v_{\rm xy}/1000 \ \kms)^2$ of
the intensity of the CMBR, while the second is of order 
$3 \times 10^{-8} \, h_{100}^{-1  } \, (v_{\rm xy}/1000 \ \kms)  $ of
the CMBR intensity. Neither signal is likely to be measurable in the
near future.
\par
Similar effects will arise in the case of the non-thermal \SZ\ effect,
but here the anisotropy of the electron distribution function is
likely to be more significant. Polarized synchrotron
radiation is also likely to be a bad contaminating signal for
observational studies of the \SZ\ effect from relativistic populations
of electrons. 
\par
No useful observational limits have yet been set on these polarization
terms, and considerable development of observational techniques would
be needed to make possible the measurement of even the largest of
these effects. 
%
%
\section{Measurement techniques\label{sec-meas}}
Three distinct techniques for the measurement of the Sunyaev-Zel'dovich
intensity effects in clusters of galaxies are now yielding reliable
results. This section reviews single-dish radiometric 
observations, bolometric observations, and interferometric
observations of the effects, 
emphasizing the weaknesses and strengths of each technique and the types of
systematic error from which they suffer. A discussion of the
constraints on observation of the non-thermal effect is contained in
the discussion of bolometric techniques. No concerted efforts at 
measuring the polarization \SZ\ effects have yet been made, and so only
intensity-measuring techniques will be addressed here. 
\subsection{Single-dish radiometer measurements \label{sec-radiom}}
The original technique used to detect the \SZ\ effects made use of
existing radio telescopes on which large tranches of observing time
could be obtained. This always meant the older single-dish telescopes,
so that the measurements were made using traditional radiometric
methods. This is exemplified by the early work of Gull \&
Northover\markcite{gn76} (1976) using the Chilbolton 25-m telescope,
or the more recent work of Uson\markcite{uson86} (1986) on the NRAO
140-foot telescope. These telescopes tend to have beam-sizes of a few
arcminutes at microwave frequencies, which is a fairly good match to
the angular sizes of the moderately distant clusters of galaxies which
X-ray astronomy was then beginning to study. With such large and
general-purpose telescopes, it was impossible to make major
modifications that would optimize them for observations of the
microwave background radiation, and much early work had to cope with
difficulties caused by the characteristics of the telescopes through
minor changes to the receiver package or careful design of the
observing strategy.
\par
The closest clusters of galaxies (at redshifts less than about $0.05$)
have larger angular sizes, and it is possible to observe the
\SZ\ effects using smaller telescopes. In such cases is has been
possible to rework existing antennas to optimize them for microwave
background observations --- both of the \SZ\ effects and primordial
structures (for example, using the OVRO 5.5-m telescope; 
Myers \etal\markcite{myers97} 1997). This is now leading to a generation of 
custom-designed telescopes for sensitive measurements of the CMBR:
some ground-based and some balloon-based systems should be in use in
the near future.
\par
A simple estimate of the sensitivity of a single-dish observation
is of interest. A good system might have a noise temperature of about
$40$~K (including noise from the atmosphere) and a bandwidth of
$1$~GHz. Then in 1~second, the radiometric accuracy of a simple
measurement will be $0.9$~mK, and a differenced measurement, between
the center of a cluster of galaxies and some reference region of blank
sky, would have an error of $1.3$~mK. Thus if problems with variations
in the atmosphere are ignored, it would appear that a measurement with
an accuracy of $10 \ \muK$ could be made in 4.4~hours. 
\par
This observing time estimate is highly optimistic, principally because
of emission from the Earth's atmosphere. 
Sensitive observations with large or small single dishes always use
some differencing scheme in order to reduce unwanted signals from the
atmosphere (or from the ground, appearing in the sidelobes of the
telescope) to below the level of the astronomical signal
that is being searched for. Consider, for example, observations at
20~GHz, for which the atmospheric optical depth may be $\sim 0.01$ in
good conditions at a good site. The atmospheric signal will then be of
order $3 \ \K$, several thousand times larger than the \SZ\ effects,
and atmospheric signals must be removed to a part in $10^5$ if precise
measurements of the \SZ\ effects are to be made.
\par
The simplest scheme for removing the atmospheric signal is simply to
position-switch the beam of an antenna between the direction of
interest (for example the center of some cluster) and a reference
direction well away from the cluster. The radiometric signals measured
in these two directions are then subtracted. If the atmospheric
signal has the same brightness at the cluster center as at the reference
position, then it is removed, and the difference signal contains only
the astronomical brightness difference between the two positions.
Thus the reference position is usually chosen to be offset
in azimuth, so that the elevations and atmospheric path lengths of the
two beams are as similar as possible.
\par
Of course, the sky in the target and reference
directions will be different because of variations in the properties
of the atmosphere with position and time, and because of the varying
elevation of the target as it is tracked across the sky. Nevertheless,
if the target and reference positions are relatively close together,
the switching is relatively fast, and many observations are made, it
might be expected that sky brightness differences between the target
and reference positions would average out with time. The choice of
switching angle and speed is made to try to optimize this process,
while not spending so much time moving the beams that the efficiency
of observation is compromised.
\par
An alternative strategy is to allow the sky to drift through the beam of the
telescope or to drive the telescope so that the beam is moved across 
the position of the target. The time sequence of sky brightnesses
produced by such a drift or driven scan is then converted to a
scan in position, and fitted as the sum of 
of a baseline signal (usually taken to be a low-order polynomial
function of position) and the \SZ\ effect signal associated with the
target. Clearly, structures in the atmosphere will cause the
baseline shape to vary, but provided that these structures lie on
scales larger than the angular scale of the cluster, 
they can be removed well by this technique. Many scans are
needed to average out the atmospheric noise, and this technique is often
fairly inefficient, because the telescope observes baseline regions
far from the cluster for much of the time.
\par
In practice these techniques are unlikely to be adequate, because of
the amplitude of the variations in brightness of the atmosphere with
position and time: at most sites the sky noise is a large contribution
to the overall effective noise of the observation.
Nevertheless, the first reported detection of a \SZ\ effect
(towards the Coma cluster, by Parijskij\markcite{par72} 1972)
used a simple drift-scanning technique, with a scan length of about
$290$~arcmin and claimed to have measured an effect of 
$-1.0 \pm 0.5$~mK.
\par
More usually a higher-order scheme has been employed. At cm
wavelengths, it is common for the telescope to be equipped with 
multiple feeds so that two or more directions on the sky can be
observed without moving the telescope. The difference between the
signals entering through these two feeds is measured many times per
second, to yield an ``instantaneous'' beam-switched sky signal. On a
slower timescale the telescope is position-switched, so that the sky
patch being observed is moved between one beam and the other. At mm
wavelengths it is common for the beam switching to be provided not by 
two feeds, but by moving the secondary reflector,
so that a single beam is moved rapidly between two positions on the
sky. This technique would allow complicated differencing 
strategies, if the position of the secondary could be controlled
precisely, but at present only simple schemes are being used.
Arrays of feeds and detectors are now in use on some telescopes, 
and differencing between signals from different elements of these 
feed arrays also provide the opportunity for new switching strategies
(some of which are already being used in bolometer work, see
Sec.~\ref{sec-bolom}).
\par
Table~\ref{tab-radiom}, which reports the critical observing
parameters used in all 
published radiometric observations of the \SZ\ effects, indicates the
switching scheme that was used. Most measurements have been made
using a combination of beam-switching (BS) and position-switching
(PS), because this is relatively efficient, with about half the
observing time being spent on target. Some observations, including all
observations by the 
Effelsberg group, have used a combination of beam-switching and drift
or driven scanning (DS). The critical parameters of these techniques
are the telescope beamwidth (the full width to half maximum, FWHM),
$\theta_{\rm h}$, the beam-switching angle, $\theta_{\rm b}$, and the
angular length of the drift or driven scan, $\theta_{\rm s}$. Some of
the papers in Table~\ref{tab-radiom} have been partially or fully
superseded by later papers, and are marked $\ast$.
\par
\begin{figure}[p]
\epsscale{0.7}
\plotone{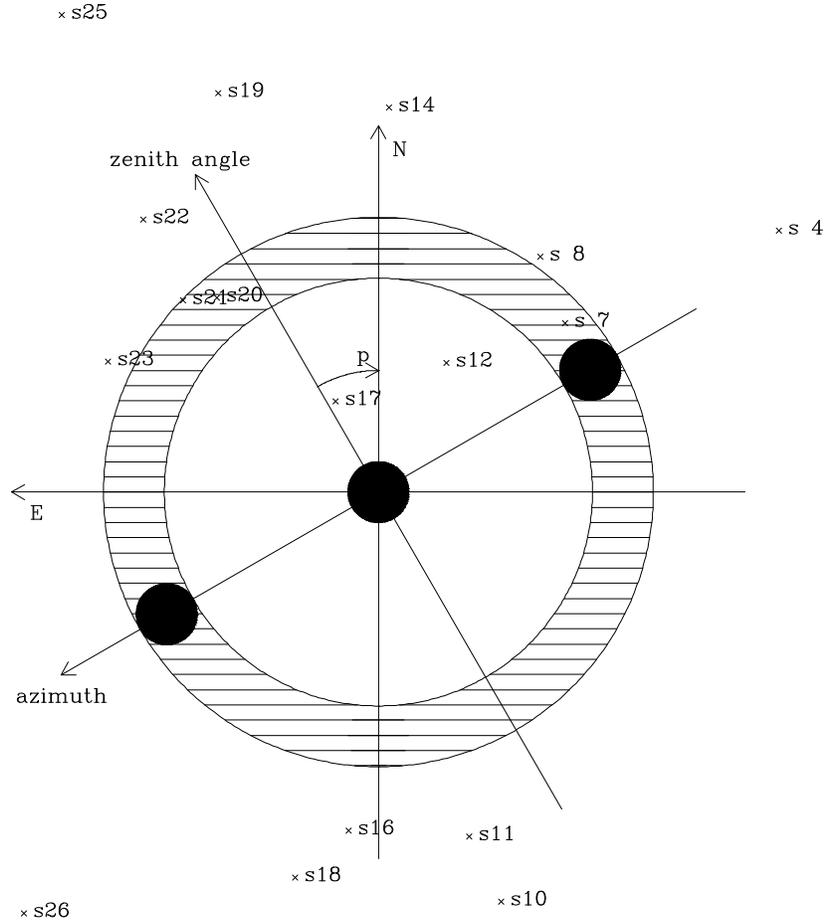}
\caption{\label{fig-switching}
A representative beam and
position-switching scheme, and
background source field: for observations of a point near the center
of Abell~665 by the Owens Valley Radio Observatory 40-m telescope.
The locations of the on-source beam and reference beams 
in this symmetrical switching experiment are shown
as solid circles. The main beam is first pointed at the center of the
cluster with the reference beam to the NW while beam-switched data are
accumulated. The position of the main beam is then switched to the SE
offset position, with the reference beam pointed at the center of the
cluster, and more beam-switched data are accumulated. Finally, the
main beam is again pointed at the cluster center. As the observation
extends in time, the offset beam locations sweep out arcs
about the point being observed, with the location at any one time
conveniently described by the parallactic angle, $p$. 
Since Abell~665 is circumpolar from the
Owens Valley, the reference arcs close about the on-source position:
however, the density of observations is higher at some 
parallactic angles because $p$ is not a linear function of time (see
equation~\ref{eq-parallactic}).}
\end{figure}

In techniques that use a combination of beam-switching and position
switching, the beam and position 
switching directions need not be the same, and need
bear no fixed relationship to any astronomical axes. However, the
commonest form of this technique (illustrated in 
Fig.~\ref{fig-switching}) has the telescope equipped with a twin-beam
receiver with the two beams offset in azimuth. Since large
antennas are usually altazimuth mounted, it is convenient also to
switch in azimuth (and so keep the columns of atmosphere roughly
matched between the beams). In any one
integration interval (typically some fraction of a second) the output
of the differential radiometer is proportional to the brightness
differences seen by the two beams (possibly with some offset because
of differences in the beam gains, losses, etc.). The configuration used by 
Readhead \etal\markcite{read89} (1989) in their observations of primordial
anisotropies in the microwave background radiation is typical. In the
simplest arrangement, where the two beams (A and B) are pointed towards
sky positions $1$ and $2$, and the signals from the feeds enter 
a Dicke switch, followed by a
low-noise front-end receiver, and then are synchronously detected by a
differencing backend, the instantaneous output of the radiometer,
$\Delta P$, was written by Readhead \etal\ as
\begin{eqnarray}
  \Delta P &=& \phantom{+} G \biggl(
                   g_{\rm A} (1 - l_{\rm A}) T_{\rm sky1}
                 - g_{\rm B} (1 - l_{\rm B}) T_{\rm sky2}
                              \biggr) \nonumber \\
           &  & + G \biggl(
                   g_{\rm A} (1 - l_{\rm A}) (T_{\rm atm1} + T_{\rm gndA})
                 - g_{\rm B} (1 - l_{\rm B}) (T_{\rm atm2} + T_{\rm gndB})
                              \biggr) \nonumber \\
           &  & + G \biggl(
                   g_{\rm A} \int \Theta_{\rm A} dl_{\rm A}
                 - g_{\rm B} \int \Theta_{\rm B} dl_{\rm B}
                              \biggr) \nonumber \\
           &  & + G \biggl(
                   (g_{\rm A} - g_{\rm B}) T_{\rm maser}
                              \biggr) \quad .
 \label{eq-radiometeroffset}
\end{eqnarray}
Here $G$ is the gain of the front-end system (a maser in the OVRO
40-m telescope configuration used by Readhead \etal), with an
equivalent noise temperature of $T_{\rm maser}$. $g_{\rm A}$ and
$g_{\rm B}$ are the back-end gains corresponding to the A and B (main
and reference) feeds, and $l_{\rm A}$ and $l_{\rm B}$ are the losses
in the feeds, waveguides, and Dicke switch associated with the two
channels. These losses are distributed over a number of components,
with temperatures $\Theta_{\rm A}$ and $\Theta_{\rm B}$ which range
from the cryogenic temperatures of the front-end system to the ambient
temperatures of the front of the feeds.
\par
It is clear that the instantaneous difference power between channels A
and B may arise from a number of causes other than the sky temperature
difference $(T_{\rm sky1} - T_{\rm sky2})$ which it is desired to
measure. In general, the atmospheric signal greatly exceeds the
astronomical signal towards any one position on
the sky ($T_{\rm atm1} \gg T_{\rm sky1}$) so that small
imbalances in the atmospheric signal between the two beams dominate
over the astronomical signals that are to be measured. Over
long averaging times, and with the same zenith angle coverage in the
two beams, it is expected that $\langle T_{\rm atm1} - T_{\rm atm2}
\rangle \approx 0$. The accuracy with which this is true will depend
on the weather patterns at the telescope, the orientation of the beams
relative to one another, and so on.
\par
The ground pickup signals through the two beams are likely to be
different, since the detailed shapes of the telescope beams are also
different. This leads to an imbalance $T_{\rm gndA}-T_{\rm gndB}$, and
an offset signal in the radiometer. The amplitude of this signal is
reduced to a minimum by tapering the illumination of the primary
antenna, so that as little power as possible arrives at the feed from
the ground. Some protection against ground signals is also achieved by
operating at elevations for which the expected spillover signal is
smallest. However, since the ground covers a large solid angle
there are inevitably reflection and diffraction effects that cause 
offsets from differential ground spillover. 
\par
The losses $l_{\rm A}$ and $l_{\rm B}$ in the two channels of the
radiometer are reduced to a minimum by keeping the waveguide lengths
to a minimum and using the best possible microwave components, 
but it is impossible to ensure that the losses are equal. Furthermore, the
temperatures of the components in which these losses occur depend
critically on where they are in the radiometer housing, so that the
radiated signals from the components, 
$\int \Theta_{\rm A} dl_{\rm A}$ and $\int \Theta_{\rm B} dl_{\rm B}$
are likely to be significantly different. Once again, this produces an
offset signal between the two sides of the radiometer.
\par
These problems are exacerbated by their time variations. It is likely
that the gains and temperatures will drift with time. The receiver
parameters are stabilized as well as possible, but are still
seen to change slowly. The atmosphere and ground pickup temperatures
change more significantly, with varying weather conditions and
varying elevations of the observation. Thus simple beam-switched
measurements of the Sunyaev-Zel'dovich effect are unlikely to be
successful, even after filtering out periods of bad weather and rapid
temperature change when the atmospheric signal is unstable.
\par
The level of differencing introduced by position-switching removes many
of these effects to first order in time and position on the sky. The
standard position-switching technique points one beam (A, the ``main
beam'') at the target position for time $\tau$, with the second beam
(B, the ``reference beam'') offset in azimuth to some reference
position, then switches the reference beam onto the target for time $2
\tau$, with the main beam offset to a reference position on the
opposite side of the target, then switches back for a final time
$\tau$ with the original beam on the target. If the total cycle time
($4\tau + s_1 + s_2$, where $s_1$ and $s_2$ are the times spent
moving) is small, then the reference positions observed by beams A~and
B do not change appreciably during a cycle, and the combination
\begin{equation}
   S =  {1 \over \tau} \int_{0}^{\tau} \Delta P dt
      - {2 \over \tau} \int_{\tau+s_1}^{3\tau+s_1} \Delta P dt
      + {1 \over \tau} \int_{3\tau+s_1+s_2}^{4\tau+s_1+s_2}
                            \Delta P dt 
 \label{eq-radiometerdata}
\end{equation}
is a much better measurement of the sky temperature difference
between the target and the average of two points to either side of it
(offset in azimuth by the beam-switching angle, $\theta_{\rm b}$) than
the estimate in (\ref{eq-radiometeroffset}). This is so even when
the move and dwell times in the different 
pointing directions change slightly, for example because of variations
in windage on the telescope. If $\tau$ is chosen to be small,
then quadratic terms in the time and position variations of
contaminating effects in (\ref{eq-radiometeroffset}) can be made very
small, but at the cost of much reduced efficiency
$4\tau/(4\tau+s_1+s_2)$ in the switching cycle. For observations with
the OVRO 40-m telescope made by Readhead \etal\markcite{read89} (1989)
and Birkinshaw \etal\markcite{bghm} (1998), $\tau$ was chosen to be
about 20~sec, and even large non-linear terms in the telescope
properties are expected to be subtracted to an accuracy of a few $\mu
\rm K$. 
\par
Even at this degree of differencing, it is important to check that the 
scheme is functioning properly. For this reason, the best work has
included either a check of regions of nominally blank sky near the
target point, or a further level of differencing involving the
subtraction of data from fields leading and following the target field
by some interval. A representative method (Herbig 
\etal\markcite{herbig95} 1995) consists of  
making a few ($\sim 10$) observations using the beam-switching plus
position switching technique described above at the target field,
referenced to the same number of observations on offset regions
before and after the target field, with the time interval arranged so
that the telescope 
moves over the same azimuth and elevation track as the target
source. The off-target data may be treated as controls, or may be
directly subtracted from the on-target data to provide another level
of switching which is likely to reduce the level of differential
ground spillover. In either case, rigorous controls of this type
necessarily reduce the efficiency of the observations by a factor~2
or~3. Alternatively, observations can be made of closer positions
(perhaps even overlapping with the reference fields of the target
point), without attempting an exact reproduction of the azimuth and
elevation track on any one day, but allowing an equal coverage to
build up over a number of days. This was the approach used by
Birkinshaw \etal\markcite{bghm} (1998).
\par
The various beam switching schemes that have been used are described
in detail in the papers in Table~\ref{tab-radiom} that discuss
substantial blocks of measurements. Quantitative estimates of their
systematic errors from differential ground spillover, residual
atmospheric effects, or receiver drifts, are also usually given. 
Whichever beam-switching technique is used, it is advisable to use the
same technique to 
observe control fields, far from known X-ray clusters, where the
expected measurement is zero. Systematic errors in the technique are
then apparent, as is the extra noise in the data caused by primordial
structures in the CMBR. It is important to realize
that the \SZ\ effect plus primordial signal at some point can
be measured to more precision than the systematic error on the \SZ\
effect that is set by the underlying spectrum of primordial
fluctuations. That is, the measurement error is a
representation of the reproducibility of the measurement, which is the
difference between the brightness of some point relative to a weighted
average of adjacent points. Noise from the spectrum of primordial
fluctuations must be taken into account if realistic
errors on physical parameters of a cluster are to be deduced from
measured \SZ\ effect data.
\par
A further difficulty encountered with single-dish observations is that
of relating the measured signal from the radiometer (in volts, or
some equivalent unit) to the brightness temperature of a
\SZ\ effect on the sky. The opacity of the atmosphere can be
corrected using tip measurements, and generally varies little during
periods of good weather, so that the principal problem is not one of
unknown propagation loss but rather one of calibration. Generally the absolute
calibration of a single-dish system is tied to observations of
planets, with an internal reference load in the radiometer being
related to the signal obtained from a planet. If that planet has solid
brightness temperature $T_{\rm p}$, then the output signal is
proportional to $T_{\rm p}$, with a constant of proportionality which
depends on the solid angle of the planet, the telescope beam pattern,
etc. Thus by measuring the telescope beam pattern and the signal from
planets, it is possible to calibrate the internal load. The accuracy
of this calibration is only modest because of
\begin{enumerate}
\item measurement errors in the planetary signal, from opacity
 errors in the measurement of the transparency of the atmosphere,
 pointing errors in the telescope, etc.,
\item uncertainties in the brightness temperature scale of the
 planets, and in the pattern of brightness across their disks, and
\item variations in the shape of the telescope beam (and hence
 the gain) over the sky.
\end{enumerate}
\noindent
Thus, for example, the recent measurements of Myers
\etal\markcite{myers97} (1997) are 
tied to a brightness temperature scale using the measurement of the
brightness temperature of Jupiter at 18.5~GHz (Wrixson, Wright \&
Thornton\markcite{wwt71} 1971). This measurement may itself be in
error by up to 6~per cent. Difficulties may also arise from changes in
the internal reference load, which will cause the 
calibration to drift with time. Relating this load back to sky
temperatures at a later date will introduce another set of ``transfer
errors''. Even if these are well controlled, it is clear that
radiometric \SZ\ effect data contain systematic uncertainties in the
brightness scale at the 8~per cent level or worse. This calibration
error has a significant effect on the interpretation of the results. 
\par
\begin{figure}[p]
\epsscale{0.7}
\plotone{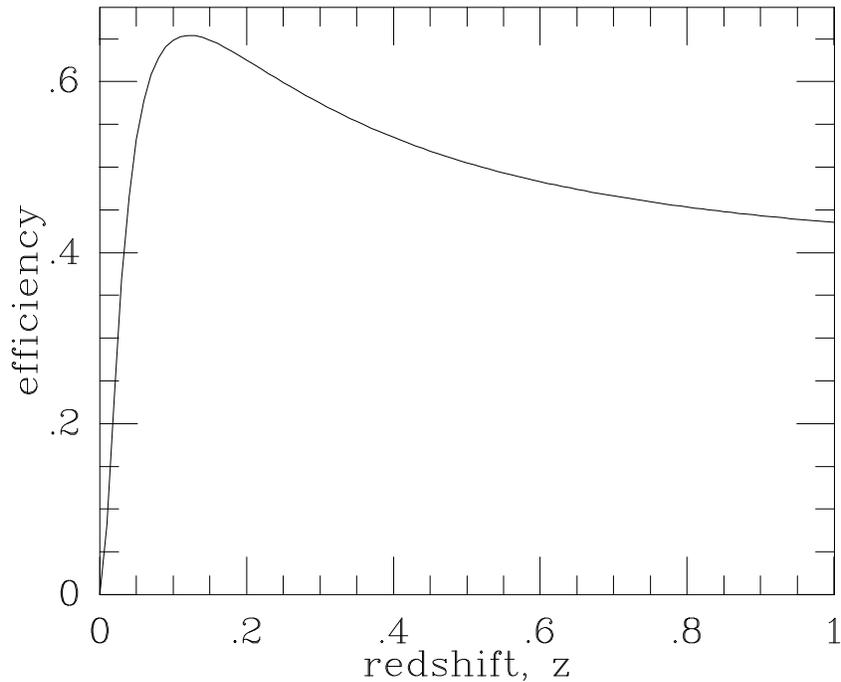}
\caption{\label{fig-zdep}
The observing efficiency factor, $\eta$,
as a function of redshift, for observations of clusters with core
radius $300 \ \kpc$ and $\beta = 0.67$, using the 
OVRO 40-m telescope at 20~GHz and assuming $h_{100} = 0.5$ and $q_0 =
{1 \over 2}$. $\eta$ is defined to be the
central effect seen by the telescope divided by the true
amplitude of the Sunyaev-Zel'dovich effect, and measures the
beam-dilution and beam-switching reductions of the cluster signal.
The decrease in $\eta$ at $z > 0.15$ is slow, so that these
observations would be sensitive to the \SZ\ effects
over a wide redshift range.}
\end{figure}

It is important to mention, at this stage, that the differencing
schemes described here have the effect of restricting the range of
redshifts for which the telescope is useful. If observations are to be
made of a cluster of galaxies at low redshift, then the angular size
of the cluster's \SZ\ effects (which are several times larger than of
the cluster's X-ray surface brightness) may be comparable to the
beam-switching angle, $\theta_{\rm b}$, and beam-switching reduces the
observable signal. Alternatively, if the cluster is at high redshift,
then its angular size in the \SZ\ effects may be smaller than the
telescope FWHM, $\theta_{\rm h}$, and beam dilution will reduce the
observable signal. The two effects compete, so that for any telescope
and switching scheme, there is some optimum redshift band for
observation, and this band depends on the structures of cluster
atmospheres and the cosmological model. An example of a
calculation of this
efficiency factor, defined as the fraction of the central
\SZ\ effect from a cluster that can be observed with the telescope, 
is shown in Fig.~\ref{fig-zdep} for the OVRO 40-m
telescope. The steep cutoff at small redshift represents
the effect of the differencing scheme, while the decrease of the
efficiency factor at large $z$ arises from the slow variation of 
angular size with redshift at $z \gtrsim 0.5$.
\par
In dealing with the variations of signal during a tracked observation
of a cluster, it is convenient to introduce the concept of parallactic
angle, the angle between the vertical circle and the declination
axis. An observation at hour angle $H$ of a source at declination
$\delta$ from a telescope at latitude $\lambda$ will occur at
parallactic angle
\begin{equation}
   p = \tan^{-1} \left( { \sin H / \cos \delta 
                          \over
                          \tan\lambda - \tan\delta \cos H} \right)
 \label{eq-parallactic}
\end{equation}
where the parallactic angle increases from negative values to positive
values as time increases (with the parallactic angle being zero at
transit; Fig.~\ref{fig-switching}) for sources south of the telescope,
and decreases from positive values to negative values for sources
north of the 
telescope. For a symmetrical beam-switching experiment, like that
depicted in Fig.~\ref{fig-switching}, the parallactic angle may be
taken to lie in 
$-90^\circ$ to $+90^\circ$. With more complicated beam-switching
schemes, which may be asymmetrical to eliminate higher-order terms
in the time or position dependence (e.g., Birkinshaw \&
Gull\markcite{bg84} 1984), the full range of $p$ may be needed.
\par
The conversion between time and parallactic
angle is particularly convenient when it is necessary to 
keep track of the radio source contamination. Many of the observations
listed in Table~\ref{tab-radiom} were made at cm wavelengths, where the
atmosphere is relatively benign and large antennas are available for long
periods. However, the radio sky is then
contaminated by non-thermal sources associated with galaxies (in the
target cluster, the foreground, or the background) and quasars, and
the effects of these radio sources must be subtracted if the
Sunyaev-Zel'dovich effects are to be seen cleanly. 
\par
\begin{figure}[p]
\epsscale{0.65}
\plotone{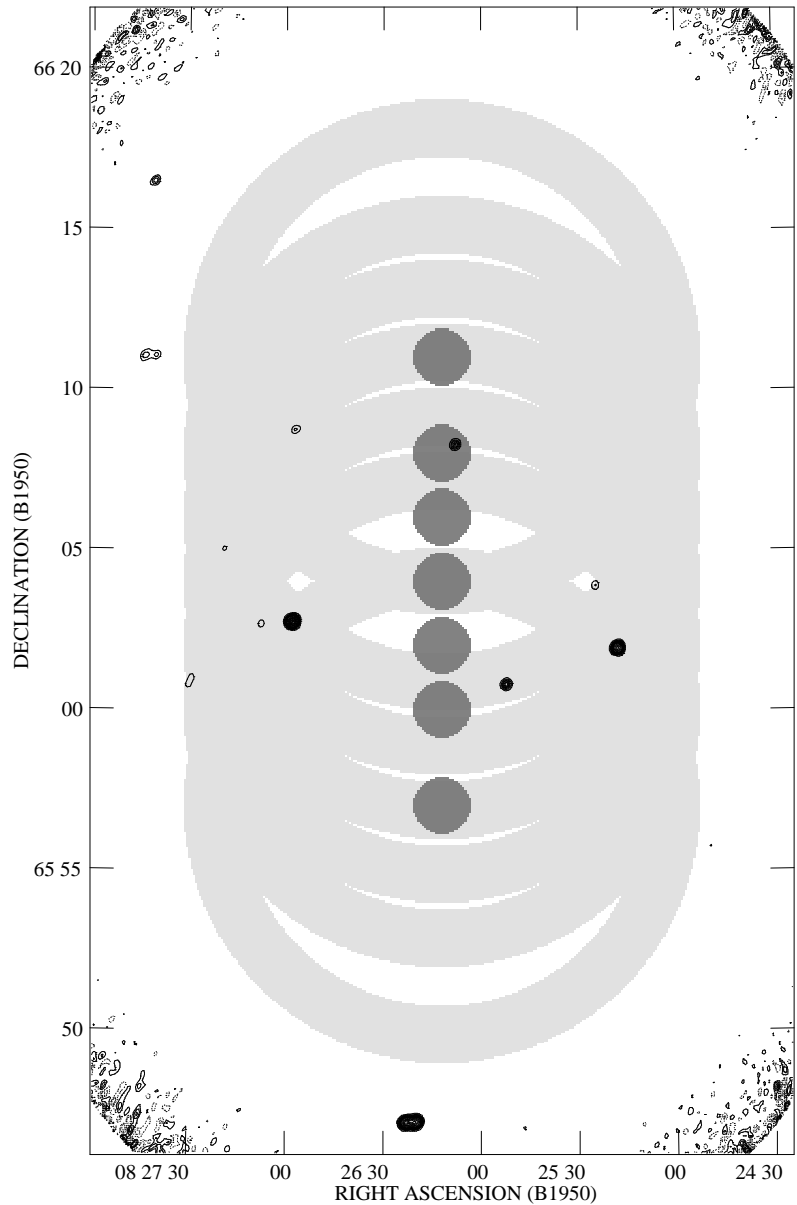}
\caption{\label{fig-a665radio}
Observing positions and radio sources in
the cluster Abell~665. The dark-grey circles represent the FWHMs of
the primary pointing positions of the Birkinshaw \etal\ (1998) \SZ\ effect
observations in the cluster, while the light-grey areas are the
reference arcs traced out by the off-position beams. A VLA 6-cm radio
mosaic of the cluster field is shown by contours. Note the appearance
of a significant radio source under the pointing position 4~arcmin
north of the cluster center. This source appears to be variable,
causing significant problems in correcting the data at that
location. Other radio sources appear near or within the reference
arcs, and cause contamination of some parts of the data.}
\end{figure}
\begin{figure}[p]
\epsscale{0.7}
\plottwo{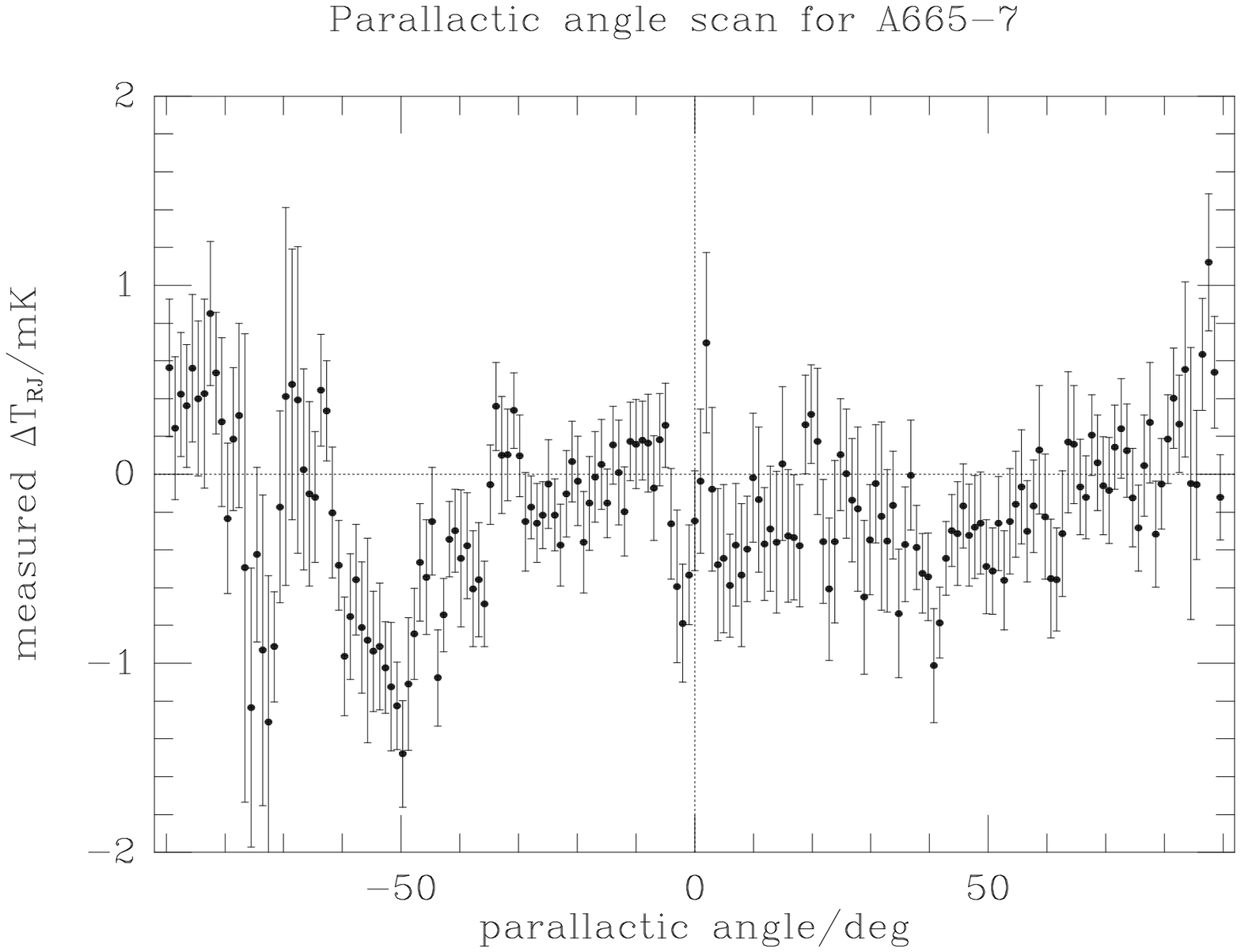}{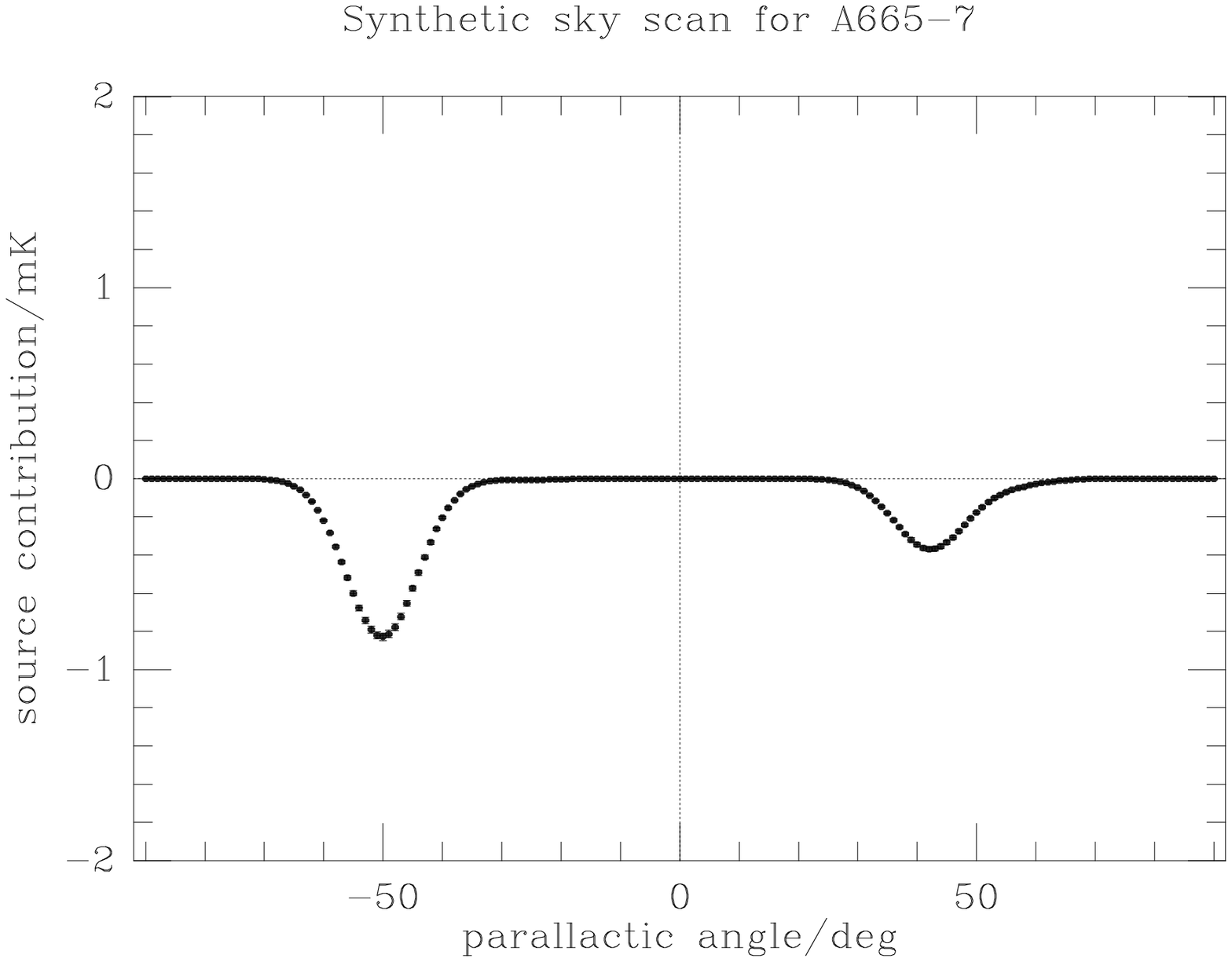}
\caption{\label{fig-a665pascan}
A comparison of the observed and
modeled parallactic angle scans for OVRO 40-m data at a point
7~arcmin south of the nominal center of Abell~665. Two features are
seen in the observed data (left). These correspond to bright radio
sources which are seen in the reference arcs at parallactic angles 
near $-50^\circ$ and $40^\circ$ (position angles 
$+40^\circ$ and $-50^\circ$ on Figs~\ref{fig-switching}
and~\ref{fig-a665radio}). The model of the 
expected signal based on VLA surveys of the cluster (Moffet \&
Birkinshaw 1989) shows features of
similar amplitude at these parallactic angles, so that moderately good
corrections can be made for the sources. The accuracy of these
source corrections is questionable because of the
extrapolation of the source flux densities to the higher frequency of
the \SZ\ effect data and the possibility that the sources are
variable.}
\end{figure}

Figure~\ref{fig-a665radio} shows a map of the radio sky near Abell~665. 
Significant radio source emission can be
found in the reference arcs of the observations at 
many parallactic angles. Such emission causes
the measured brightness temperature difference between the center and
edge of the cluster to be negative: a fake \SZ\ effect is
generated. Protection against such fake effects is implicit in the
differencing scheme. Sources in the reference arcs affect
the \SZ\ effect measurements only for the range of parallactic angles that
the switching scheme places them in the reference beam. A plot of
observational data arranged by parallactic angle therefore shows
negative features at parallactic angles corresponding to radio source
contamination (e.g., Fig.~\ref{fig-a665pascan}), and data near these
parallactic angles can be corrected for the contamination using radio
flux density measurements from the VLA, for example.
\par
This procedure is further complicated by issues of source
variability. At frequencies above 10~GHz where most radiometric
observations are made (Table~\ref{tab-radiom}), many of the brightest
radio sources are variable with timescales of months being
typical. Source subtraction based on archival data is therefore
unlikely to be good enough for full radiometric accuracy to be
recovered. Simultaneous, or near-simultaneous, monitoring of variable
sources may then be necessary if accurate source subtraction is to be
attempted, and this will always be necessary for variable sources
lying in the target locations. Variable sources lying in the
reference arcs may also be simply eliminated from consideration by
removing data taken at the appropriate parallactic angles: thus in
Fig.~\ref{fig-a665pascan}, parallactic angle ranges near $-50^\circ$
and $+40^\circ$ might be eliminated on the basis of variability or of
an imprecise knowledge of the contaminating sources. However, sources
which are so strongly variable that they appear from below the flux
density limit of a radio survey will remain a problem without adequate
monitoring of the field.
\par
Despite difficulties with radio source contamination, calibration, and
systematic errors introduced by the radiometer or spillover, 
recent observations of the \SZ\ effects using radiometric techniques
are yielding significant and 
highly reliable measurements. The detailed results and critical
discussion appear in Sec.~\ref{sec-data}, but a good example is the
measurement of the Sunyaev-Zel'dovich effect of the Coma cluster by 
Herbig \etal\markcite{herbig95} (1995), using 
the OVRO 5.5-m telescope at 32~GHz. Their result, an antenna
temperature effect of $-175 \pm 21 \ \rm \mu K$, corresponds to a
central Sunyaev-Zel'dovich effect
$\Delta T_{\rm T0} + \Delta T_{\rm K0} = -510 \pm 110 \ \rm \mu K$,
and is a convincing measurement of the Sunyaev-Zel'dovich effect from
a nearby cluster of galaxies for which particularly good X-ray and
optical data exist (e.g., White, Briel \& Henry\markcite{wbh93} 1993). 
\par
\begin{figure}[p]
\epsscale{0.7}
\plotone{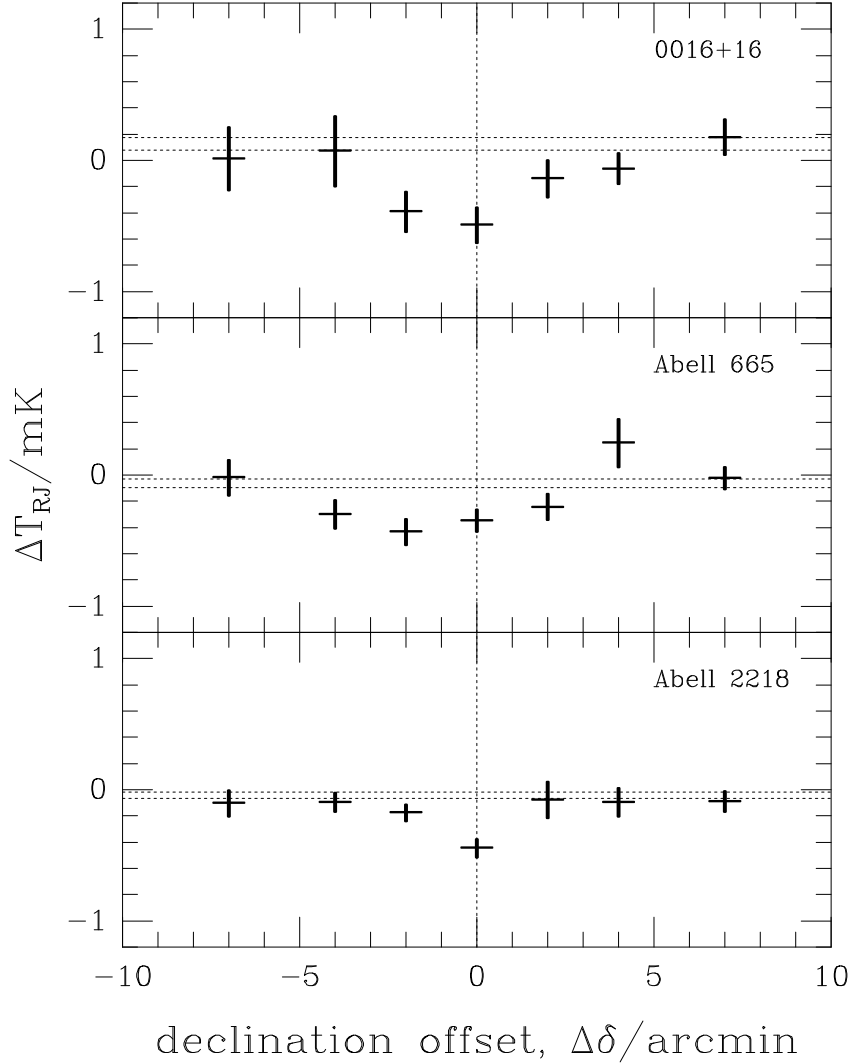}
\caption{\label{fig-sz3cluster}
Measurements of changes in the apparent brightness temperature of
the microwave background radiation as a
function of declination near the clusters CL~0016+16, Abell~665
and Abell~2218 (Birkinshaw \etal\ 1998). The largest
Sunyaev-Zel'dovich effect is seen at the point closest to
the X-ray center for each cluster (offset from the scan
center in the case of Abell~665), and the apparent angular
sizes of the effects are consistent with the predictions of
simple models based on the X-ray data. The horizontal lines delimit
the range of possible zero levels, and the error bars
include both random and systematic components.}
\end{figure}

For a few clusters, single-dish measurements have been used not only
to detect the central decrements, but also to measure the angular
sizes of the effects. This is illustrated in Fig.~\ref{fig-sz3cluster}
which, for the 
three clusters CL~0016+16, Abell~665, and Abell~2218 shows the \SZ\ effect
results of Birkinshaw \etal\markcite{bghm} (1998). The close agreement
between the centers of the \SZ\ effects and the X-ray images of the
clusters is a good indication that the systematic problems of
single-dish measurements have been solved, although 
observing time limitations and the need to check for systematic
errors restricts this work to a relatively coarse measurement
of the cluster angular structure. Much better results should be
obtained using two-dimensional arrays of detectors, as should be
available on the Green Bank Telescope when it is completed.
\subsection{Bolometric methods \label{sec-bolom}}
The principal advantage of a bolometric
system is the high sensitivity that is achieved, but these
devices are also of interest because of their frequency range: at
present they provide the best sensitivity for observing the microwave
background outside the Rayleigh-Jeans part of the spectrum, and hence
for separating the thermal and  kinematic components of the
\SZ\ effect using their different spectral shapes. 
Furthermore, the best systems consist of several detectors arranged in
an array, and some 
provide simultaneous operation in several bands. A suitable choice of
differencing between elements of the array reproduces many
of the sky-noise subtraction properties of radiometric observing, and
the multiband capability holds out the hope of rapid spectral
measurements. Bolometric measurements of the \SZ\ effects are now
becoming more common, as reliable technology becomes more widely
available (Table\ref{tab-bolom}). 
\par
A bolometer such as SCUBA on the James Clerk Maxwell Telescope (JCMT)
at a wavelength of $850 \ \rm \mu m$ (near the peak of the thermal
effect in intensity terms) has a
sensitivity of $80 \ \rm mJy \, Hz^{-1/2}$, with a 13-arcsec pixel
size. The equivalent sensitivity in the Rayleigh-Jeans brightness
temperature change of the thermal \SZ\ effect, $\DTRJ$, is
about $60$~mK in one second in each pixel, or 
about 13~mK in a 1~arcmin beam created by averaging over detector
elements (compare the radiometric sensitivity of a typical radio
telescope in Sec.~\ref{sec-radiom}). A few hours of observation should
then suffice to detect the thermal \SZ\ effect at high sensitivity,
and by using several bands (perhaps simultaneously), a coarse spectrum
of the effect could be measured. Deviations from the spectrum of the
thermal effect could then set limits to the velocities of clusters of
galaxies --- if the sensitivity of the bolometer is similar at
frequencies near the zero of the thermal effect, then a velocity
accuracy of about $6 \times 10^4 \ \kms$ can be achieved in an hour of
observation in any one 13-arcsec pixel. 
This can be reduced to $10^3 \ \kms$ or less with modern
bolometers if the measurements are averaged over the entire face of a
cluster (as, for example, in Holzapfel \etal\markcite{holz97b} 1997b).
The fundamental limit of this technique for measuring cluster peculiar
velocities may be set not by sensitivity, but rather by the background
fluctuations in the CMBR which arise from primordial
anisotropies. This depends on the angular spectrum of anisotropies
(see Sec.~\ref{sec-cobe}).
\par
Although the raw sensitivity of bolometer systems is high because of
the large bandpasses and sensitive detector elements, a problem with
the technique is the extremely high sky
brightness against which observations must be made. Coupled with the
varying opacity of the sky, this implies that telescopes on high, dry,
sites are essential for efficient observing --- balloon operations are
possible, and the CalTech Submillimeter Observatory (CSO) on Mauna Kea
has been used successfully. Antarctic operations are also an
interesting future possibility, as is space operation with bolometer
arrays. At present, the best results are obtained by differencing out
atmospheric signals using bolometer arrays. This involves the use of
small differencing angles, and introduces limitations on the selection
of clusters that are similar to those that apply to radiometric work
(Sec.~\ref{sec-radiom}). The small angular 
separations of the beams often causes the minimum redshift cutoff to
be rather high, and the peak observing efficiency to be low
(as in Chase \etal\markcite{cjra87} 1987, for which the fraction of
the central decrement that was observable was only 0.38 for cluster
CL~0016+16).
\par
This technique is exemplified by the recent work
of Wilbanks \etal\markcite{wil94} (1994), who used the Caltech
Submillimeter Observatory (CSO) on Mauna Kea with a three-element
array to detect the Sunyaev-Zel'dovich effect from Abell~2163, a
cluster of galaxies with an exceptionally hot atmosphere (Arnaud
\etal\markcite{arn92} 1992) and a bright radio halo source (Herbig \&
Birkinshaw\markcite{hb98} 1998). The combination of drift-scanning and
element-to-element differencing used by Wilbanks \etal\ achieved an
excellent separation of the atmospheric signal from the
Sunyaev-Zel'dovich effect and provided a measurement of the
angular structure of the effect. At the wavelength of operation
($\lambda = 2.2$~mm) radio source confusion is not a problem. This is
not the case at microwave frequencies, where observations of the \SZ\ effect
in Abell~2163 are severely affected by the radio environment near the
cluster center, which includes a variable and inverted-spectrum radio
source as well as the radio halo (Herbig \& Birkinshaw\markcite{hb98}
1998). Nevertheless, recent observations at 18~GHz with the OVRO 40-m
telescope have succeeded in detecting the effect near the cluster
center, at about the level seen by Wilbanks \etal\markcite{wil94}
(1994).
\par
\begin{figure}[p]
\epsscale{0.8}
\plotone{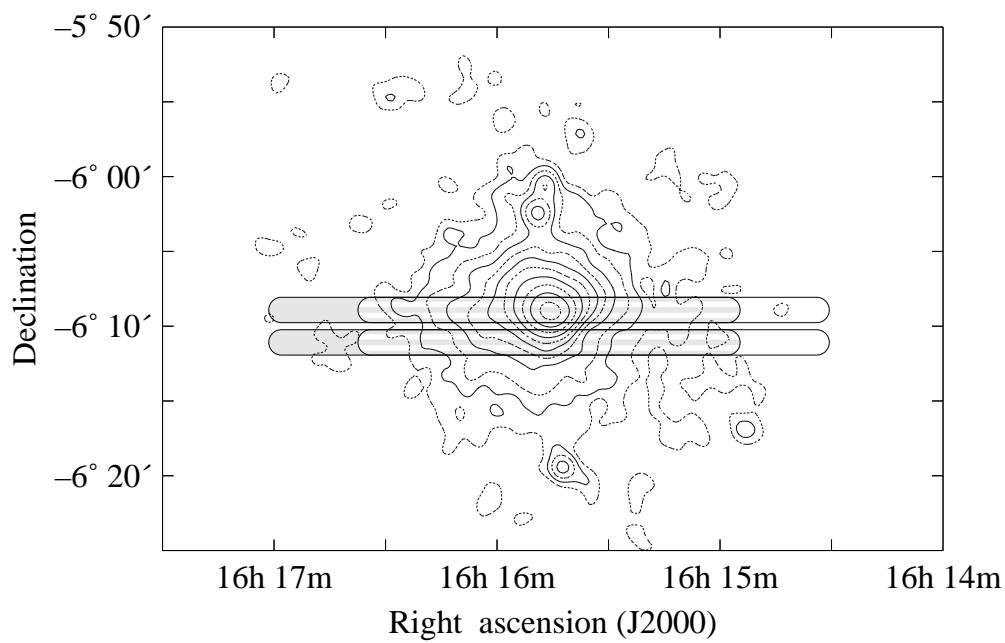}
\caption{\label{fig-suziegeom}
Sample SuZIE drift scans across 
Abell~2163 superimposed on an X-ray contour map of the cluster
(Holzapfel \etal\ 1997a). The two rows of SuZIE
detectors are separated 
by 2.2~arcmin, so that when the upper detectors pass over the X-ray
center, the lower detectors pass south of the center. Two sets of
scans are shown for each row of detectors, since the observations were
alternately begun 12 and 18~arcmin ahead of the cluster center.}
\end{figure}
\begin{figure}[p]
\epsscale{0.8}
\plotone{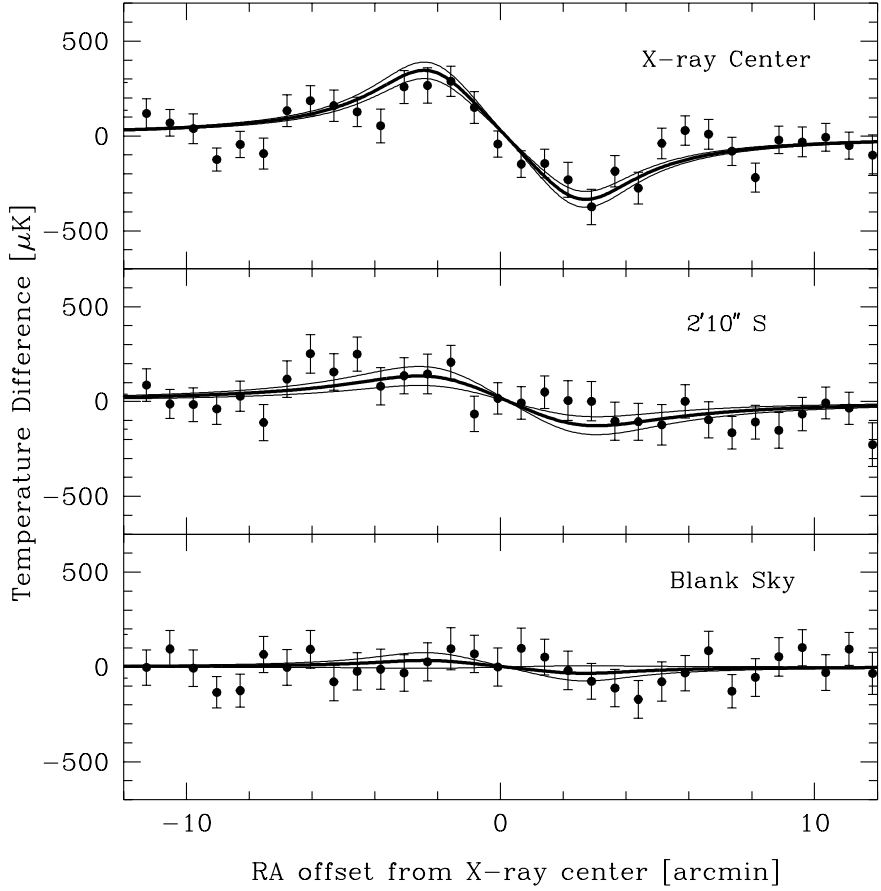}
\caption{\label{fig-suziescan}
Sample 4.6-arcmin difference data from
the 1994 SuZIE observations of Abell~2163 (Holzapfel
\etal\ 1997a). The upper panel shows data taken
across the cluster center 
together with the best-fitting (non-isothermal) model of the cluster
\SZ\ effect based on the X-ray data (heavy line) and the same model
with $\pm 1\sigma$ errors on the amplitude (upper and lower light
lines). The middle panel compares the predictions of the same model
with the data taken 2.2~arcmin south of the cluster center, while the
bottom panel shows the corresponding model fit to a region of blank
sky well separated from the cluster but at a similar declination.}
\end{figure}

The most sensitive observations with bolometers (with SuZIE, the
Sunyaev-Zel'dovich Infrared Experiment on the CSO) have been made
using a drift-scan mode (Holzapfel \etal\markcite{holz97a} 1997a), as
illustrated in Fig.~\ref{fig-suziegeom}, in order to reduce
microphonic and sidelobe spillover effects to the minimum possible
level. The SuZIE array consists of two rows of three elements, with
the rows separated by 2.2~arcmin and the elements in each array
separated by 2.3~arcmin. Array elements within a row are
electronically differenced to produce continuous measurements of the
brightness differences that they see. During a drift-scan each
difference voltage is then proportional to the brightness difference
on the sky between two locations which vary as the sky rotates past
the detectors. The array is oriented with the long axis parallel to
right ascension, so that the time series can be interpreted as a right
ascension scan (as in Fig.~\ref{fig-suziescan}). Repeated drift-scans,
with the angle of the array changed from scan to scan, then allow
repeated measurements of brightness differences at the same points on
the sky.
\par
A simple isothermal model of Abell~2163 (Holzapfel
\etal\markcite{holz97a} 1997a) has $\beta = 0.62 \pm 0.03$ and $\thc =
1.2 \pm 0.1$~arcmin (in equation~\ref{eq-nebeta}). 
The 2.3 and 4.6-arcmin difference signals that SuZIE produces
then correspond to peak observing efficiencies (fractions of the
central \SZ\ effect seen by each 1.75-arcmin FWHM array element) of
0.31 and 0.51, respectively. With this type of observing, the signals
returned by SuZIE are close to being measurements of the gradient of
the \SZ\ effect on the sky, as can be seen in the data shown in
Fig.~\ref{fig-suziescan}. 
\par
Just as for radiometric work, it is important to check that the
observing technique being used does not suffer from baseline effects
from parasitic signals from the sky, the telescope, or the
electronics. Control observations of regions of blank sky are
used to provide such checks, as in the example of
Fig.~\ref{fig-suziescan}. In all plots in this figure, a 
best-fitting linear baseline has been removed, and then the data have
been fitted using a model of the cluster Abell~2163: only small
residual baseline effects remain, and the fits are of reasonable
quality. 
\par
As with the radiometer data, it is important to remove from the data
periods when the sky is opaque, or has rapidly-varying opacity, and
the data must also be corrected for the line-of-sight opacity through
the atmosphere. At the best millimetric wavelengths these corrections
are small, just as they are for most cm-wave observations. Also, as
radiometer data must be cleaned of radio interference, so bolometer
data must be cleaned of cosmic ray hits. In both cases, this does not
create additional difficulties because the effects are generally large
and obvious.
\par
A final similarity with radiometric work is the problem of calibrating
the data into absolute temperature (or intensity) units. Again, the
calibration is usually made by reference to the brightness of planets,
and again the difficulty is that the planetary temperature scale is
good to 6~per cent at best. Additional errors from the beam-pattern
of the detectors, the bandpasses of the detector elements, and the
opacity of the atmosphere add to this error, so that the 
intensity scale of any measurement is not known to better than about
8~per cent. The effect of this on the interpretation of the data will
become apparent later. 
\par
No radio source or Galactic contamination signals are thought to be
significant at the frequencies and angular resolutions at which
bolometric data are taken on clusters (Fischer \& Lange\markcite{fl93}
1993), and dusty galaxies within the clusters should also be
weak. Nevertheless, such signals are present (Smail, Ivison \&
Blain\markcite{sib97} 1997), and may be enhanced by 
emission from distant (background) dusty, star-forming galaxies 
gravitationally lensed by clusters --- especially by the massive
clusters which produce the strongest \SZ\ effects
(Blain\markcite{bla98} 1998). If 
the bolometer array that is used has sufficient angular resolution, it
should be possible to reduce this contamination by removing the
individual pixels in the map that are affected, but at present only
low-resolution bolometric observations of the \SZ\ effect exist (e.g.,
from SuZIE, with 1.7~arcmin resolution, Table~\ref{tab-bolom}).
Higher-resolution observations of clusters (e.g., with SCUBA on the
JCMT) are now possible and should allow checks for the presence of
confusing sources, and then their subtraction from the
lower-resolution data. 
\par
Since differencing in bolometric work usually involves switching over
angles which are only a small multiple of the FWHM of the array
elements, the observing efficiencies are low. This has led to the
results from these experiments usually being 
quoted in terms of fitted central \SZ\ effects (or, equivalently, the
$y$ parameter) rather than the beam-averaged central \SZ\ effect that
is usually quoted in radiometric measurements. Quoting
the results as central $y$ values 
has the virtue of encapsulating the combined
statistics of the observational errors and the angular structure data
(Fig.~\ref{fig-suziescan}) into a single 
number, but it also has the drawback of not allowing the data to be
re-interpreted later, as improved structural information becomes available. 
From the data in Fig.~\ref{fig-suziescan}, Holzapfel
\etal\markcite{holz97a} (1997a) 
find that Abell~2163 has a central Comptonization parameter $y = (3.7
\pm 0.4) \times 10^{-4}$ if the cluster gas follows a simple 
isothermal model. The corresponding central
Rayleigh-Jeans brightness temperature change is $-1.6 \pm 0.2$~mK: a
remarkably large \SZ\ effect, presumably
because of the high temperature of the atmosphere in this cluster,
although uncertainties in the model, which is based on X-ray
data, cause additional $\sim 10$~per cent uncertainties
in the values of $y$ and the central \SZ\ effect that are derived.
\par
\begin{figure}[p]
\epsscale{1.2}
\plotone{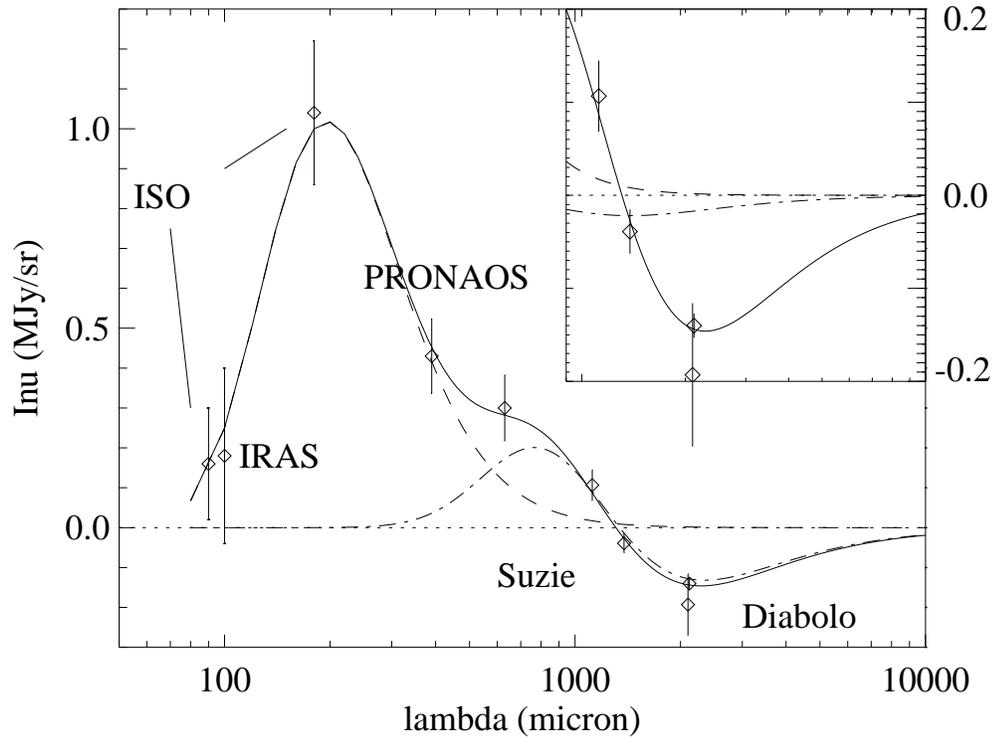}
\caption{\label{fig-a2163mm}
The mm to far-IR spectrum of Abell~2163 (from Lamarre \etal\
1998). The solid line shows a best-fit model composed of dust
emission and \SZ\ effects. The dashed line shows the dust contribution
to the overall spectrum. The dash-dotted line shows the \SZ\ thermal
effect. The insert shows the contribution of the kinematic \SZ\ effect
in the mm to cm part of the spectrum.}
\end{figure}

An interesting recent result on the spectrum of the \SZ\ effect from
Abell~2163 is shown in Fig.~\ref{fig-a2163mm} (Lamarre
\etal\markcite{lam98} 1998). Lamarre \etal\ combined data taken using
several instruments into a single spectrum which 
shows the relative sizes of the \SZ\ effect and far-IR dust-like
emission (which dominates from $100 - 1000 \ \mu \rm m$). This
shorter-wavelength emission may arise from the lensed population of
background starburst galaxies, from Galactic dust which happens to be
brighter near the centre of the cluster, or from dust in Abell~2163
itself. If the spectrum in Fig.~\ref{fig-a2163mm} is characteristic of
other clusters of galaxies, then the interpretation of sub-mm data 
will need to take careful account of such contamination. This might
particularly affect the measurement of the kinematic \SZ\ effect. 
\subsection{Interferometric methods \label{sec-interf}}
The two techniques discussed provided most of 
the existing data on the Sunyaev-Zel'dovich effect until very
recently. Both techniques are excellent for large-scale surveys of
clusters of galaxies which are well matched to the beam-switching
technique being used, but provide only modest angular resolutions on
the sky (although higher-resolution and two-dimensional bolometer
arrays are now becoming available) and hence are suitable only for
simple mapping (as in Figs~\ref{fig-sz3cluster}
and~\ref{fig-suziescan}). Radio interferometry
is a powerful method for making 
detailed {\it images} of \SZ\ effects. Such images are
valuable for making detailed comparisons with X-ray images, and can
also measure accurate \SZ\ effects while avoiding some of the
systematic difficulties of the other techniques. Perhaps for these
reasons, interferometry is the most rapidly-growing area for
observation of the \SZ\ effects (Table~\ref{tab-interf}).
\par
The extra resolution that is available using interferometers is
also a handicap. Interferometers work by measuring some range of
Fourier components of the brightness distribution on the sky: the
correlation of signals from a pair of antennas produces a response
which is (roughly) proportional to a single Fourier component of the
brightness of the source. For ``small'' sources, observed with narrow
bandwidths and short time constants, the measured source visibility is
\begin{equation}
   {\cal V}(u,v) \propto \int_{-\infty}^{\infty} \, d\xi \,
                         \int_{-\infty}^{\infty} \, d\zeta \,
                         B(\xi,\zeta) \,
                         G(\xi,\zeta) \,
                         e^{-2 \pi i (u\xi + v\zeta)}
 \label{eq-fouriermap}
\end{equation}
where $B(\xi,\zeta)$ is the brightness distribution of the sky, 
$G(\xi,\zeta)$ represents the polar diagram of the antennas of the
interferometer, $(u,v)$ are the separations of the antennas, measured
in wavelengths, $(\xi,\zeta)$ are direction cosines relative to the
center of the field of view, and the constant of proportionality
depends on the detailed properties of the interferometer (see
Thompson, Moran \& Swenson\markcite{tms86} (1986) for a detailed
explanation of the meaning of this expression and the assumptions that
go into it). An image of the sky brightness distribution,
$B(\xi,\zeta)$, can be recovered from the measurements ${\cal
V}(u,v)$, by a back Fourier transform and division by the polar
diagram function: alternatively, estimation techniques can be used to
measure $B(\xi,\zeta)$ directly from the ${\cal V}(u,v)$. 
\par
Most interferometers were originally designed to achieve high angular
resolution. The finiteness of interferometer measurements means that
not all $(u,v)$ values are sampled: in particular, the design for high
resolution means that the antennas are usually placed so that their
minimum separation is many wavelengths (and always exceeds the antenna
diameter by a significant factor). The Fourier relationship
(\ref{eq-fouriermap}) means that the short baselines contain
information about the large angular scale structure of the source, and
so there is some maximum angular scale of structure that is sampled
and imaged by interferometers. The \SZ\ effects of clusters of
galaxies have  angular sizes of several arcminutes --- most
interferometers lose (``resolve out'') signals on these or larger
angular scales, and hence would find extreme difficulty in detecting
\SZ\ effects.
\par
\begin{figure}[p]
\epsscale{0.9}
\plotone{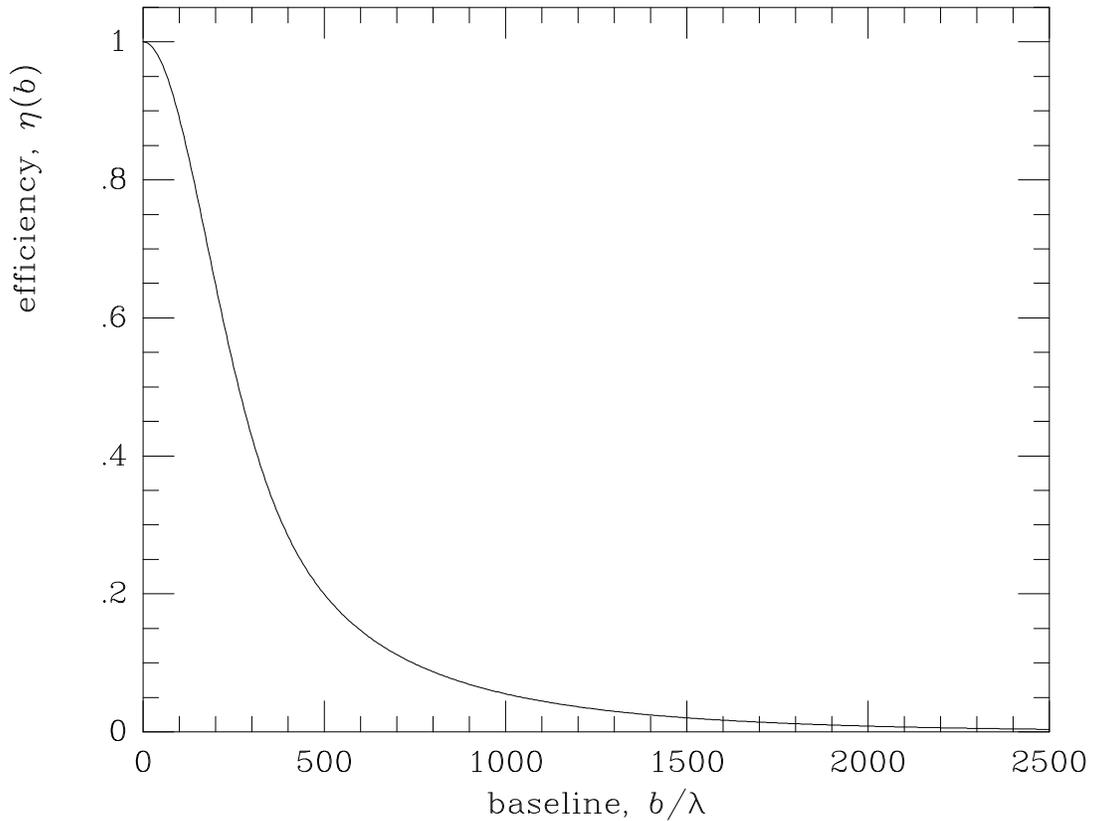}
\caption{\label{fig-vlaresponse}
The interferometer response that
would be expected from a 6-cm VLA observation of the \SZ\ effect from
cluster CL~0016+16, normalized to the effect with zero baseline. The
minimum separation of VLA antennas is 650~wavelengths, but projection
effects mean that the minimum observable baseline is roughly equal to
the antenna diameter, or about 420~wavelengths. Thus on the shortest
one or two baselines, for a brief interval, the VLA can observe about
25~per cent of the total available \SZ\ effect (which corresponds to
about $-0.9 \ \mJy$ if the central \SZ\ effect is about $-1$~mK).
Consequently, the VLA is a poor
instrument for observing the \SZ\ effect in this cluster.}
\end{figure}

Figure~\ref{fig-vlaresponse} illustrates this effect for model Very
Large Array (VLA) observations of cluster CL~0016+16 at $\lambda =
6$~cm. Since the VLA antennas shadow one another at baselines less
than the antenna diameter (of 25~m), no information about the
amplitude or shape of the 
visibility curve can be recovered at baselines less than
$420\lambda$. Most of the VLA baselines are much larger than the
minimum baseline, even in the most compact configuration
(D~array). Hence the VLA's effective sensitivity to the \SZ\ effect in
CL~0016+16 is low. But CL~0016+16 is a cluster at redshift $0.5455$, 
has a small angular size, and so represents one of the best candidate
clusters for observation with the VLA --- the VLA is therefore not a
useful instrument for measuring the \SZ\ effects of any clusters
unless those clusters contain significant small-scale substructure in
the \SZ\ effect, or the clusters can have significantly smaller
angular sizes and substantial \SZ\ effects.
Thus, for example, the VLA observations of Partridge
\etal\markcite{part87} (1987) suffered from this effect: in their data
the \SZ\ effect signal from Abell~2218 was strongly suppressed because of 
the excessive size of the array.
\par
Smaller interferometers would allow the \SZ\ effects to be
measured. What is needed is an array of antennas whose individual
beam-sizes are significantly larger than the angular sizes of the
cluster \SZ\ effects, so that many antenna-antenna baselines can be
arranged to be sensitive to the effects. A first attempt to customize
a telescope for this experiment was the upgrade of the 5-km
telescope at Cambridge, UK into the Ryle telescope (Birkinshaw \&
Gull\markcite{bg83b} 1983b; Saunders\markcite{sau95} 1995). In its new
configuration, the five central
12.8-m diameter antennas can occupy a number of parking points which
provide baselines from 18~m to 288~m. At the prime operating
wavelength of 2~cm, the maximum detectable \SZ\ effect signal is
about $-1.3 \ \mJy$, and several baselines should see effects in excess
of $-0.1$~mJy.
\par
The choice of operating wavelength for mapping the \SZ\ effect is
constrained to some extent by confusion, in the same way that the
radiometric observations are affected. Some clusters of
galaxies (particularly clusters of galaxies with strong \SZ\
effects; Moffet \& Birkinshaw\markcite{mb89} 1989) contain cluster
halo sources, with similar angular size to the cluster as a whole and
whose non-thermal radio emission can swamp the \SZ\ effects at low
frequencies (although their non-thermal \SZ\ effects are probably
small; Sec.~5). Such sources 
have steep spectra, and so are avoided by working at higher
frequencies. Clusters of galaxies also contain a population of radio
sources, many of which are extended (the wide angle tail sources,
narrow angle tail sources, etc.). These extended sources are also
avoided by working at high frequency, where their extended emission is
minimized and where the small-scale emission can be recognized by its
different range of Fourier components. Background, flat-spectrum,
radio sources can also affect the data, but can be
recognized by their small angular size. 
\par
Interferometers with a wide range of baselines are useful in this
respect: the longer baselines are sensitive to the small-angular scale
radio sources which dominate the radio confusion signal (and which
affect the radiometric data: see Fig.~\ref{fig-a665radio}), while the shorter
baselines contain both the radio source signal and the \SZ\ effect
signal. Thus the longer-baseline data can be analysed first to locate
the confusing radio sources, and then these sources can be subtracted
from the short-baseline data, so that a source-free map of the sky can
be constructed and searched for the \SZ\ effect. Furthermore, by
tuning the range of baselines that are included in the final map, or
by appropriately weighting these baselines, a range of image
resolutions can be produced to emphasize any of a range of angular
structures. 
\par
Of course, this technique depends on there being a good separation of
angular scales between the radio sources and the \SZ\ effects in the
clusters: extended, cluster-based, radio sources cannot be removed
reliably using this technique, and there are a number of clusters in
which no good measurements of (or limits to) the \SZ\ effects can be
obtained without working at a higher frequency with a smaller
interferometer (to avoid resolving out the \SZ\ effect). A good choice
of operating frequency might be 90~GHz, with antenna baselines of a
few metres: a design which also commends itself for imaging primordial
fluctuations in the background radiation.
\par
Since many of the brightest radio sources at the frequencies for which
interferometers are used are variable (with timescales of months being
typical), the subtraction technique must sometimes be applied to
individual observing runs on a cluster, rather than to all the data 
taken together. The brightest sources may also subtract imperfectly
because of dynamic range problems in the mapping and analysis of the
data: generally interferometric or radiometric observations of
clusters are only attempted if the radio source environment is
relatively benign. Any source contamination at a level $\gtrsim
10$~mJy is likely to be excessive, and to cause difficulties in
detecting the \SZ\ effects, let alone mapping them reliably. 
Nevertheless, interferometric work has the advantage over radiometric
work that the sources (in particular the variable, and hence small
angular size sources) are monitored simultaneously with the \SZ\
effect, and so interferometer maps should show much better source
subtraction.
\par
Although the interferometric technique is extremely powerful, in
taking account of much of the radio source confusion, and in allowing
a map of the \SZ\ effect to be constructed, it does suffer from some
new difficulties of its own. First, the range of baselines over which
the \SZ\ effect is detected may be highly restricted, so that the
``map'' is little more than an indication of the location of the most
compact component of the \SZ\ effects. This problem can only be solved
by obtaining more short baselines, which may not be possible because
of excessive antenna size (as with the VLA, for example).
\par
The source subtraction may also cause problems, since strong sources
outside the target clusters 
often lie towards the edges of the primary beam of the antennas of the
interferometer. Small pointing errors in the antennas can then cause
the amplitude of these sources to modulate significantly, adding to
the noise in the map and reducing the accuracy with which the
contaminating source signal can be removed from the \SZ\ effect.
The problem is worst for sources lying near the half-power point
of the primary beam, but significant difficulties can be caused by
sources lying even in distant sidelobes, although this extra noise
does not usually add to produce a coherent contaminating signal at the
map center, where the \SZ\ effect is normally expected.
\par
Careful attention must also be paid to the question of correlator
errors, which can produce large and spurious signals near the
phase-stopping center (see Partridge \etal\markcite{part87} 1987). In
order to avoid 
excessive bandwidth smearing for contaminating sources which must be
identified and removed successfully, it is also normal to observe using
bandwidth synthesis methods (which split the continuum bandpass of the
interferometer into a number of channels). The combination of these
individual channel datasets back into a continuum map of the \SZ\
effect may sometimes be complicated by steep (or strongly-inverted)
sources on the image which have different fluxes in the
different channels.
\par
One major advantage of using an interferometer is that the effects of
structures in the atmosphere are significantly reduced. Emission from
the atmosphere is important only in its contribution to the total noise power
entering the antennas, since this emission is uncorrelated over
baselines longer than a few metres and does not enter into the
(correlated) visibility data. Furthermore, there are no background
level problems: an interferometer does not respond to a constant
background level, and so a well-designed interferometer will not
respond to constant atmospheric signals, the uniform component of the
microwave background radiation, large-scale gradients in galactic
continuum emission, or ground emission entering through the telescope
sidelobes.
\par
\begin{figure}[p]
\epsscale{0.8}
\plottwo{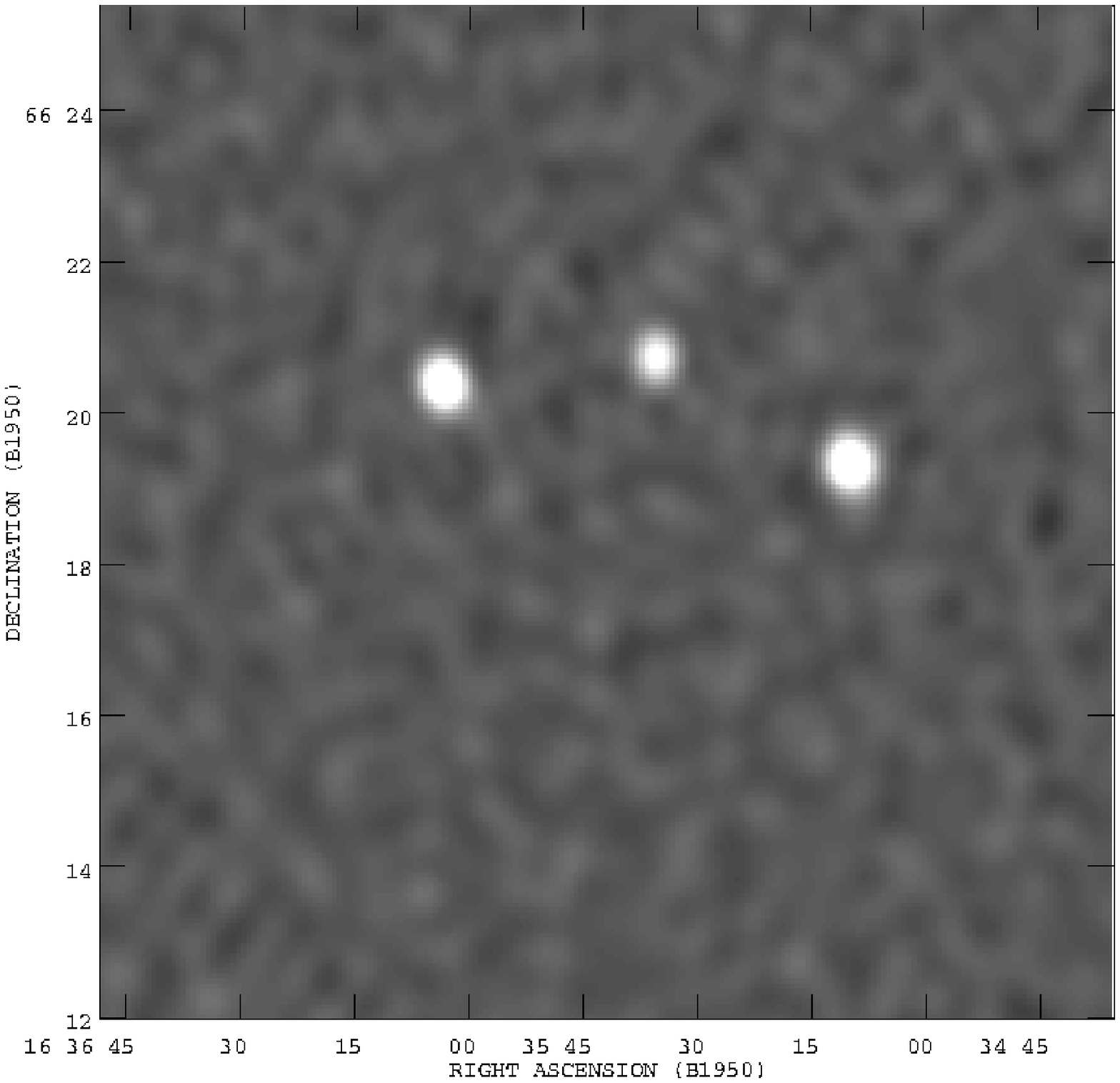}{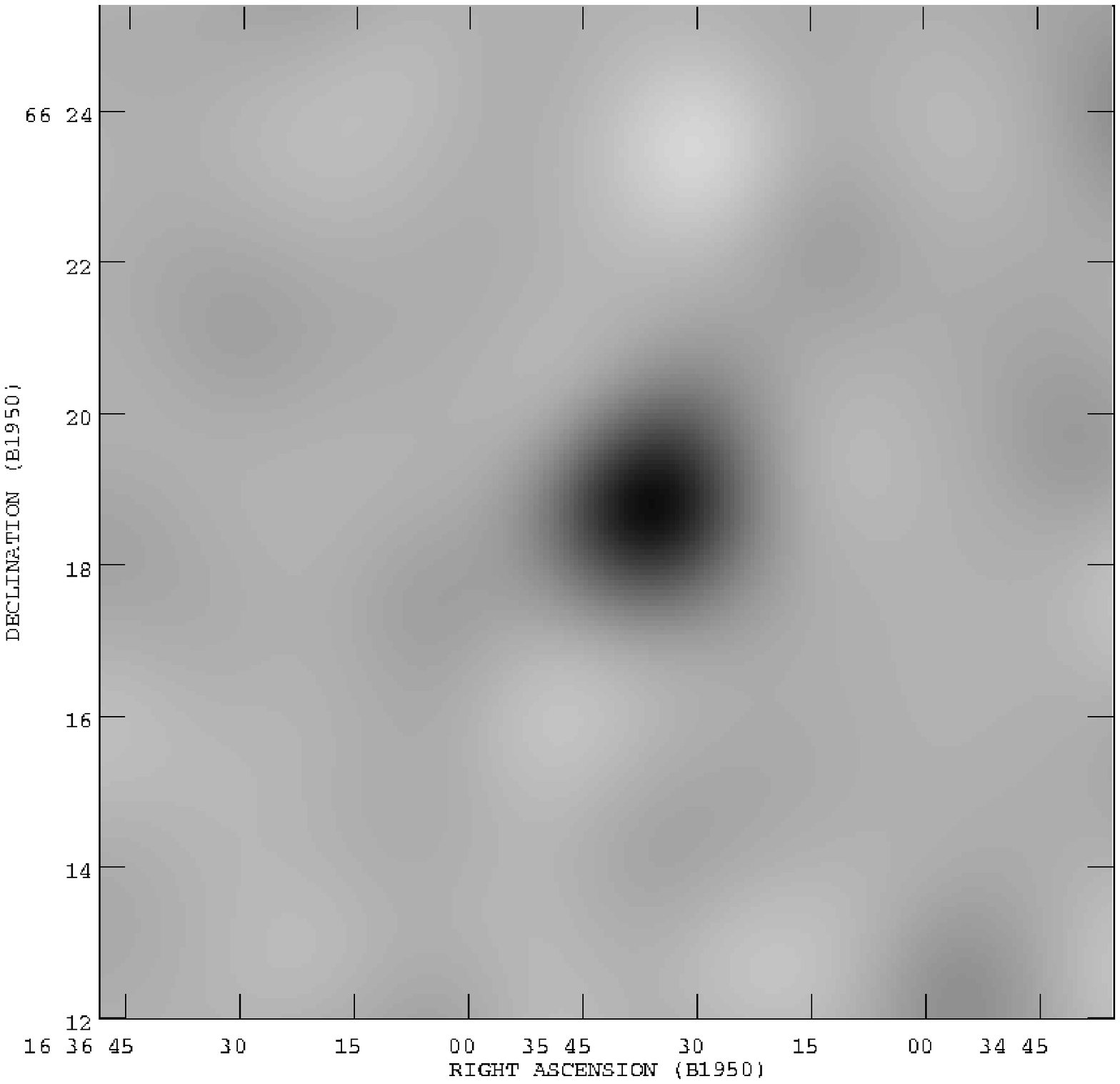}
\caption{\label{fig-rylea2218}
Interferometric maps of Abell~2218, made
with the Ryle telescope of the Mullard Radio Astronomy Observatory
(Jones \etal\ 1993). Left: an image made with the longer-baseline
data, which is sensitive chiefly to small angular scales. Three faint
radio sources dominate the image. Right: an image made with the
short-baseline data, after subtraction of the sources detected on the
long-baseline image. Here the image is dominated by the \SZ\ effect
from the cluster.}
\end{figure}

The first cluster for which interferometric techniques were used
successfully is Abell~2218, which had been shown to have a strong \SZ\
effect with a small angular size using single-dish measurements
(Birkinshaw, Gull \& Hardebeck\markcite{bgh84} 1984). Jones
\etal\markcite{jon93} 
(1993) used the Ryle interferometer at 15~GHz, with baselines
from 18~to~108~m, to locate sources and to map the diffuse
Sunyaev-Zel'dovich effect. The images that they obtained are shown in
Fig.~\ref{fig-rylea2218}. Using baselines from 36~to 108~m, and 27
12-hour runs, a high signal/noise map of the cluster radio sources was
made (Fig.~\ref{fig-rylea2218}, left). Using only the 18-m baseline,
and subtracting the signals from 
these sources, a map with effective angular resolution about 2~arcmin
was then made (Fig.~\ref{fig-rylea2218}, right). This clearly shows a
significant negative signal, of
$-580 \pm110 \ \rm \mu Jy$, centered at 
$\rm 16^h 35^m 47^s$ $+66^{\circ} 12' 50''$ (J2000).  The
corresponding value for the central \SZ\ effect in the cluster cannot
be determined without knowing the shape of the efficiency curve
(e.g., Fig.~\ref{fig-vlaresponse}, which is effectively a visibility
curve) on baselines 
less than those that were observed.  The Ryle interferometer data
could be fitted with models of the form (\ref{eq-ycirc}), with a 
parameter space extending from $\beta \simeq
0.6$, $\thc \simeq 0.9$~arcmin, $\Delta T_0
\simeq -1.1$~mK, to $\beta \simeq 1.5$, $\thc \simeq 2.0$~arcmin,
$\Delta T_0 \simeq -0.6$~mK. The random error on the detection of an
\SZ\ effect is, therefore, much smaller than the systematic error in
the central measurement of the effect --- a better range of baselines,
and a detection of the \SZ\ effect on more than a single baseline,
would be needed to improve this situation.
\par
Much analysis of the \SZ\ effect can usefully be carried out in the
data, rather than the map, plane --- by fitting the model
${\cal V}(u,v)$ to the measured visibilities. Indeed, the most
reliable indication of the reality of a \SZ\ effect may be its
presence first in visibility plots (like Fig.~\ref{fig-vlaresponse}),
and such plots are invaluable for assessing the extent of the missing
visibility data in $(u,v)$, and hence the fraction of the full \SZ\
effect of a cluster that is being detected by the interferometer. Of
course, similar calculations are needed for radiometric and bolometric
observations of the \SZ\ effects, but the efficiency factors $\eta(b)$
are often lower in interferometric work, and so the sampling of the
full \SZ\ effect is more critical to its interpretation.
\par
More recently, excellent imaging data on the clusters CL~0016+16 and
Abell~773 has been published by Carlstrom, Joy \&
Grego\markcite{cjg96} (1996). These 
authors used the Owens Valley Millimeter Array (OVMMA) at 1~cm: by
equipping an array designed for operation at 3~mm and shorter
wavelengths with cm-wave receivers, they were assured of 
accurate pointing and a relatively large primary beam, so that the
interferometer should not over-resolve the \SZ\ effects on short
baselines. The total negative flux
density of CL~0016+16 in this operating configuration is near $-13$~mJy
if the cluster has a central decrement of $-1$~mK, 
so that the cluster should be relatively strong (negative) source. With the
OVMMA, Carlstrom \etal\markcite{cjg96} (1996) detected a total
negative flux density 
of $-3.0$~mJy after 13~days of observation: their map of the cluster
is shown in Fig.~\ref{fig-ovro0016}. 
\par
The power of radio interferometric mapping of a cluster is apparent in
Carlstrom \etal's map of
the \SZ\ effect from CL~0016+16. The \SZ\ decrement
is extended in the same position angle as the X-ray
emission (Fig.~\ref{fig-0016pspc})
and the distribution of optical galaxies (and close to the
position angle from the cluster to a companion cluster; Hughes
\etal\markcite{hbh95} 1995). The small-scale structure seen in this
image is close to that predicted from the X-ray image, and corresponds
closely with the predicted amplitude based on earlier radiometric
detections of the \SZ\ effect of the cluster (Uson\markcite{uson86}
1986; Birkinshaw\markcite{b91} 1991). 
\par
\begin{figure}[p]
\epsscale{0.8}
\plotone{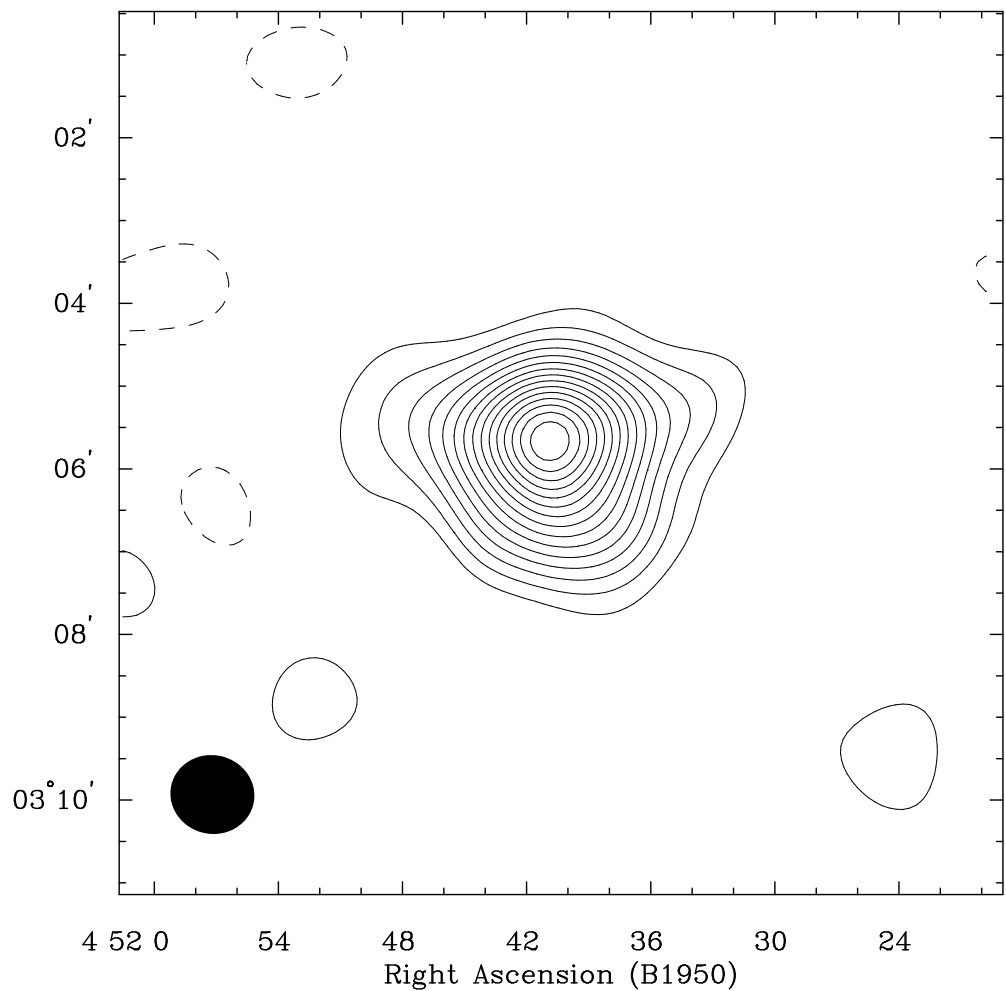}
\caption{\label{fig-ms0451}
An image of the \SZ\ effect from the cluster MS~0451.6-0305 at
$z=0.55$, as measured by Joy \etal\ (in preparation) using the OVMMA. The
beamshape of the image is shown in the lower left 
corner. This cluster was first detected in the Einstein
Medium-Sensitivity Survey (Gioia \etal\ 1990b), and therefore can be
regarded as a part of an X-ray complete sample.}
\end{figure}

The success of recent interferometric mapping campaigns, which have
produced results such as Fig.~\ref{fig-ms0451} has
amply justified demonstrated the potential
of this technique to improve on single-dish
observations of the \SZ\ effect. The critical elements of this
breakthrough have been the development of small interferometers
dedicated to \SZ\ effect mapping over long intervals, and the
existence of stable, low-noise receivers with exceptionally wide
passbands. 
%
%
\section{\SZ\ effect data\label{sec-data}}
The techniques discussed in Section~\ref{sec-meas} have been
used to search for the 
thermal and kinematic \SZ\ effects towards a large number of clusters,
and the non-thermal \SZ\ effects towards a few radio galaxies. Over
the past few years this work has been increasingly successful, because
of the high sensitivity that is now being achieved, and the careful
controls on systematic errors that are used by all groups. The most
impressive results are those obtained from radio interferometers,
which are producing images of the cluster \SZ\ effects that can be
compared directly with images of cluster X-ray structures. In the
present section I collect all published results on \SZ\ effects of
which I am aware, and review the reliability of the measurements.
\subsection{Cluster data\label{sec-clusterdata}}
Table~\ref{tab-szdata} contains the final 
result measured in each series of observations for each of the 
clusters that has been observed in the \SZ\ effect. Not all
papers in Tables~\ref{tab-radiom}, \ref{tab-bolom}, and
\ref{tab-interf} are represented in Table~\ref{tab-szdata}, since 
I have excluded interim reports where they have been superseded by later
work (which often involves improved calibrations and assessments of
systematic errors). The column marked ``O/C'' reports whether the
quoted value of $\DTRJ$ is as observed or as deconvolved, by the
observers, into some central estimated \SZ\ effect. As explained in
Secs~\ref{sec-bolom} and~\ref{sec-interf}, model-fitting to produce a
central decrement is commonly used when only a small fraction of the
central decrement can be recorded by the telescope.
\par
The overall set of clusters for which \SZ\ effects have been sought does
not constitute a well-defined sample in any sense. Early work on the
\SZ\ effects concentrated on clusters with strong X-ray sources, or
for which the radio source contamination was known to be small.
Abell~426 (the Perseus cluster) is an example of a cluster observed
for the first reason, despite its strong radio sources (Lake \&
Partridge\markcite{lp80} 1980). Abell~665, on the other hand, was observed
principally because it was known to be largely free of strong radio
sources, but also because it is the richest cluster in the Abell
catalogue (Birkinshaw \etal\markcite{bgn78a} 1978a). With more sensitive X-ray
surveys, X-ray images, and X-ray spectroscopy, several clusters with
exceptional X-ray properties have also been observed. Examples are the
high-luminosity cluster CL~0016+16 (Birkinshaw \etal\markcite{bgm81}
1981a), and the high-temperature cluster Abell~2163 (Holzapfel
\etal\markcite{holz97b} 1997b).
\par
More recently, there has been some effort to observe complete samples
of clusters of galaxies selected on the basis of their X-ray or
optical properties, since the interpretation of cluster \SZ\ effects
in cosmological terms may be biased by the use of the {\it ad hoc}
samples that have been assembled to date. Initial steps in these
directions have been taken by, for example, Myers
\etal\markcite{myers97} (1997). At present, though, it is not possible
to use the sample of clusters contained in Table~\ref{tab-szdata} to
make reliable statistical statements about the effects of clusters on
the CMBR. Attempts to normalize a \SZ\ effect cluster luminosity
function (e.g., Bartlett \& Silk\markcite{bs94a} 1994a) based on 
these clusters may not be safe. 
\par
Extreme care is needed in interpreting the results given in this
table. First, the datum that is recorded, $\DTRJ$, is the
measured \SZ\ effect from the cited paper at the most significant
level observed (code O), or the central \SZ\ effect in the cluster, 
as fitted based on some model of the cluster gas (code C), and which 
would be seen in the Rayleigh-Jeans limit if the cluster were observed
with infinitely good angular resolution. That is, for C~codes,
\begin{equation}
 \DTRJ = - 2 \, \trad \, y_0 \quad .
 \label{eq-centraldecrement}
\end{equation}
It is not simple to convert from the measured effects to the central
effects, since proper account must be taken of the method used to
observe the cluster and the efficiency factor $\eta$ (see
Fig.~\ref{fig-zdep}, for example). For some observations, for
example with multichannel bolometer systems, it may have been
necessary for the observers to undertake a significant fitting
exercise to extract the central $\DTRJ$, $\DTRJzero$, with the result
depending on the model of the cluster gas adopted
(Sec.~\ref{sec-bolom}). Clusters with only poor X-ray images are
therefore difficult to assess, but in cases in which there is good
X-ray data this fitting step is relatively reliable. Thus it can be 
shown, for example, that recent results for Abell~2218 are in much
better agreement than is apparent from Table~\ref{tab-szdata} (see
later). 
\par
Many of the observations made with bolometers express their results in
terms of $y_0$, the central value of $y$ through the target
cluster. In those cases (e.g., Holzapfel \etal\markcite{holz97b}
1997b), I have converted the results to central decrements using
(\ref{eq-centraldecrement}). In cases where the peak beam-averaged
value of $\DTRJ$ is stated (e.g., Chase \etal\markcite{cjra87} 1987),
that value is preferred in the table. 
\par
For the interferometric data, the measured flux densities on 
the most appropriate (usually lowest-resolution) maps have been
converted into measured brightness temperatures using the synthesized
beamsize quoted. That is, it is assumed that the synthesized beam is
an elliptical Gaussian, with solid angle $\Omega_{\rm ab}$ (calculated
from the full widths to half-maximum in two directions, 
$h_{\rm a} \times h_{\rm b}$), and the brightness temperature is
obtained from 
\begin{equation}
    \Delta S_\nu = 2 \, \boltz \, \DTRJ \, 
                   \left( {\Omega_{\rm ab} \over \lambda^2} \right)
\end{equation}
which in convenient units, becomes
\begin{equation}
    ( \DTRJ / \muK ) = 340
          \, (\Delta S / \muJy \ {\rm beam^{-1}}) 
          \, (\nu/\GHz)^{-2}
          \, (h_{\rm a}/\arcmin)^{-1}
          \, (h_{\rm b}/\arcmin)^{-1}.
\end{equation}
In the case of the Partridge \etal\markcite{part87} (1987) data, I
have estimated the error on the central \SZ\ effect from their
visibility curves, taking rough account of the systematic errors in
the data caused by correlator offsets.
\par
For the radiometric results, which are the bulk of the entries in
Table~\ref{tab-szdata}, the values of $\DTRJ$ are taken directly from
the papers. The results from Rudnick\markcite{rud78} (1978) are given
for a 2-arcmin FWHM structure at the cluster center, since this is the
closest match to the resolution of the telescope used. Rudnick also
quotes more sensitive results for $\DTRJ$ at a number of larger
angular scales by convolving the data. These larger scales may be more
appropriate for some clusters. 
\par
In many cases the radiometric data have been adjusted for the effects
of cluster and background radio sources. These adjustments are not
necessarily consistent between the different papers: as further radio
work has been done on the clusters, some have shown that substantial
radio source corrections are needed (see, for example,
Abell~2507). Sometimes the detections of these radio sources led to
the cluster observations being abandoned (e.g., for Abell~426). For
other clusters, later work may have used better source corrections and
is often more reliable on these grounds alone. Many of the clusters
with radiometric \SZ\ results reported here have had 
little supporting work on the radio source environment. This makes it
difficult to assess the extent to which the results are affected by
radio source contamination.
\par
A number of trends are clear in Table~\ref{tab-szdata}. Early
observations were dominated by single-dish radiometers (e.g.,
Birkinshaw \etal\markcite{bgn81} 1981b). More recently,
the bolometric technique has been used, specially because of the
interest in detecting the effect near 190~GHz, where the kinematic
effect is more obvious (e.g., Holzapfel \etal\markcite{holz97b}
1997b). Finally, the completion of the Ryle array and the use of the
OVMMA and BIMA for \SZ\ effect measurements has produced a series of
sensitive maps of clusters (e.g., Jones \etal\markcite{jon93} 1993;
Carlstrom \etal\markcite{cjg96} 1996), where some evidence of the
cluster structure is seen (e.g., for CL~0016+16; Carlstrom
\etal\markcite{cjg96} 1996; Sec.~\ref{sec-interf}). 
\par
Despite the increasing use of these new techniques, single-dish
radiometry is still used --- principally for survey work,
to locate target clusters with significant \SZ\ effects that might be
the subjects of detailed mapping later. Thus observations at OVRO
with the 40-m telescope at present are concentrating on a sample of
clusters selected because of their excellent exposures by the ROSAT
PSPC. Myers \etal\markcite{myers97} (1997) are making a
survey of another sample of clusters with the OVRO 5.5-m telescope.
\par
The results in Table~\ref{tab-szdata} span more than 20~years of work
on the \SZ\ effect, and involve a number of different techniques with
different observing characteristics. Thus it is difficult to compare
the results of different groups for any one cluster without taking
detailed account of the structure of the cluster and the details of
the method used. This causes the apparent disagreements between
different groups' results to be accentuated. Nevertheless, there are
clusters for which the data (particularly the more recent data) are
largely in agreement, and clusters for which the situation is less clear. 
\par
Consider, for example, the cluster Abell~2218, for which a
particularly large number of measurements are available. First,
consider the history of results for Abell~2218 obtained by the group
with which I have been working. The published results from 1976 to
1996 are given in Table~\ref{tab-a2218consist1}. These results are not
independent: later results from the Chilbolton 25-m telescope included
the data used in earlier papers, and the OVRO 40-m results also
changed as more data were accumulated, and as the radio source
corrections and data calibrations were better understood.
\par
The internal consistency of the early data is clearly poor. The final
result based on the Chilbolton data is only marginally consistent
with the first published result, suggesting that the 
later data were quite inconsistent with the earlier data. Since a
number of changes in the configuration of the Chilbolton system
occurred during the period that data were taken, it is likely
that this inconsistency arose from unrecognized systematic
errors, possibly involving strong ground signals entering through
distant sidelobes.
\par
Later data, from the OVRO 40-m telescope, appear more consistent ---
the 10.7-GHz result and the 20.3-GHz results seem to be indicating
that the value of $\DTRJ$ towards the center of the cluster is about
$-0.35$~mK. However, the observing characteristics of these
observations was very different, and the low-significant detection at
10.7~GHz is due almost completely to a correction for contaminating
radio sources near the center of the cluster.
\par
If it is assumed that the atmosphere of Abell~2218 follows the model
(\ref{eq-nebeta}), and is isothermal, then the structural parameters
$\beta = 0.65 \pm 0.05$ and $\thc = 1.0 \pm 0.1$~arcmin derived from
X-ray observations (Birkinshaw \& Hughes\markcite{bh94} 1994) may be
used to calculate the efficiencies with which the cluster was observed
by any telescope. For observations of the \SZ\ effect of Abell~2218
with the Chilbolton 25-m telescope, the OVRO 40-m telescope at
10.7~GHz, and the OVRO 40-m telescope at 20.3~GHz, these efficiencies
are about $0.35$, $0.49$, and $0.60$, respectively. The inferred
central \SZ\ effects from the cluster according to the final results
from these three telescope configurations are therefore $-3.0 \pm
0.6$, $-0.77 \pm 0.38$, and $-0.67 \pm 0.08$~mK. The result from the
Chilbolton 25-m telescope is clearly inconsistent with the other two
measurements. Only a very contrived structure for the cluster
atmosphere could cause such differences and be consistent with the
other \SZ\ effect data and the X-ray image and spectrum. Thus an
economical assumption is that the early data were badly contaminated
by systematic errors, and should be discarded, and that the true
central decrement from Abell~2218 is near $-0.7$~mK.
\par
Another effect that can be seen in Table~\ref{tab-a2218consist1} is
the strong variation in the errors quoted for the 20.3-GHz data as a
function of time. The smallest error ($\pm 0.03$~mK, in Birkinshaw \&
Moffet\markcite{bm86} 1986) represents the error on the data
accumulated at that time if all the data are considered to be drawn
from a single, static, Gaussian distribution. The largest error, $\pm
0.13$~mK, in Birkinshaw \& Gull\markcite{bg84} (1984), is based on the
smallest amount of data, under the same assumptions. On the other hand, the
entry for Birkinshaw\markcite{b86} (1986) is based on substantially
more data than in Birkinshaw \& Moffet\markcite{bm86} (1986), but
includes a generous allocation for possible systematic errors. Later
entries in the Table include further data, and were derived with
detailed analyses for systematic errors. It should be noted that the
final result in the table, $-0.40 \pm 0.05$~mK, contains no
contribution from the background CMBR anisotropies, so that the error
represents the reproducibility of the measurement rather than the
external error that would be achieved if Abell~2218 could be observed
against another patch of the background radiation. 
\par
Of course, Table~\ref{tab-a2218consist1} illustrates principally the
difficulty in measuring the \SZ\ effect signals in the presence of
systematic errors with unknown characteristics: reductions in the error are
principally achieved by stronger controls against
systematic errors (for example by observing multiple regions of blank
sky, performing checks for radio source contamination, and so
on). More rigorous controls against systematic error are obtained by
comparing the results from different groups who observe the same
cluster in different ways. The most frequently-observed cluster is
Abell~2218, and Table~\ref{tab-a2218consist2} lists the central
decrements for Abell~2218 deduced from 16~independent measurements
using the same model atmosphere as in discussion of
Table~\ref{tab-a2218consist1}.
\par
It is at once apparent from Table~\ref{tab-a2218consist2} that the
individual results are inconsistent: the early data are often
scattered with dispersion 
several times their nominal error about the later data. In some of the
early papers, large parasitic signals from ground spillover have been
removed (e.g., Perrenod \& Lada\markcite{pl79} 1979), but there
remains a suspicion that residual systematic errors are present in the
data. Overall, the later data are in much better agreement. A notable
exception is the result of Klein \etal\markcite{kle91} (1991), where
the measured decrement is consistent with predictions based on other
data, but its location on the sky is far from the X-ray center of the
cluster so that the implied central decrement in Table~6 is
unrealistically large. If an average is taken over these data, and the most 
obviously discordant results are excluded, then the central decrement in
Abell~2218 is found to be $-0.74 \pm 0.07$~mK. The error here has been
increased to take some crude account of the remaining discordance in
the data (the value of $\chi^2 = 15$  with 10 degrees of freedom).
\par
The \SZ\ effect results for Abell~2218 are generally in better
agreement now than they were for the first few years of reported
measurements. This suggests that several groups are now able to
measure reliable \SZ\ effects, and based on this
conclusion, I have collected into Table~\ref{tab-reliableSZ} the set of
all \SZ\ effects that I believe are both significant (at $> 4\sigma$)
and reliable. These objects constitute a set for which a simultaneous
analysis of the \SZ\ effect data and the X-ray data may provide useful
constraints on the cluster atmospheres (Sec.~\ref{sec-cluster}), and
possibly a measurement of the Hubble constant
(Sec.~\ref{sec-cosmo}). Of the thirteen clusters in the 
table, seven were first detected using single-dish radiometers,
two using bolometers, and four using interferometers. Only four of
these detections have independent confirmations at significance $>
4\sigma$. Much work remains to be done to measure the \SZ\ effects in
these clusters, and all three measurement techniques still have their
place in \SZ\ effect research, although bolometer measurements are
becoming more important, and interferometric maps of the effect are
probably the most reliable.
\par
Detections at lower significance exist for more objects, including the
lines of sight towards two high-redshift quasars (PHL~957, Andernach
\etal\markcite{and86} 1986; PC 1643+4631, Jones \etal\markcite{jon97}
1997). These 
detections may 
arise from distant clusters of galaxies along the lines of sight, or
from the host clusters of the quasars themselves, or from some other
cause. However, if the \SZ\ effects arise from line-of-sight objects,
then observations towards ``blank'' sky regions should show \SZ\
effects as often as observations towards the quasars --- it is not yet
clear whether this is the case, so the interpretation of these \SZ\
effects and the limits from observations of other quasars (Jones
\etal\markcite{jon97} 1997) or blank fields (Richards
\etal\markcite{rich97} 1997) is at present obscure. 
\par
Further complications in the interpretation of these results have
arisen as deep optical and X-ray followups have been made. Thus for
the PC~1643+4631 field, Saunders \etal\markcite{sau97} (1997) find no
cluster that might be responsible for a \SZ\ effect in deep optical
images, and Kneissl, Sunyaev \& White\markcite{ksw98} (1998) find no
X-ray emission associated 
with hot gas. The interpretation of the CMBR anisotropy as a \SZ\
effect has become difficult because of the high redshift needed for a
relatively massive cluster that could hold a detectable amount of hot
gas (Bartlett, Blanchard \& Barbosa \markcite{bbb98}
1998). Alternative models involving kinematic effects from colliding 
QSO winds (Natarajan \& Sigurdsson\markcite{ns97} 1997), extreme
Rees-Sciama effects, etc. are being considered, but seem
implausible. Independent observational confirmation of the reality of
these microwave background structures is therefore a priority: early
results are yielding a mixed verdict.
\subsection{Non-thermal \SZ\ effects}
Only McKinnon \etal\markcite{moe90} (1990) have yet made direct
attempts to measure the non-thermal \SZ\ effect, and their results are
reported in Table~\ref{tab-nonthermalSZ}. 
\par
As explained in Sec.~\ref{sec-nonthermal}, the aim of observations of
the non-thermal \SZ\ effect is to set limits on the electron
population in radio source lobes. The constraints that McKinnon
\etal\markcite{moe90} (1990) derived 
based on the data in Table~\ref{tab-nonthermalSZ} are far (two orders
of magnitude) from achieving this aim. With the best techniques
available, it should be possible to improve the sensitivity by roughly
a factor of 10 over McKinnon \etal's results in a
modest allocation of observing time: a further
improvement would be gained by working on radio sources with steep
spectra and for which the telescope beam is a small fraction of the
radio source size.
\par
Perhaps the best possibility of effecting these improvements is with
modern bolometer arrays, observing radio sources with large
lobes and sufficiently steep radio spectra that radio emission is not
an issue. In view of the possible impact on radio source theory, such
observations should certainly be attempted.
\par
Beam-filling non-thermal \SZ\ effects may already have been observed,
however, as part of the signals from some clusters of galaxies. In the
Coma cluster, for example, Herbig \etal\markcite{herbig95} (1995)
detected a strong \SZ\ effect, but a bright radio halo source is known
to exist (Hanisch\markcite{han82} 1982; Kim \etal\markcite{kim90}
1990), and it may be responsible for some part of the 
observed signal if the relativistic electron population has a
significant lower-energy component. Another case where this is true is
Abell~2163, where a strong \SZ\ effect has been measured (Holzapfel
\etal\markcite{holz97a} 1997a), and a powerful radio halo source
exists (Herbig \& Birkinshaw\markcite{hb98} 1998). However, if we
attempt to interpret the results for Abell~2163 in terms of a
contamination of the measured \SZ\ effect by a non-thermal component,
we rapidly recognize that the non-thermal \SZ\ effect is principally a
test of the lower energy cutoff of the power-law distribution of
electrons responsible for the radio halo source's synchrotron emission
rather than of equipartition. Roughly,  
\begin{equation}
   \DTRJ = -50 (\gamma_1/100)^{-26/9} \quad \muK 
\end{equation}
which depends strongly on the lower Lorentz factor cutoff of the
electron spectrum, $\gamma_1$ (and weakly on the value of the Hubble
constant). If we assume that about half the central \SZ\ effect (of
$-1.62 \pm 0.22 \ \mK$; Holzapfel \etal\markcite{holz97b} 1997b) is
produced by this non-thermal process, then $\gamma_1 \approx 30$. This
corresponds to the radio-emitting plasma in the cluster contributing a
small fraction of the gas pressure (if the radio source is close to
equipartition). However, a non-thermal \SZ\ effect of this size would
have a severe effect on the location of the zero of the spectrum of
the combined thermal, kinematic, and non-thermal effects from the
cluster, and Holzapfel \etal\markcite{holz97b} (1997b) find that the
spectrum shows no signature of a cluster peculiar velocity, and hence
no zero shift. Since $\gamma_1$ depends on $\DTRJ$ only weakly,
changing the fraction of the central $\DTRJ$ to ensure consistency
with Holzapfel \etal\ results in a limit $\gamma_1 \gtrsim 50$ which
is little changed from the value above. An even higher value for
$\gamma_1$ is likely, since low-energy electrons 
suffer rapid ionization losses (Rephaeli \& Silk\markcite{rs95}
1995), and might not be expected to be present unless there is a fast
local acceleration mechanism.
\par
Thus although the sizes of the non-thermal effects from radio halo
sources are likely to be a small fraction of the thermal 
effects, they may exert an interesting influence on the 
spectrum of the combined signal by shifting the location of
the zero of the spectrum away from the location expected on the basis
of the thermal and velocity effects alone, if the spectrum of
relativistic electrons in the cluster extends down to moderate Lorentz
factors. This non-thermal \SZ\ effect is a source of
systematic error that should be considered when 
measuring cluster peculiar velocities, and argues that several
spectral bands, and detailed spectral fitting, are required to set
rigorous limits to cluster velocities.
%
%
%
\section{The \SZ\ effects analysed in terms of cluster
         properties \label{sec-cluster}}
The \SZ\ effects provide a window on cluster properties which differs
significantly from that afforded by optical, X-ray, or
conventional radio data. The present section of this review
concentrates on these implications of the measurement of the effects
for the understanding of cluster properties.
\subsection{Cluster gas properties}
The original purpose of measuring the \SZ\ effects of clusters was to
test whether cluster X-ray emission was thermal in origin, or came
from non-thermal processes such as inverse-Compton emission from
relativistic electrons and the cosmic background radiation (e.g.,
Harris \& Romanishin\markcite{hr74} 1974). This use of the effects was
rapidly made moot by the detection of line emission from clusters of
galaxies (e.g., Serlemitsos \etal\markcite{serl77} 1977).
\par
Until recently there were few high-sensitivity measurements of the
\SZ\ effects from clusters, so that little information could be
obtained that was not already available from X-ray images and spectra.
Thus, for example, the structural information from cluster
\SZ\ effects based on radiometric data (e.g., Fig.~\ref{fig-sz3cluster})
has much lower signal/noise than the X-ray images of those same
clusters (e.g., Fig.~\ref{fig-0016pspc}). This is less true with 
imaging of the quality that should be available from interferometers,
but at present interferometers measure only a fraction of the
Fourier information needed for a full reconstruction of the microwave
background structure generated by clusters of galaxies, and hence 
model-fitting to these interferometer images is usually based on existing
X-ray data (see Sec.~\ref{sec-interf}).
\par
The \SZ\ effects do differ significantly from the X-ray data in their
sensitivity to different properties of the atmospheres. If a cluster
is at rest in the Hubble flow, then in the non-relativistic limit
the low-frequency, thermal, \SZ\ effect from that cluster on a
particular line of sight is
\begin{equation}
   \DTRJzero = -2 y \trad
\end{equation}
where $y$ is the Comptonization parameter, which depends on
the line-of-sight electron density and temperature as
\begin{equation}
   y = \int n_{\rm e}({\bf r}) \, \sigma_{\rm T} 
            \, {\boltz \te({\bf r}) \over m_{\rm e} c^2} \, dl
 \label{eq-ygeneral}
\end{equation}
(\ref{eq-yclus}), and is thus proportional to the line-of-sight
integral of the electron pressure.
By contrast, the X-ray surface brightness on that
line of sight depends on these same quantities as
\begin{equation}
   b_{\rm X}(E) = {1 \over 4\pi (1 + z)^3} \,
                   \int n_{\rm e}({\bf r})^2 \, \Lambda(E,T_{\rm e}) \, dl
\end{equation}
(\ref{eq-bxclus}), where $\Lambda(E,T_{\rm e})$ is the X-ray spectral
emissivity, which is a function of the energy of the X-ray
observation, $E$, the electron temperature of the gas, $T_{\rm e}$,
the metallicity of the gas, and the redshift, $z$. The emissivity
depends on temperature roughly as 
$\Lambda \propto T_{\rm e}^{1/2}$ if the X-ray pass-band is
sufficiently broad, so that the X-ray
surface brightness is proportional to the line-of-sight integral of
$n_{\rm e}^2 T_{\rm e}^{1/2}$ while the \SZ\ effect is proportional to
the line-of-sight integral of $n_{\rm e} T_{\rm e}$.
The \SZ\ effect and X-ray surface
brightness of a cluster of galaxies are then likely to have different angular
structures (if we rule out the possibility of coincidences in the
density and temperature structures), and the 
difference between the X-ray and \SZ\ effect images should
provide information on the runs of temperature and density in the
cluster gas.
\par
Once again, this has largely been superseded by improvements in X-ray
technology. The newer generation of X-ray observatories provides
some spatially-resolved X-ray spectra of clusters
of galaxies and hence direct measurements of 
variations in the thermal structures of clusters. \SZ\ effect
data could still be an important probe of structure in the outer parts of
clusters, since at low densities the \SZ\ effect drops off less
rapidly ($\propto n_{\rm e}$) than the X-ray surface brightness ($\propto
n_{\rm e}^2$). This region of the gas distribution might be expected
to show the clearest evidence of deviations from the remarkably
successful isothermal-$\beta$ model, but the current sensitivity of
\SZ\ effect measurements is too low, relative to the sensitivity of
X-ray images and spectra, for useful comparisons to be made.
Where the cluster contains a radio source (particularly a
radio halo source), the thermal \SZ\ effect is of particular interest
since it provides a direct measurement of the electron pressure near
that radio source, and so can be used to test whether the dynamics of
the radio emitting plasma are strongly affected by the external gas
pressure.
\par
The remaining area where information about the \SZ\ effect provides
unique information about the structure of the cluster gas is on the
smallest scales, where structures in the X-ray gas are unresolved by
X-ray or radio telescopes. In this case, the structures are better
described by a (possibly position-dependent) clumping of the
gas, and unless the density and temperature changes in the clumps 
conspire, the \SZ\ effect and X-ray surface
brightness scale differently. For example, if clumping is isobaric,
with the pressure in clumps the same as outside, then the \SZ\ effect
will show no variations in regions where the gas is strongly clumped,
while the X-ray emissivity will increase as $n_{\rm e}^{3/2}$. No
useful results on the clumping of cluster gas have been reported in
the literature to date: it is more usual to see clumping referred to
as one of the limiting factors in the use of the \SZ\ effects to
measure the Hubble constant (Sec.~\ref{sec-cosmolpar}), although 
clumping in the intracluster medium is also a biasing factor in the
use of the X-ray data to determine gas densities and masses from X-ray
images and spectra. 
\par
A direct use of the thermal \SZ\ effect is as a probe of the gas
mass enclosed within the telescope beam (Myers \etal\markcite{myers97}
1997). For an isothermal model of the form (\ref{eq-nebeta}), the
surface mass density in gas along a given line of sight is
\begin{equation}
 \Sigma_{\rm g} = \int \, dl \, n_{\rm e}({\bf r}) \, \mu_{\rm e}
\end{equation}
where $\mu_{\rm e}$ is the mean mass of gas per electron, while the 
thermal \SZ\ effect at low frequency is proportional to the
Comptonization parameter (eq.~\ref{eq-ygeneral}).
Thus the surface mass density in gas can be related to the \SZ\ effect
(as measured through the Comptonization parameter) as
\begin{equation}
    \Sigma_{\rm g} = \mu_{\rm e} \, 
        \left( {m_{\rm e} c^2 \over \boltz \te } \right) \,
        {y \over \sigma_{\rm T}}
    \label{eq-sigma}
\end{equation}
if the electron temperature of the gas is constant. 
\par
For clusters such as Abell~2218 which have both a rich population of
arcs (Sarantini, Petrosian \& Lynds\markcite{spl96} 1996) and a strong
\SZ\ effect, the measure (\ref{eq-sigma}) of the gas surface density
could be compared directly with mass estimates produced by the
study of gravitational arcs to estimate 
the fraction of the lensing mass that is contained in gas. Although
this study is possible using the X-ray emission from a
cluster,  X-rays provide a less direct measure of gas mass, being
biased by uncertainties in the clumpiness of the gas. The
\SZ\ effect should be less susceptible to errors of
interpretation, and give a clean estimate of the ratio of baryonic and
dark matter within the arcs, which relates to the baryon problem in
clusters (White \& Fabian\markcite{wf95} 1995). To make the best use of this
comparison, the \SZ\ effect data should be taken with resolutions
better than the radii of the gravitational arcs. Unfortunately
observations with high brightness temperature sensitivity and angular
resolutions of 10~arcsec or better are very difficult, and this limits
the utility of this comparison at present.
\par
Myers \etal\markcite{myers97} (1997) show that for three clusters of galaxies,
the ratio of baryonic mass to total gravitating mass (here derived not
from gravitational lensing, but rather from cluster dynamics)  is in
the range $0.06 h_{100}^{-1}$ to $0.17 h_{100}^{-1}$. These values are
larger than the baryonic mass fraction $(0.013 \pm 0.002)
h_{100}^{-2}$ expected from calculations of big-bang nucleosynthesis
if $\Omega_0 = 1$ (Smith \etal\markcite{smi93} 1993). As a result, we
can infer that the Universe is open, with $\Omega_0 \approx 0.2
h_{100}^{-1}$, or that clusters show a baryon segregation effect, with
excess baryons in their X-ray luminous cores and excess dark matter
further out.
\subsection{Cluster velocities\label{sec-clustervels}}
Although the \SZ\ effects have not revealed much new information about
the detailed structures of cluster atmospheres, the kinematic \SZ\
effect (Sec.~6) can provide a direct measurement of the peculiar
velocity of a cluster of galaxies relative to the Hubble flow --- a
measurement that cannot be made with comparable accuracy by any other
means, and which is of great importance in the study of the formation
of structure. Here the \SZ\ effect is particularly good, since it
could be measured at any redshift provided that the cluster to be
observed has a significant electron scattering optical depth (i.e., a
well-developed atmosphere), and that the telescope used has high
observing efficiency.
\par
The first application of this technique to set useful limits on cluster
velocities was made by Holzapfel \etal\markcite{holz97b} (1997b), who
measured the \SZ\ effects from Abell~1689 and~2163 using the SuZIE 
array detector on the Caltech Submillimeter Observatory. After
decomposing the CMBR anisotropy into thermal and kinetic parts,
and using an isothermal model for the cluster gas based on the X-ray
image of the cluster from ROSAT, Holzapfel \etal\ find
line-of-sight peculiar velocities for the clusters of 
\begin{eqnarray}
  v_{\rm z}({\rm Abell \ 1689}) &=& +170^{+760}_{-570} \ \kms \\
  v_{\rm z}({\rm Abell \ 1689}) &=& +490^{+910}_{-730} \ \kms
\end{eqnarray}
which only limits cluster peculiar velocities at $z \approx 0.2$ to
less than about $2000 \ \kms$. However, this is not too far (in
terms of required observational sensitivity) from the result of Lauer \&
Postman\markcite{lp94} (1994) that clusters in the local Universe
exhibit a bulk velocity of $730 \pm 170 \ \kms$. Small improvements in
the accuracy of the measurement of the \SZ\ effects should allow
useful velocity measurements to be made, although uncertainties at the
level of $200 \ \kms$ may be unavoidable because of the background of
primordial anisotropies against which the clusters are observed. Since
the kinematic \SZ\ effects and the primordial anisotropies have the same
spectrum, they can be separated only through their different angular
structures --- but at present there is no direct evidence about the
amplitude of the primordial anisotropies as observed with a
cluster-shaped filter on these angular scales.
\par
Transverse velocity components could be measured through higher-order
\SZ\ effects (see Sec.~\ref{sec-polarization}), or through
measurements of the Rees-Sciama terms (Birkinshaw\markcite{b89} 1989),
although the latter are more subject to confusion with primordial
anisotropies. Even the noisy measurements of the three-dimensional
velocity field of clusters as a function of redshift which
might be measured in this way are likely to be useful in studies of
the formation of large-scale structure in the Universe, and
observations of CMBR anisotropies induced by clusters, even though of
limited power to measure individual cluster velocity vectors, are
likely to prove important for this reason.
%
%
%
\section{The \SZ\ effect interpreted in cosmological terms\label{sec-cosmo}}
The simplest cosmological use of the \SZ\ effect is to prove
that the CMBR is genuinely a cosmological phenomenon: the appearance
of an effect from a cluster of galaxies at $z = 0.5455$
(CL~0016+16) proves that the CMBR originates at $z > 0.54$, 
higher-redshift detections push this limit even further.
However, it is as a probe of cosmological parameters, and as
a distance-independent probe of earlier phases of the Universe that
the \SZ\ effect has attracted most interest, and such uses of the
effect are the focus of this section.
\subsection{Cosmological parameters\label{sec-cosmolpar}}
The basis of the use of the \SZ\ effect as a tracer of cosmological
parameters was given in Sec.~\ref{sec-szcluster}. The essence of the
idea is the same as for other distance-measuring techniques
that depend on a comparison of the emission and absorption of
radiation from gas: the surface brightness of the gas in emission
is proportional to the line-of-sight integral of some density squared,
\begin{equation}
   E \propto \int n_{\rm e}^2 dl
\end{equation}
while the absorption of some background source of radiation is
proportional to the optical depth
\begin{equation}
   A \propto \int n_{\rm e} dl \quad . 
\end{equation}
Thus if both the emission from the gas, $E$, and its absorption, $A$,
can be measured, the quantity $A^2/E$ is a density-weighted measure of
the path-length through the gas. If the structure of the gas is known,
and its angular size, $\theta$, can be measured, then the angular
diameter distance of the gas can be estimated from $A^2/(E\theta)$.
\par
Although this technique may eventually be applied using only X-ray
data (Krolik \& Raymond \markcite{kr88} 1988), it is currently used
for the measurement of distances using a combination of X-ray and
the \SZ\ effect data (Gunn\markcite{gunn78} 1978; Silk \&
White\markcite{sw78} 1978; Birkinshaw\markcite{b79} 1979;
Cavaliere, Danese \& De Zotti\markcite{cdd79} 1979). The emission of
gas in a cluster of galaxies is measured by its X-ray surface
brightness,
\begin{equation}
   b_X = {1 \over 4\pi ( 1 + z)^3} \, \int n_{\rm e}^2 \, 
         \Lambda_{\rm e} \, dl
\end{equation}
where $\Lambda_{\rm e}(E,T_{\rm e})$ is the X-ray spectral emissivity
of the cluster gas (Sec.~\ref{sec-szcluster}), while the absorption by
the gas is measured by the thermal \SZ\ effect, which can be expressed
as an intensity change
\begin{equation}
   \Delta I(x) = I_0 \, \int \, n_{\rm e} \sigma_{\rm T} \,
                 \Psi(x,T_{\rm e}) \, dl
\end{equation}
at dimensionless frequency $x = h \nu / \boltz \trad$, where 
$I_0 = {2 h \over c^2} \left( {\boltz \trad \over h} \right)^3$ is a
scale intensity and $\Psi(x,T_{\rm e})$ is the dimensionless form of
the frequency-dependent, relativistic, spectrum of the effect (from 
equation~\ref{eq-deltaIintegral}), 
\begin{equation}
  \Psi(x,T_{\rm e}) = 
                  \int\limits_{-\infty}^{\infty} P_1(s) \, ds \,
                  \left( {x_0^3 \over e^{x_0} - 1}
                  -
                         {x^3   \over e^{x  } - 1}
                  \right)
\end{equation}
with $s = \ln(x/x_0)$ (see Secs~\ref{sec-elpopscat}
and~\ref{sec-effectspectrum}; this form is used by Holzapfel
\etal\markcite{holz97a} 1997a).
\par
Since the technique compares the angular size of a cluster of galaxies
with a measure of the line-of-sight size of the cluster, it is
important to have a model for the structure of the gas so that the
relationship between the projected quantities $b_{\rm X}$ and 
$\Delta I$ can be calculated. It is convenient to express the electron
concentration and temperature in terms of reference values (chosen as
the central values here, although the values at any fiducial point can
be used) and dimensionless form factors describing the angular
structure of the gas in density, $f_n(\theta,\phi,\zeta)$, and
temperature, $f_T(\theta,\phi,\zeta)$. The angular variables are
$\theta$, the angle from the reference line of sight through the
cluster center, $\zeta = l/D_{\rm A}$, an angular
measure of distance down the line of sight, and $\phi$, an azimuthal
angle about the line of sight. $D_{\rm A}$ is the angular diameter
distance of the cluster. Then the electron density and
temperature at some location, $\bf r$, are 
\begin{eqnarray}
     n_{\rm e}({\bf r}) &=& n_{\rm e0} \, f_n(\theta,\phi,\zeta) 
     \label{eq-formne} \\
     T_{\rm e}({\bf r}) &=& T_{\rm e0} \, f_T(\theta,\phi,\zeta)
     \label{eq-formte}
\end{eqnarray}
and the energy loss and spectrum functions may be written
in terms of similar form factors which depend on $f_n$ and $f_T$ in
complicated ways,
\begin{eqnarray}
 \Lambda_{\rm e}(E,T_{\rm e}) &=& \Lambda_{\rm e0} \,
                                 f_\Lambda(\theta,\phi,\zeta) 
 \label{eq-formlambda} \\
 \Psi(x,T_{\rm e})            &=& \Psi_0 \, f_\Psi(\theta,\phi,\zeta)
 \label{eq-formpsi}
\end{eqnarray}
(Birkinshaw \etal\markcite{bha91} 1991; Holzapfel
\etal\markcite{holz97a} 1997a). The X-ray surface 
brightness and the thermal \SZ\ effect intensity change can then be
expressed in terms of physical constants and angular structure
factors, as
\begin{eqnarray}
 b_{\rm X}(\theta,\phi) &=& {\Lambda_{\rm e0} \, n_{\rm e0}^2 \,
                            D_{\rm A} \over 4 \pi (1 + z)^3} \,
                            \Theta^{(1)}(\theta,\phi) \nonumber \\
                        &\equiv& N_{\rm X} \,
                            \Theta^{(1)}(\theta,\phi) 
                            \label{eq-bxhofit} \\
 \Delta I(\theta,\phi)  &=& \Psi_0 \, I_0 \, n_{\rm e0} \,
                            \sigma_{\rm T} \, D_{\rm A} \,
                            \Theta^{(2)}(\theta,\phi) \nonumber \\
                        &\equiv&
                            N_{\rm SZ} \, \Theta^{(2)}(\theta,\phi)
                            \label{eq-dthofit}
\end{eqnarray}
with the structural information for the cluster contained in the
angles 
\begin{eqnarray}
  \Theta^{(1)}(\theta,\phi) &=& \int f_n^2 f_\Lambda d\zeta \\
  \Theta^{(2)}(\theta,\phi) &=& \int f_n   f_\Psi    d\zeta 
\end{eqnarray}
which describe the shapes of the X-ray and \SZ\ effects that the model
gas distribution would produce.
\par
An absolute distance for a cluster is then found by fitting the X-ray and
\SZ\ effect data to models of the form (\ref{eq-bxhofit})
and (\ref{eq-dthofit}) to deduce $N_{\rm X}$ and $N_{\rm SZ}$, and
calculating the angular diameter distance using
\begin{equation}
 D_{\rm A} = \left( {N_{\rm SZ}^2 \over N_{\rm X}} \right) \,
             \left( {\Lambda_{\rm e0} \over 4 \pi (1 + z )^3
                    I_0^2 \, \Psi_0^2 \, \sigma_{\rm T}^2 } \right)
 \label{eq-daho}
\end{equation}
(Holzapfel \etal\markcite{holz97a} 1997a), or equivalently from the
form given by Birkinshaw \etal\markcite{bha91} (1991) in their
equation~39, if brightness temperature rather than intensity is used
as the \SZ\ effect observable. The value of the Hubble constant is
then obtained from the measured redshift of the cluster and the value
of $D_{\rm A}$ under some assumption about the value of $q_0$ using
equation (\ref{eq-angdia}).
\par
This is a {\it direct} method of measuring the distance of a cluster
of galaxies and the value of the Hubble constant: it can be applied at
large cosmological distances without any intervening chain of distance
estimators (as in the usual distance ladder). The distance
estimate relies on simple physics --- the properties of a fully-ionized
gas held nearly in hydrostatic equilibrium in the gravitational
potential well of a cluster of galaxies. The basis of this distance
estimate can therefore be tested by making a detailed study of the
properties of the cluster being used as a cosmological tracer and the
population of similar clusters. It is also important that
in this method each cluster of galaxies is treated as an individual
--- the evolutionary peculiarities of a distant cluster need not
affect the distance estimate provided that the physical state of the
intracluster gas is understood. Of course, if the cluster gas has much
small-scale density and temperature structure, it may be difficult
to obtain good models for the form factors (\ref{eq-formne}) --
(\ref{eq-formpsi}), and there 
may be a substantial systematic error in the distance estimate. Some
protection against this systematic error can be obtained by cross-checking
the independent results that are obtained from a number of clusters.
\par
The measurement of the values of $N_{\rm X}$ and $N_{\rm SZ}$ from the
X-ray and \SZ\ effect data not only requires knowledge of the form factors
$f_n$, $f_T$, $f_\Lambda$, and $f_\Psi$ but also the 
fiducial electron temperature of the cluster, $T_{\rm e0}$, 
since $T_{\rm e0}$ is an implicit variable in (\ref{eq-daho}), where
it enters in both $\Lambda_{\rm e0}$ and $\Psi_0$. Even with the help
of the resolved X-ray spectroscopy that will
become available on the next generation of X-ray telescopes (such as
AXAF), it is not possible to use the X-ray or \SZ\ effect
data to measure these three-dimensional
form factors. Therefore, the calculation proceeds by adopting some
parameterized models for the electron concentration and 
temperature as functions of position which are consistent with the
X-ray image and spectroscopy and the \SZ\ effect data. The 
normalizations $N_{\rm X}$ and $N_{\rm SZ}$ that are found are then 
dependent on the unknown structural parameters
of the model atmosphere after any adjustable parameters have been
determined. 
\par
Rephaeli \& Yankovitch\markcite{ry97} (1997) have recently pointed out
that for good accuracy in calculating cluster distances in this way,
it is important to the full relativistic formalism
(Gould\markcite{gould80} 1980) to calculate the value of $\Lambda_{\rm
e0}$ and $f_\Lambda$ for X-ray emission from the cluster gas, just as
the relativistic expression for the \SZ\ effect
(\ref{eq-deltaIintegral}) must be used. Hughes \&
Birkinshaw\markcite{hub98} (1998) have shown that the size of the
relativistic correction in Rephaeli \& Yankovitch's work is excessive,
apparently because of their use of an equation containing a
typographical error in Gould\markcite{gould80} (1980). Even so, 
the size of the relativistic corrections is appreciable (5~per cent or
so) for the hot clusters for which \SZ\ effects have been measured.
\par
A convenient form that has been used to describe the structure of
cluster atmospheres is the spherical isothermal beta model
(equation~\ref{eq-nebeta}), with constant electron temperature and a
concentration form factor 
\begin{equation}
   f_n = \left( 1 + {\theta^2+\zeta^2 \over \thc^2} 
                  \right)^{\!-{3 \over 2}\beta} \quad .
   \label{eq-fnsimple}
\end{equation}
The quantity $\thc = \rc/D_{\rm A}$ is the angular equivalent of the
core radius of the atmosphere, $\rc$.
This model leads to simple
expressions for the angles $\Theta^{(1)}$ and $\Theta^{(2)}$,
\begin{eqnarray}
  \Theta^{(1)} &=& \sqrt{\pi} \,
                  {\Gamma(3\beta - {1 \over 2}) \over
                   \Gamma(3\beta              )       } \,
                   \thc \, 
                   \left( 1 + {\theta^2 \over \thc^2}
                   \right)^{{1 \over 2} - 3\beta}
  \label{eq-theta1} \\
  \Theta^{(2)} &=& \sqrt{\pi} \,
                  {\Gamma({3\over 2}\beta - {1 \over 2}) \over
                   \Gamma({3\over 2}\beta              )       } \,
                   \thc \, 
                   \left( 1 + {\theta^2 \over \thc^2}
                   \right)^{{1 \over 2} - {3\over 2}\beta} 
 \label{eq-theta2}
\end{eqnarray}
which must then be convolved with the responses of the telescopes to
calculate the structures that would be seen in practice. 
Values of $\beta \approx 0.7$, and $\rc \approx 150 h_{100}^{-1}$~kpc 
are typically obtained in fitting X-ray images of clusters to
the structure defined by (\ref{eq-bxhofit}) and (\ref{eq-theta1}).
\par
\begin{figure}[p]
\epsscale{0.8}
\plotone{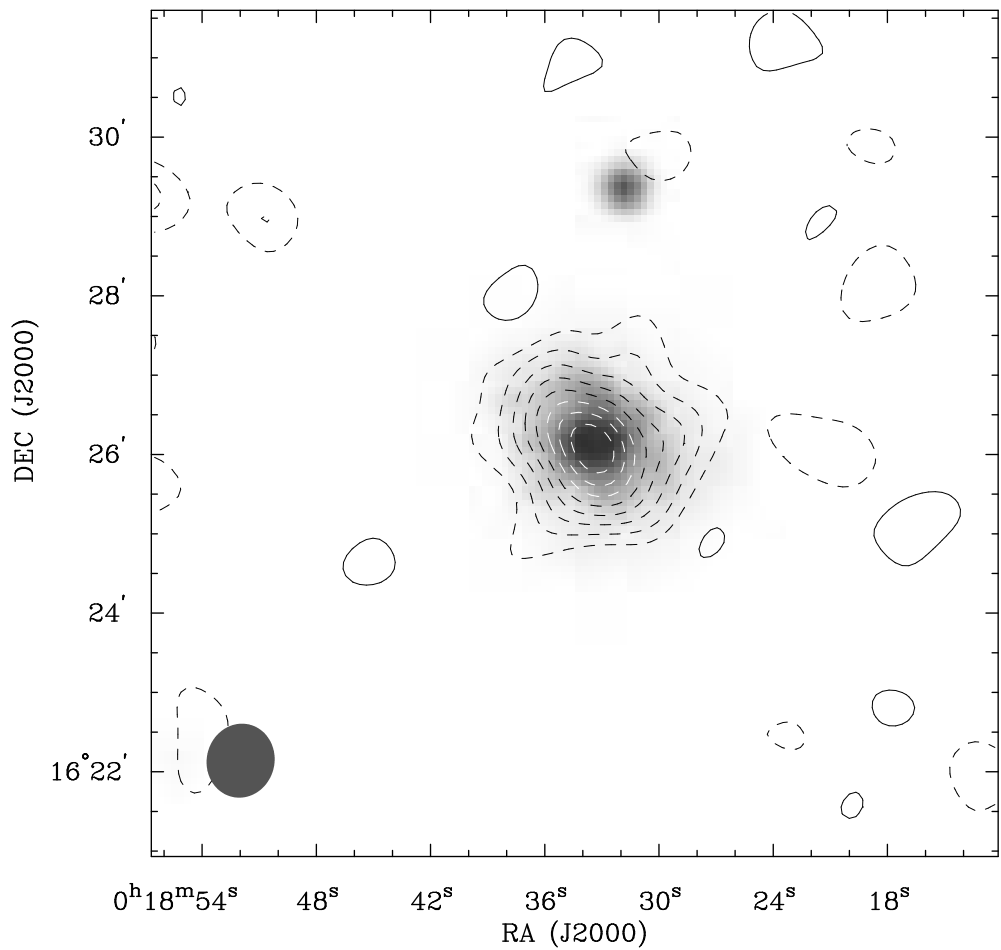}
\caption{\label{fig-ovro0016}
An interferometric map of CL~0016+16,
from Carlstrom \etal\ (1996), superimposed on a grey-scale
representation of the X-ray emission of the cluster (from the ROSAT
PSPC). The radio data on which this image is based
were taken with the Owens Valley Radio Observatory Millimeter Array
operated at 1~cm, contain antenna baselines from 20 to 75~m, and have
a synthesized beam of about $55$~arcsec (as shown in the lower left
corner).}
\end{figure}

Values of the Hubble constant based on this distance estimation
technique are now available for nine clusters. For the clusters
with \SZ\ effects shown in Fig.~\ref{fig-sz3cluster},
\ref{fig-suziescan}, \ref{fig-rylea2218} and \ref{fig-ovro0016}, a
detailed discussion of the fitting procedures used is given by
Birkinshaw \etal\markcite{bha91} (1991), Birkinshaw \&
Hughes\markcite{bh94} (1994), Jones\markcite{jon95} (1995), Holzapfel
\etal\markcite{holz97a} (1997a), and Hughes \&
Birkinshaw\markcite{hub98} (1998). The distances estimated for all
nine clusters are displayed as luminosity distances as a function of
redshift in Fig.~\ref{fig-szhubble}. The error bars on the distance
estimates are symmetrized errors taken from the individual papers and
include systematic errors as well as random errors from uncertainties
in the data.
\par
If the results in Fig.~\ref{fig-szhubble} are taken at face value,
the measurements suggest a Hubble constant near $60 \ \kmsMpc$, and
have a scatter of about $\pm 20 \ \kmsMpc$ (see the similar 
analysis of Furuzawa\markcite{futhesis} 1996). However, we
cannot use this to conclude that $H_0 = 60 \pm 10 \ \kmsMpc$, as seems 
reasonable based on nine~measurements, since those measurements are not
truly independent. In particular, only three different telescopes were
used in the measurement of $N_{\rm SZ}$ and only two in the
measurement of $N_{\rm X}$, so that there are only about two
independent X-ray calibrations and three independent \SZ\ effect
calibrations in the set of results for $H_0$. An improvement in the
precision of the determination of $H_0$, even in the absence of any
other problems, must depend on convincing absolute calibrations of
the \SZ\ effect and X-ray data.
\par
There are a number of other systematic problems in using this technique.
The most serious may be a selection effect, which causes the value of
$H_0$ to be biased low. If the model (\ref{eq-fnsimple}) for $f_n$ is
modified to make the cluster atmosphere prolate or oblate, then the
apparent X-ray and \SZ\ effect images of a cluster will be
ellipsoidal, or circular if the symmetry axis lies along the line of
sight. In the latter case it is clear that it will not be possible to
tell that the cluster is aspherical based on the images: indeed, if
the core radius of the gas distribution on the line of sight is larger
by a factor $Z$ than the core radii in the other two directions, then
the density form factor becomes 
\begin{equation}
   f_n = \left( 1 + {\theta^2+(\zeta^2/Z^2) \over \thc^2} 
                  \right)^{\!-{3 \over 2}\beta} \quad .
 \label{eq-fnflattened}
\end{equation}
and the expressions for $\Theta^{(1)}$ and $\Theta^{(2)}$
(\ref{eq-theta1} and \ref{eq-theta2}) remain valid, while the
normalizations $N_{\rm X}$ and  
$N_{\rm SZ}$ both increase by a factor $Z$. The result is that a prolate gas
distribution, with the symmetry axis along the line of sight, tends to
give a higher central surface brightness than other gas
distributions in which the same mass of gas 
is distributed spherically or with the symmetry axis perpendicular to
the line of sight. This causes clusters elongated along the line of
sight to be easier to detect in the X-ray or in the \SZ\ effect. Such
clusters also give biased estimates of distance, since the true
angular diameter distance is
\begin{equation}
   D_{\rm A}({\rm true}) = {D_{\rm A}({\rm estimated}) \over Z}
\end{equation}
if the distance is estimated using (\ref{eq-daho}) not knowing that the
cluster is elongated on the line of sight.
\par
\begin{figure}[p]
\epsscale{0.8}
\plotone{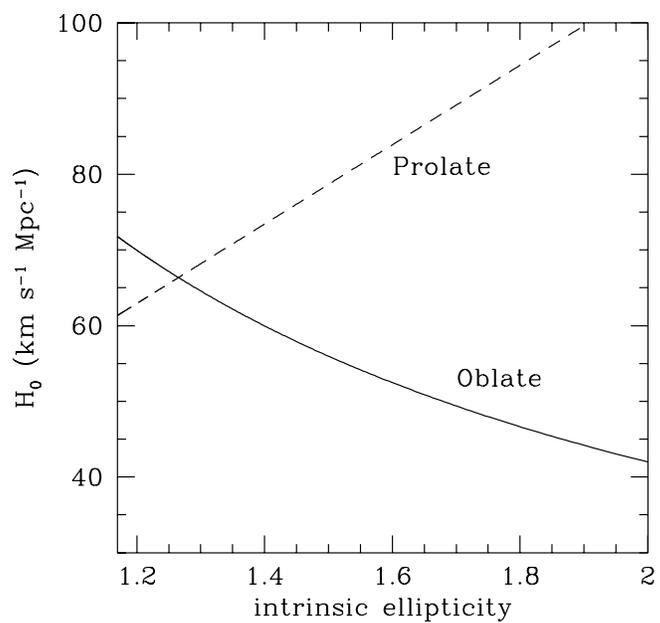}
\caption{\label{fig-0016oblate}
The dependence of the Hughes \& Birkinshaw's (1998) estimate
of the value of the Hubble constant on assumptions about the
oblateness or prolateness of CL~0016+16 in the extreme case where the
cluster symmetry axis lies in the plane of the sky.}
\end{figure}

An indication of the importance of this effect is shown in
Fig.~\ref{fig-0016oblate}, where the estimated value for the Hubble
constant from CL~0016+16 is shown as a function of the intrinsic
ellipticity (axial ratio) of an ellipsoidal model for the gas
distribution. An ellipsoidal model is clearly preferred because of 
the non-circular X-ray and \SZ\ effect isophotes
(Figs.~\ref{fig-0016pspc}, \ref{fig-ovro0016}). The value of the
Hubble constant derived by fitting the 
cluster by a spherical isothermal model is $68 \ \kmsMpc$: it
can be seen from the figure that by allowing ellipsoidal models with 
axial ratios as large as
2:1, values over the range $40 - 100 \ \kmsMpc$ can be obtained. 
\par
In order to avoid the selection bias in favor of clusters which are
elongated along the line of sight, and hence of high surface
brightness, and for which low estimates of the Hubble constant are
produced, this technique must be applied to a sample of clusters
selected without regard to their central surface brightness ---
perhaps clusters with total X-ray luminosities or flux densities above
some limiting value. Such a selection is now possible using the
high-sensitivity survey data recently returned by ROSAT (e.g., Ebeling
\etal\markcite{eb96} 1996). A corollary is that clusters which are
intrinsically hard to study in the X-ray or the \SZ\ effect (and
including \SZ\ effect non-detections) must be included in the set used
to measure $H_0$: the clusters with the weakest \SZ\ effects for their
measured X-ray brightnesses are exactly those which imply larger 
values of $H_0$ (albeit with larger observational errors). By
contrast, the clusters in Fig.~\ref{fig-szhubble} were often selected
based on having particularly strong \SZ\ effects, and are therefore
likely to show an orientation bias. The size of this bias is not known
at present, but is probably less than 30~per cent based on the
distribution of X-ray axial ratios seen in other cluster samples. 
\par
In addition to this bias, there is a further contribution to the error
in the estimated distance from the unknown intrinsic shape of cluster
atmospheres. The range of observed shapes suggests an error of order
20~per cent is possible (Hughes \& Birkinshaw\markcite{hub98} 1998),
and calculations of the evolution of cluster atmospheres confirm that
this error estimate is reasonable (Roettiger, Stone \&
Mushotzky\markcite{rsm97} 1997; Yoshikawa, Itoh \&
Suto\markcite{yis98} 1998).
\par
A major component of the error in the estimates of the normalizations
often arises from uncertainties in the parameters of the model
(equation~\ref{eq-fnflattened} or some more complicated function). This is
particularly evident when the fits are based on older X-ray data (as,
for example, Birkinshaw \& Hughes\markcite{bh94} 1994). The more
recent X-ray imaging data from ROSAT substantially reduce the
allowable range of parameters 
$\beta$ and $\thc$, so that this component of the error in the Hubble
constant may be reduced. However, there is an intrinsic uncertainty in
the types of gas model that are chosen to describe the atmosphere, and
the extent to which they fail to represent aspects of the density and
thermal structure of the gas that affect the distance estimate.
\par
Modeling the gas appropriately is important because it is not the
same gas that is responsible for the X-ray and \SZ\ effect signals
that are used to determine the distance. The
X-ray surface brightness is dominated by the densest parts of the
cluster, since the X-ray emissivity of the gas is proportional to
$n_{\rm e}^2 T_{\rm e}^{1/2}$, while the \SZ\ effect is dominated by
the lower-density and hotter parts of the gas where the path lengths
are longest. This effect is particularly important where single-dish
measurements of the \SZ\ effect are used, while interferometer maps
tend to resolve out structures on the largest angular scales.
Uncertainties in the relationship between the contributions of low
and high-density regions to the X-ray surface brightness
and the \SZ\ effect can be avoided by making deep X-ray images,
which trace the gas to sufficiently large radii that 90 or 95~per
cent of the gas responsible for the \SZ\ effect is included. This
means, however, tracing the cluster X-ray emission out to at least
10~core radii, at which the surface brightness has fallen to less than
$10^{-3}$ of its central value, which often requires long integration
times and careful treatment of the background in the X-ray detectors.
\par
Thermal structure in the cluster atmosphere is harder to measure, and
to achieve good accuracy in the distance estimates it is necessary to
know about the temperature of the cluster gas out to 10~core
radii. This is difficult, not only because of the 
low surface brightnesses of clusters at such radii, but also because
of the lower angular resolution 
of X-ray detectors with useful spectral response. There is 
little clear information on the changes in temperature of cluster
gas as a
function of radius outside a few core radii, and an isothermal model
(or sometimes a temperature model based on a bright nearby cluster,
such as Coma; equation~\ref{eq-hugheste}) are usually
assumed. Systematic errors at the ten per cent level are likely from
this uncertainty, and larger errors are possible for more extreme
temperature profiles --- hydrodynamical models of the evolution of
cluster atmospheres (Roettiger \etal\markcite{rsm97} 1997; Yoshikawa
\etal\markcite{yis98} 1998) suggest that systematic errors of as much
as 30~per cent and random errors of order 10~per cent in the Hubble
constant may arise because of departures from isothermality.
\par
A different type of density and temperature structure is often found
in the central parts of clusters, where the high X-ray emissivity
causes the cooling time of the gas to be short. The consequent
decrease in central pressure causes a ``cooling flow'' to be
established, with a slow inward drift of the atmosphere, an increase
in the central X-ray surface brightness, and a decrease in the central
gas temperature (e.g., Fabian \etal\markcite{fnc84} 1984). Since the
central region in which there is a large change of gas properties is
fairly small, it is still possible to use a model of the form
(\ref{eq-fnsimple}) to describe the gas distribution, provided that
the central X-ray brightness spike is excluded from the X-ray fit, and
a corresponding change is made to the fitting for the \SZ\ effect. The
\SZ\ effect will show less modification than the X-ray surface
brightness in the presence of a cooling flow because the path length
through the cooling region is relatively small, and there is only a
small change of electron pressure in that region. However, the cooling
gas may partly ``fill in'' the cm-wave microwave background diminution
with free-free emission (Schlickeiser\markcite{schl91} 1991), so that
excluding the central region of a cluster from the fit may be important.
\par
Even smaller-scale structure in cluster atmospheres can have an
effect on the derived distance. If the intracluster gas is isothermal,
but shows density clumping on a scale less than the resolution of the
images, then the X-ray emissivity of a small element of gas is
enhanced by a factor
\begin{equation}
   C_{\rm n} = {\langle  n_{\rm e}^2  \rangle   \over
                \langle  n_{\rm e}    \rangle^2       }
\end{equation}
while the value of $\langle n_{\rm e} \rangle$ is
unchanged. Thus the cluster generates more X-ray emission than would
be expected based on a uniform atmosphere, and hence the true
angular diameter distance is
\begin{equation}
   D_{\rm A}({\rm true}) = C_{\rm n} \, D_{\rm A}({\rm estimated}) 
\end{equation}
so that with $C_{\rm n} > 1$, the true value of the Hubble constant is
smaller than the value estimated based on (\ref{eq-daho}) without
knowledge of the small-scale clumping. 
\par
Unlike the orientation bias, where averaging over a large number of
clusters in random orientations with a known distribution of cluster
shapes can correct the distance estimate, all cluster
atmospheres are expected to be clumpy to some degree, and it is
necessary to estimate the value of the clumping in the ``average''
cluster atmosphere, or to measure it in each cluster, in order to be
sure that the distance estimate is not seriously in error. A
theoretical estimate of the degree of clumping of the intracluster
medium would be difficult, since it must take into account the
processes that cause clumping (such as gas injection from the
galaxies and energy input from galaxy motions) and that erase clumping
(thermal conduction, gas-dynamical processes, and so on). 
If the clumping is strong and non-isothermal, then detailed X-ray
spectroscopy may be able to measure the distribution of 
temperatures within a cluster, but it is unlikely that full account
could be taken of a distribution of $C_{\rm n}$ (with an associated
form factor, $f_C$) over the cluster volume, nor that the full range
of types of clumping could be tested in this way. At present it
appears that the clumping of the intracluster medium is relatively
weak, since if $C_{\rm n}$ is often large, then it would be expected 
to show significant variation from cluster to cluster, and the Hubble
diagram (Fig.~\ref{fig-szhubble}) 
would show stronger scatter than it does. However, the errors on the
distance estimates in Fig.~\ref{fig-szhubble} at present cannot
exclude values of $C_{\rm n} \sim 1.5$, with consequent large
systematic error in $H_0$.
\par
A variety of other potential problems with this method can be
imagined. The \SZ\ effect signal could be contaminated by a background
primordial anisotropy in the microwave background radiation (e.g.,
Cen\markcite{cen98} 1998), or by the non-thermal \SZ\ effect of a
cluster radio halo source, or by the kinematic \SZ\ effect, or by
diffuse radio emission from cool gas (perhaps clumped
into a population of spiral galaxies) towards the edge of the
cluster. The X-ray signal could also be contaminated, perhaps by
the inverse-Compton emission of relativistic electrons
in the cluster radio halo source.
Some of these effects are one-sided biases in the distance
estimate, others would increase scatter in the Hubble diagram, but in
general they should provide additional errors at the level of 10~per
cent or less in the distance estimate
(Birkinshaw \etal\markcite{bha91} 1991; Holzapfel
\etal\markcite{holz97a} 1997a).
\par
The potential of this method for measuring the Hubble constant is only
now starting to be realized, as better \SZ\ effect data become
available. I expect a large increase in the number of clusters on a
future Hubble diagram like Fig.~\ref{fig-szhubble}, and that useful
cosmological results will be obtained, especially as the maximum
redshift at which an \SZ\ effect cluster is detected increases above
$0.55$. However, in view of the likely presence of residual systematic
effects in the data and the low accuracy of any one measurement, I 
believe that it is premature to use them to estimate the values 
of the deceleration parameter and cosmological constant, as has been
attempted recently by Kobayashi, Sasaki \& Suto\markcite{kss96}
(1996).
\subsection{Contributions to the CMBR spectrum}
The cosmological effects of the \SZ\ effect fall into two categories:
the integrated effect on the spectrum (discussed in this section) and
the angular fluctuation pattern that is created (Sec.~11.3).
Both the gas in clusters of galaxies and the distributed hot
intergalactic medium between clusters will contribute to these
effects: indeed, at a  general level we can consider the cluster gas
to be merely a strongly clumped fraction of the hot intergalactic
medium. The cosmological \SZ\ effects then measure the 
projected electron pressure distribution since recombination.
\par
It is convenient in discussing the effect of the intergalactic medium
(IGM) on the CMBR to work in terms of the fraction of the critical
density that this gas comprises. This is described by the quantity
\begin{equation}
   \Omega_{\rm IGM} = { \rho_{\rm IGM} \over \rho_{\rm crit} }
\end{equation}
where the critical density, $\rho_{\rm crit}$
(equation~\ref{eq-rhocrit}) just closes the Universe. Limits to the
contribution of {\it neutral} gas to $\Omega_{\rm IGM}$ are already
stringent, because of the absence of neutral hydrogen absorption
features in the spectra of high-redshift quasars (the Gunn-Peterson
test; Gunn \& Peterson\markcite{gp65} 1965), with a recent 
limit on the optical depth $\tau_{\rm GP} < 0.07$ at redshifts near 4.3
based on a spectrum of a quasar at $z = 4.7$ (Giallongo
\etal\markcite{gia94} 1994). Further limits on the contribution of hot
gas to $\Omega_{\rm IGM}$ can be set 
based on the X-ray background, most of which can be accounted for by
the integrated emission of active galaxies and quasars (Comastri
\etal\markcite{cszh95} 1995).  At low energies it has been suggested
that the bremsstrahlung of hot gas in clusters and groups of galaxies may make
a significant contribution to the X-ray background, or even
over-produce the background under some models for cluster evolution
(Burg \etal\markcite{burg93} 1993), while the possibility that a
diffuse intergalactic medium is responsible for much of the X-ray
background was suggested by Field \& Perrenod\markcite{fp77} (1977). 
\par
If some significant contribution to the X-ray background does come
from distributed gas, then the assumption that the gas is fully
ionized out to some redshift $z_{\rm ri}$ (at time $t_{\rm ri}$)
leads to an optical depth for inverse-Compton scatterings between
ourselves and the epoch of recombination of 
\begin{eqnarray}
 \tau_{\rm e} &=& \int_{t_{\rm ri}}^{t_{\rm 0}} \, 
                 \sigma_{\rm T} \, n_{\rm e}(z) \, c \, dt \nonumber \\
              &=& {c \sigma_{\rm T} n_{\rm e0} \over H_0} \,
                 \int_0^{z_{\rm ri}} \, dz {(1+z) \over (1 + \Omega_0
                 z)^{1/2}}
\end{eqnarray}
where $n_{\rm e0}$ is the electron density today and I have assumed a
Friedmann-Robertson-Walker cosmology with zero cosmological
constant. If the thermal history of this intergalactic plasma is
parameterized by a redshift-dependent electron temperature, $T_{\rm
e}(z)$, then the Comptonization parameter is
\begin{eqnarray}
   y &=& \int_{t_{\rm ri}}^{t_{\rm 0}} \, 
                   \sigma_{\rm T} \, n_{\rm e}(z) \, c \, 
                   {\boltz T_{\rm e}(z) \over m_{\rm e} c^2} \, dt
                   \nonumber \\
     &=& {c \sigma_{\rm T} n_{\rm e0} \over H_0} 
                   \, {\boltz \over m_{\rm e} c^2}
                   \int_0^{z_{\rm ri}} \, dz \,
                   T_{\rm e}(z) \, {(1+z) \over (1 + \Omega_0
                   z)^{1/2}}.
     \label{eq-ycosmological}
\end{eqnarray}
\par
For re-ionization redshifts $\lesssim 30$, and any $\Omega_0 < 1$, the
scattering optical depth is less than about $2.6 \, \Omega_{\rm IGM}
\, h_{\rm 100}$, and when the integral in (\ref{eq-ycosmological}) is
performed for plausible thermal histories of the intergalactic medium 
(e.g., Taylor \& Wright\markcite{tw89} 1989; Wright
\etal\markcite{wr96} 1996), then the recent COBE FIRAS limit $y < 15
\times 10^{-6}$ (Fixsen \etal\markcite{fix96} 1996) leads to a limit
on the electron scattering optical depth (averaged over the sky) of
less than $3 \times 10^{-4}$ (Wright \etal\markcite{wr94} 1994). This
corresponds to an electron density that is $\approx 100$ times less
than the density needed to produce a significant fraction of the X-ray
background by thermal bremsstrahlung, which in turn suggests that a
uniform, hot, IGM produces less than $10^{-4}$ of the X-ray
background, and that a significant fraction of the X-ray background
can only arise from thermal bremsstrahlung if the gas has a filling
factor $< 10^{-4}$ on the sky. 
\par
Direct calculations of the effects of clusters of galaxies on the
spectrum of the CMBR have been made by Markevitch
\etal\markcite{mark91} (1991) and Cavaliere, Menci \&
Setti\markcite{cms91} (1991). An integration like that in
(\ref{eq-ycosmological}) must now be performed over an evolving
population of clusters of galaxies, with varying space density, size,
gas properties, etc. Markevitch \etal\ used self-similar models for
the variations of cluster properties with redshift. These models
are characterized by a power-law index $n$, which defines the
relationship between the redshift and density, size, mass, and
comoving number density scales of a population of
clusters. Specifically, the mass scale of the population is 
\begin{equation}
  M^\ast \propto (1 + z)^{-6/(n+3)}
  \label{eq-massscale}
\end{equation}
if $\Omega_0 = 1$ and a more complicated expression for other values
of $\Omega_0$ (White \& Rees\markcite{wr78} 1978; Kaiser
\markcite{kai86} 1986). For the physical range $-3 < n < 1$, slower
evolution of $M^\ast$ is obtained for larger values of $n$. 
\par
Markevitch \etal\markcite{mark91} (1991) normalized the properties of
a population of clusters using present-day observed density,
temperature, and structure based on X-ray data, and integrated over
this population as it evolved to calculate the mean Comptonization
parameter that would result. The important parameters of this
calculation are $n$, $\Omega_0$, and $z_{\rm max}$, the maximum
redshift for which clusters can be said to follow the evolution model
(\ref{eq-massscale}). Using the most recent limits on the
Comptonization parameter from the analysis of the COBE FIRAS data
(Fixsen \etal\markcite{fix96} 1996), the numerical results obtained by
Markevitch \etal\ can be interpreted as implying that 
$z_{\rm max} \lesssim 10$ for a non-evolving cluster population, and
that $\Omega_0 \gtrsim 0.1$ if the cluster population evolves with
$-1 \le n \le 1$. Similar conclusions can be drawn from the results
given by Cavaliere \etal\markcite{cms91} (1991).
\par
The closeness of the COBE FIRAS limit to the Comptonization parameter
to the prediction from these models for the change of cluster
properties with redshift indicates the power of the FIRAS data in 
constraining models for the evolution of clusters, and
perhaps the value of $\Omega_0$ (Markevitch \etal\markcite{mark91}
1991; Wright \etal\markcite{wr94} 1994). It should now be possible to
take into account all the constraints on the population of clusters
containing dense atmospheres, including the controversial ``negative
evolution'' of the population of X-ray clusters (Edge
\etal\markcite{edge90} 1990; Gioia \etal\markcite{gio90a} 1990a), to
place strong restrictions on the range of acceptable models of cluster
evolution.
\subsection{Fluctuations in the CMBR}
The \SZ\ effect not only causes changes in the integrated spectrum of
the CMBR, but also induces fluctuations in its brightness which appear
superimposed on the fluctuations arising from the formation of
structure in the early Universe (Sec.~1.3). The angular
scale of these new structures in the CMBR will depend on their origin,
and on the large-scale structure of the Universe.  Constraints on
both the manner in which clusters evolve and $\Omega_0$ have been
obtained by the limits to the fluctuation power from arcminute-scale
experiments. 
\par
The present review concentrates on the fluctuations induced
in the CMBR by clusters and superclusters of galaxies, but a diffuse
ionized intergalactic medium with density and velocity irregularities,
such as those created as large-scale structure develops, will
also produce significant CMBR fluctuations. The best-known of these
is the Vishniac effect (Vishniac\markcite{vish87} 1987), which is due
to the kinematic \SZ\ effect of a perturbation in the electron density
in the (re-ionized) diffuse intergalactic medium. Discussions of this,
and other, structures that are superimposed on the primordial spectrum
by inhomogeneities in the re-ionized intergalactic medium are given by
Dodelson \& Jubas\markcite{dj95} (1995) for $\Omega_0 = 1$, by
Persi\markcite{per95} (1995) for open Universes, and in the review of
White \etal\markcite{wss94} (1994). 
\par
Cluster \SZ\ effects can have a strong influence on the CMBR because,
unlike ``normal'' astrophysical sources, the surface brightness of the
\SZ\ effect from a cluster is independent of redshift, and does not suffer
$(1+z)^{-4}$ fading. This is because the effect is a fractional change in
the brightness of the CMBR, and the CMBR's energy density itself
increases with redshift as $(1+z)^4$, cancelling out the dimming
effect of cosmology. The integrated flux density of a cluster at
observed frequency $\nu$,
\begin{equation}
   S_\nu = j(x) \int d\Omega
           \int {\boltz \te \over m_{\rm e} c^2}
           \,  \sigma_{\rm T} \, n_{\rm e} \, dl
 \label{eq-clusterflux}
\end{equation}
in the Kompaneets approximation, where $j(x)$ is the Kompaneets
spectral function (defined by $\Delta n = y j(x)$ in equation
\ref{eq-kompaneetsdeltan}), 
$x = {h\nu / \boltz \trad}$ is the usual dimensionless frequency,
and the first integral is over the
solid angle of the cluster. Equation (\ref{eq-clusterflux})
can be written as an integral over the cluster volume
\begin{equation}
   S_\nu = j(x) \int d^3x
              \,  {\boltz \te \over m_{\rm e} c^2}
              \,  {\sigma_{\rm T} \, n_{\rm e} \over d_A^2}
\end{equation}
which for a constant electron temperature over the cluster can be
written simply in terms of the total number of electrons in the
cluster, $N_{\rm e}$, and the angular diameter and
luminosity distances, $D_{\rm A}$ and $D_{\rm L}$, as
\begin{equation}
    S_\nu = j(x) 
           \,  {\boltz \te \over m_{\rm e} c^2}
           \,  {\sigma_{\rm T} \, N_{\rm e} \over D_{\rm A}^2}
           = j(x) 
           \,  {\boltz \te \over m_{\rm e} c^2}
           \,  \sigma_{\rm T} \, N_{\rm e}
           \, {(1+z)^4 \over D_L^2} \quad .
\end{equation}
This indicates that the cluster's apparent luminosity increases as
$(1+z)^4$ --- or, alternatively, that its flux
density is a function of intrinsic properties and angular diameter
distance only.
\par
As a result, a population of clusters with the same $N_{\rm e}$
and $T_{\rm e}$, observed at different redshifts, will exhibit
a minimum flux density at the redshift of maximum angular diameter
distance in that cosmology (Korolyev, Sunyaev \&
Yakubtsev\markcite{kor86} 1986). Although this might provide a
cosmological test for $\Omega_0$, in practice clusters exhibit a wide
range of properties and change significantly with redshift so it might
be difficult to distinguish the effects of cosmology, cluster
populations, and cluster evolution. The realizable cluster source
counts (the histogram of sky brightnesses observed by a particular
telescope of given properties) will then depend on a complicated mix of
observational characteristics of the telescope used, the cosmological
parameters, and the evolution of the cluster
atmospheres. Nevertheless, Markevitch \etal\markcite{mark94} (1994)
suggest that a study of source counts at the $\muJy$ level (at
cm-wavelengths) or at the $\mJy$ level (at mm-wavelengths) can
constrain the spectrum of cluster masses (which determines the value of
$y$), the cosmological parameter $\Omega_0$, and the redshift of cluster
formation. Using the results of arcminute-scale measurements of the 
anisotropy in the CMBR (the OVRO
RING experiment; Myers\markcite{myers90} 1990, Myers
\etal\markcite{myers93} 1993), Markevitch \etal\ were able to rule out
slowly-evolving ($n=1$ in eq.~\ref{eq-massscale}) models in an open
Universe with $\Omega_0 < 0.3$.
\par
More detailed treatments of the effects of foreground clusters on the
CMBR express their results in the formalism of Sec.~\ref{sec-cobe}
that is used to describe primordial fluctuations. A number of
different assumptions about the cosmology and evolution of large scale
structure have been used to calculate the
amplitude and angular pattern of the foreground fluctuations 
(Rephaeli\markcite{rep81} 1981; Cavaliere \etal\markcite{cav86} 1986;
Cole \& Kaiser\markcite{ck88} 1988; Schaeffer \& Silk\markcite{ss88}
1988; Thomas \& Carlberg\markcite{tc89} 1989; Markevitch
\etal\markcite{mark92} 1992; Makino \& Suto\markcite{ms93} 1993;
Bartlett \& Silk\markcite{bs94a} 1994a,\markcite{bs94b} 1994b;
Ceballos \& Barcons\markcite{cb94} 1994; Colafrancesco
\etal\markcite{col94} 1994; see also the review by
Rephaeli\markcite{rep95b} 1995b). A uniform result of the calculations
is that the distribution of sky brightness fluctuations that result 
is strongly non-Gaussian and asymmetrical since it is composed of
negative or positive sources (depending on the frequency of
observation, and the sign of $j(x)$) with varying numbers of sources 
on any line of sight or contained in a particular telescope beam
(e.g., Markevitch \etal\markcite{mark92} 1992). However, the
amplitude and angular scale of the cluster-generated fluctuations 
depend strongly on the pattern of cluster evolution and the cosmology
assumed. If the negative evolution of cluster atmospheres is strong
(negative $n$ in the self-similar model used by Markevitch \etal;
eq.~\ref{eq-massscale}), then the distribution will be dominated by
low-redshift clusters and the value of $\Omega_0$ will not be
important. For slow evolution or no evolution, the value of $\Omega_0$
becomes important, since the variation of $D_{\rm A}$ with redshift
dictates the appearance of the microwave background sky.
\par
The angular pattern of fluctuations that results generally shows
significant power in the two-point correlation function
(eq.~\ref{eq-correlation}) at the level $10^{-6} \lesssim \langle
\Delta T / T \rangle \lesssim 10^{-5}$ on sub-degree scales (e.g.,
Colafrancesco \etal\markcite{col94} 1994), but some models for the evolution
of clusters (and cluster atmospheres) can be ruled out from the
absence of large anisotropies in the OVRO data of Readhead
\etal\markcite{read89} (1989), Myers \etal\markcite{myers93} (1993),
or other experiments, and some 
cosmological parameters can be excluded under particular models for
the evolution of cluster atmospheres. Since different models can make quite
different predictions for the angular pattern and the amplitude of
fluctuations, there is a potential for studying the processes that
lead to the accumulation of cluster atmospheres through a study of the
microwave background radiation on the range of angular
scales (arcminute to degree) on which the cluster signal should
be significant. 
\par
If the evolution of clusters is to be studied in this way, then
observations of the cluster-induced \SZ\ fluctuation pattern would
need to be made over a wide range of angular scales in order to
validate or falsify any one of the models unambiguously. This range of
angular scales overlaps that occupied by the stronger ``Doppler
peaks'' in the primordial spectrum of fluctuations, so that the
cluster signal may be hard to detect (see, e.g., the review of
Bond\markcite{bond95} 1995). The cluster signal is also an important
contaminant of the Doppler peaks, which are expected to be a useful
cosmological indicator and whose characterization is an important aim
of the coming generation of CMBR satellites (MAP and Planck). 
Fortunately, measurements of the anisotropy pattern at several
frequencies can be used to separate \SZ\ effects imposed by clusters
and the primordial fluctuation background (Rephaeli\markcite{rep81}
1981): the sensitivity required to achieve clean separations is
formidable, but achievable with the current baseline design of the
satellites' detectors. 
\par
\begin{figure}[p]
\epsscale{0.8}
\plotone{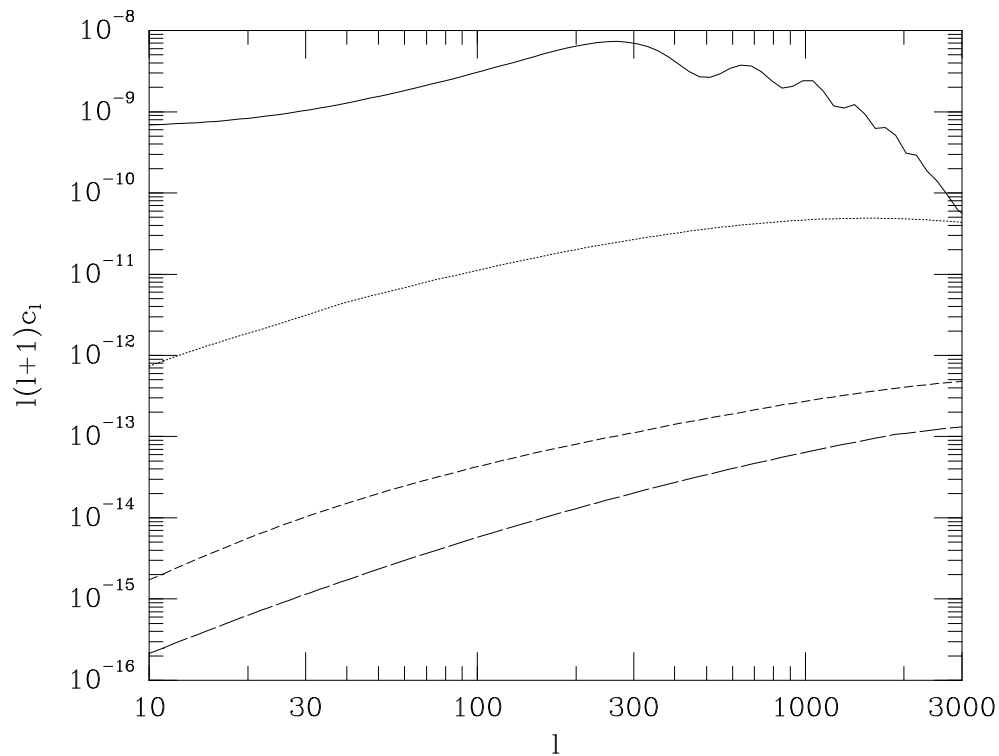}
\caption{\label{fig-lcdm}
The zero-frequency power spectrum of primordial microwave background
anisotropies (solid line; calculated using the CMBFAST code of
Zaldarriaga \etal\ 1998), the thermal \SZ\ effect
(dotted line), the kinematic \SZ\ effect (short dashed line), and the
Rees-Sciama effect from moving clusters (long dashed line) predicted
in the $\Lambda$-CDM cosmology discussed by Bahcall \& Fan
(1998). Figure from Molnar (1998).}
\end{figure}

An illustration of these results is given in Fig.~\ref{fig-lcdm},
which shows the relative strengths of the power spectra of primordial
fluctuations, the thermal and kinematic \SZ\ effects, and the
moving-cluster Rees-Sciama effect in a $\Lambda$-CDM cosmology
(involving a significant cosmological constant and cold dark matter)
with an evolving cluster population (Molnar\markcite{mo98} 1998). Although
the details of the power spectra depend on the choice of cosmology
and the physics of cluster evolution (compare, e.g., Aghanim
\etal\markcite{agh98} 1998), the general features are similar 
in all cases. For $l \simless 3000$, the power spectrum is dominated
by the signal from primordial structures. The kinematic \SZ\ and
Rees-Sciama effects from the cluster population are a factor
$\simgreat 10^2$ less important than the thermal \SZ\ effect. Thus the
evident detectability of the thermal \SZ\ effect (Sec.~\ref{sec-data})
is principally due to its strongly non-gaussian nature and its
association with clusters known from optical or X-ray observations,
and not to its intrinsic power. Future work, for example the all-sky
surveys that MAP and Planck will perform, will have the spectral
discrimination to detect the thermal \SZ\ effect on a statistical
basis, and should measure the power spectra of the thermal \SZ\ effect
on small angular scales ($l \simgreat 300$; Aghanim
\etal\markcite{agh97} 1997; Molnar\markcite{mo98} 1998).
\par
\begin{figure}[p]
\epsscale{0.8}
\plotone{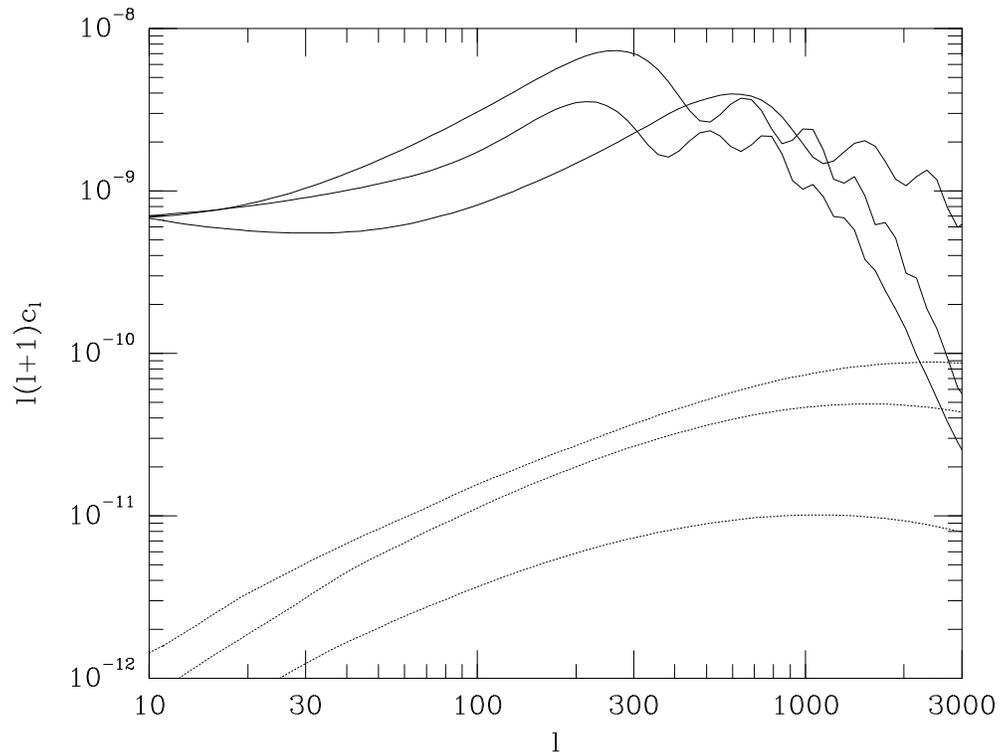}
\caption{\label{fig-threecosm}
The zero-frequency power spectrum of primordial microwave background
anisotropies (solid lines; calculated using the CMBFAST code of
Zaldarriaga \etal\ 1998), and the Sunyaev-Zel'dovich
effects of an evolving population of clusters (dotted lines) predicted
in three cosmological models consistent with the COBE anisotropies:
the open CDM, $\Lambda$-CDM, and ``standard'' CDM models discussed by
Bahcall \& Fan (1998). Figure from Molnar (1998).
}
\end{figure}

The sensitivity of the \SZ\ effect power spectrum to cosmology is
illustrated in Fig.~\ref{fig-threecosm} (Molnar\markcite{mo98}
1998). Variations of a factor $> 10$ in the power of fluctuations
induced by the thermal \SZ\ effect are evident at $l > 300$: although
this might be used as a cosmological test, the locations and strengths
of the Doppler peaks in the primordial anisotropy power spectrum are
more powerful. However, the amplitude of the \SZ\ effect power
spectrum depends on how clusters evolve, and measurements of this
power spectrum over a wide range of $l$ should provide an important
test of models of the formation of structure in the Universe.
\par
Superclusters of galaxies, and the gas pancakes from
which superclusters may have formed, are expected to make only a
minor contributions to the fluctuation spectrum
(Rephaeli\markcite{rep93} 1993; SubbaRao \etal\markcite{sub94}
1994). Once again, the angular scales on which the supercluster
signals appear are similar to those of the Doppler peaks, and both
good frequency and angular coverage will be needed to distinguish the
primordial and foreground signals.
\subsection{Quasars and the \SZ\ effects}
The intergalactic medium near a quasar must be strongly ionized by the
quasar's radiation. These hot gas bubbles are likely to be
overpressured, and to expand into their surrounding intergalactic 
medium. Thus both thermal and kinematic \SZ\ effects may arise near
quasars, and we might expect a contribution from quasars in the
spectrum of fluctuations in the CMBR (Aghanim \etal\markcite{agh96}
1996). Aghanim \etal\ find that the kinematic effect dominates, and
can cause local changes of $\approx 300 \ \muK$ in the brightness
temperature of the CMBR on scales up to $\approx 1$~deg. Whether such
structures are indeed present in the CMBR will be tested by the next
generation of CMBR surveys.
\par
\SZ\ effects may also be seen from the Lyman $\alpha$ absorption
clouds seen in quasar spectra (Loeb\markcite{loeb96} 1996). The
expected effects are much smaller, typically only a few $\muK$ and
with angular sizes of less than an arcminute, from the varying numbers
of Ly$\alpha$ systems on different lines of sight. Here again the
dominant contribution to the signal is from the kinematic \SZ\ effect,
and relies on large velocities acquired by the Ly$\alpha$ absorbing
clouds as large-scale structure forms.
\par
Either of these effects, or possibly a \SZ\ effect from a
quasar-related cluster with a deficiency of bright galaxies, or a
kinematic effect from colliding QSO winds (Natarajan \&
Sigurdsson\markcite{ns97} 1997), might explain the observations of
CMBR anisotropies towards the quasars PHL~957 (Andernach
\etal\markcite{and86} 1986) and PC~1643+4631 (Jones
\etal\markcite{jon97} 1997). However, the reality of these detections
remains in some dispute until they are independently confirmed.
\par
%
%
%
\section{Continuing research and the future of the \SZ\ effect}
Developments in the technologies of microwave background observation
are continuing, so that there is every reason to expect that all
clusters of galaxies with luminous X-ray emitting atmospheres will
eventually be detected in their \SZ\ effects. Cm-wave measurements,
with traditional single-dish telescopes and radiometers, are unlikely
to be as effective, in the long run, as mm-wave measurements using
bolometers simply because many strong X-ray clusters also contain
bright radio sources whose extended emission will not easily be
avoided at cm wavelengths. Nevertheless, radiometric surveys will be
increasingly good at locating \SZ\ effects as arrays of receivers
become more common and the bandwidths and noise temperatures of
radiometers continue to improve. Over the next few years I expect the
most spectacular improvements in type of \SZ\ effect work to emerge
from spectral measurements of the \SZ\ effects (with the principal aim of
setting limits to the velocities of clusters of galaxies) and from 
interferometric mapping of clusters, and indeed of the CMBR itself,
using optimized interferometers.
\par
A possible design for such an optimized array, tuned for work on
clusters at redshifts $\gtrsim 0.1$, would provide $\muK$
sensitivity, a full-resolution synthesized beam $\approx 30$ arcsec,
and good sensitivity to angular scales $\gtrsim 5$~arcmin. For
operation at cm wavelengths, this requires antennas of 10~m diameter
or less, baselines from 10-100~m, and sufficient antennas
simultaneously present that high sensitivity is attained rapidly and
so that radio source contamination can be well mapped. Such a system
is similar to BIMA or the OVMMA operated at cm-wavelengths, as done by
Carlstrom \etal\markcite{cjg96} (1996) and Patel \etal\markcite{pat97}
(1997), or to the planned VSA and CBI instruments. Alternatively,
smaller antennas and baselines (and smaller fractional bandwidths)
could be used at a wavelength of 3~mm with a dedicated microwave
background mapping array. This would have the advantages of better
rejection of signals from radio sources, and more leverage on the
spectrum of the \SZ\ effects with moderate changes in operating
frequency, but would need a good site if it is to operate
efficiently.
\par
Survey work, as is presently carried out from ground-based antennas,
could be done more efficiently from satellite systems, but with a
large cost. A good initial aim for a major survey would be to
provide $10 \ \muK$ or better sensitivity on a large set of clusters
selected without orientation bias, and hence suitable for statistical
interpretation of the \SZ\ effects for cluster properties and
cosmological parameters. Many clusters are likely to be detected in
such an unbiased fashion in the all-sky CMBR surveys that will be
produced by the next generation of mapping satellites (MAP and
Planck). Long-duration balloon projects (such as SOAR) should also be
able to produce excellent surveys of clusters. Cross-correlation
studies between CMBR maps of large fractions of the sky and cluster
(extended) X-ray sources from the ROSAT survey should give good
indications of the distribution of cluster properties. 
\par
It is likely to be space-based or balloon-based operation of bolometer
arrays that will produce the best measurements of \SZ\ effect spectra
of clusters and hence should measure the peculiar velocities of clusters
(or at least the peculiar velocities of cluster gas, which might not
be the same in all cases). Combined structural and spectral
measurements of a cluster, coupled with X-ray spectral and mapping
information, should allow the effects of primordial structure
contamination of the velocity signal in the CMBR to be minimized, since it is
unlikely that the primordial perturbations behind a cluster will be
distributed with an angular structure that is a close match to the 
cluster's gas distribution. The use of a 
matched filter based on the X-ray data may not be effective in all
cases, however, if the structure of cluster atmospheres is found to be
complicated by density and temperature inhomogeneities (as is
particularly likely at higher redshifts).
\par
\begin{figure}[p]
\epsscale{0.8}
\plotone{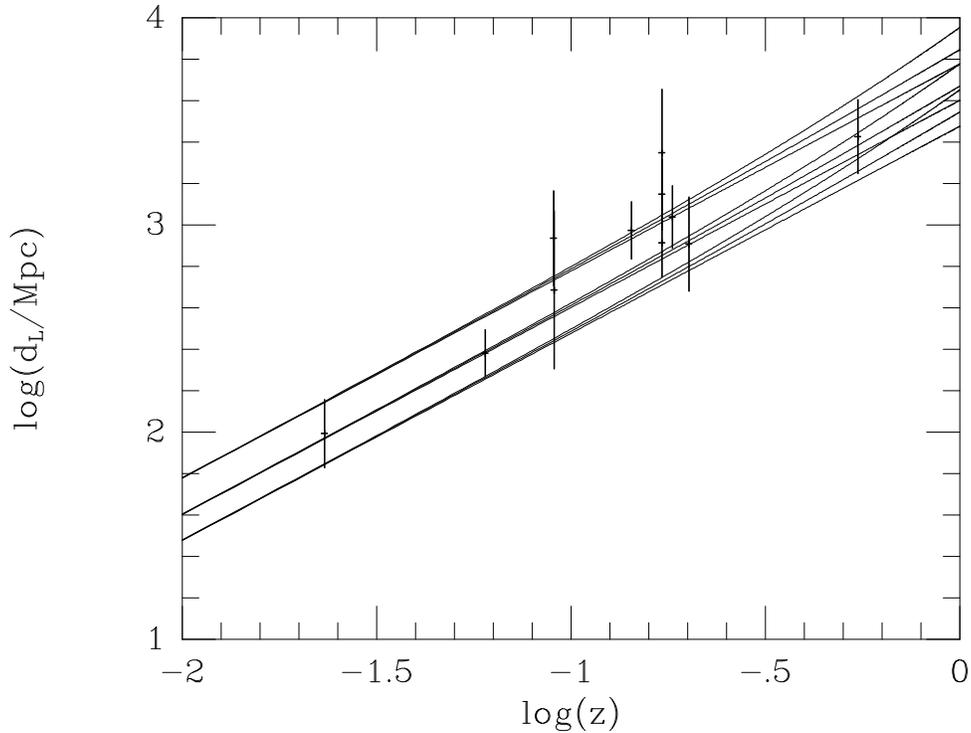}
\caption{\label{fig-szhubble}
A Hubble diagram based on the distances measured for the nine
clusters Abell~1656, 2256, 478, 2142, 1413, 2163, 2218, and 665 and
CL~0016+16 (Herbig \etal\ 1995; Myers \etal\ 1997;
Grainge 1996; Holzapfel \etal\ 1997a; McHardy \etal\ 1990;
Birkinshaw \& Hughes 1994; Jones 1995; 
Birkinshaw \etal\ 1991; and Hughes \& Birkinshaw 1996). Three values are
shown for the distance of Abell~2218 (from Birkinshaw \& Hughes,
Jones, and McHardy \etal).
The Hubble relation is drawn for $H_0 = 50$, $75$, and
$100 \ \kmsMpc$, with $q_0 = 0$, $1\over2$, and $1$. The current best
fit is for a Hubble constant of about $60 \ \kmsMpc$, with no strong
constraint on $q_0$, but no convincing error can be given because the
distance estimates contain correlated systematic errors arising from
the calibrations used (see text).}
\end{figure}

Obtaining these \SZ\ effect data at high signal/noise will not be
useful without matching high-quality X-ray data. Fortunately, such
X-ray data will be available shortly. We are already
obtaining large samples of clusters of galaxies from ROSAT (Ebeling
\etal\markcite{eb96} 1996), and with AXAF we will be able to obtain detailed
(arcsec-resolution) X-ray images of these clusters and spatially-resolved X-ray
spectra. \SZ\ interferometric maps would then be a powerful indicator
of structural inhomogeneities in the gas or anomalous heating (e.g.,
regions of clumping, perhaps in galaxy wakes). \SZ\ and X-ray data
together should provide good distance measurements over a wide range
of redshifts, leading to a substantial increase in the number of
clusters in the Hubble diagram (Fig.~\ref{fig-szhubble}), but the
estimation of reliable Hubble constant and deceleration parameter
demands an improvement in the level of systematic errors in that
diagram, especially through improvements in 
the calibration of the \SZ\ effect data (i.e., much better
absolute calibrations of the planets, and better transfer of these
calibrations to secondary sources) and the X-ray detectors.
\par
Other CMBR data on clusters of galaxies may also become available
soon. The detection of the kinematic \SZ\ effect and the Rees-Sciama
effects from the transverse motions of clusters of galaxies would
provide a full three-dimensional velocity field of clusters, allowing
the study of the evolution of this velocity field with redshift, and
providing fundamental constraints on the physics of galaxy
clustering. Observations of \SZ\ and other effects from clusters of
galaxies (or the puzzling cluster-like structures in regions of blank
sky) are likely to provide much powerful information for cosmology and
studies of clusters over the next decade or two.
\par
%
%
\acknowledgements
This review was partially supported by NASA grants NAGW-3825 and
NAG5-2415, NASA contract NAS8-39073, and a research grant from
PPARC. My research on the \SZ\ effects over the years has benefited
from many collaborators, especially S.F. Gull, J.P. Hughes, 
A.T. Moffet, and S.M. Molnar, and the generous
assistance of observatory staff at the 
Owens Valley Radio Observatory and the Very Large Array. I am also
grateful to J.E. Carlstrom, M. Jones, M. Joy, J.-M. Lamarre,
A.E. Lange, and R.D.E. Saunders for providing figures and information
about their continuing observations of the \SZ\ effects, 
and to P. Lilje, E. Linder, Y. Rephaeli and the referee for comments
on the text and other assistance.
\par
%
%

%
%
%
%
%
\clearpage
%
%
\clearpage
%
%
\clearpage
%
%
\clearpage
%
%
\clearpage
%
%
\clearpage
%
%
\clearpage
%
%
\clearpage
%
%
\clearpage
%
%
\clearpage
%
%
\clearpage
%
%
\clearpage
%
%
\clearpage
%
%
\clearpage
%
%
\clearpage
%
%
\clearpage
%
%
\clearpage
%
%
\clearpage
%
%
\clearpage
%
%
\clearpage
%
%
\clearpage
%
%
\clearpage
\clearpage
%
%
\clearpage
%
%
\clearpage
\clearpage
\clearpage
%
%
\begin{deluxetable}{lcccccl}
\tablecolumns{7}
\tablecaption{\label{tab-radiom}
 Radiometric measurements of the \SZ\ effects}
\tablehead{
 \colhead{Paper} & \colhead{Technique} 
                 & \colhead{$\nu$} 
                 & \colhead{$\theta_{\rm h}$}
                 & \colhead{$\theta_{\rm b}$}
                 & \colhead{$\theta_{\rm s}$}
                 & \colhead{} \\
 \colhead{}      & \colhead{}
                 & \colhead{$(\GHz)$}
                 & \colhead{$(\arcmin)$}
                 & \colhead{$(\arcmin)$}
                 & \colhead{$(\arcmin)$}
                 & \colhead{}
}
\startdata
   Parijskij\markcite{par72} 1972 
     & DS    &   7.5 & $1.3 \times 40$ & \nodata & 290           & \nl
   \tablevspace{2pt}
   Gull \& Northover\markcite{gn76} 1976    
     & BS+PS &  10.6 &  4.5            & 15.0 & \nodata    & $\ast$ \nl
   \tablevspace{2pt}
   Lake \& Partridge\markcite{lp77} 1977
     & BS+PS &  31.4 &  3.6            &  9.0 & \nodata    & $\ast$ \nl
   \tablevspace{2pt}
   Rudnick\markcite{rud78} 1978
     & BS+DS &  15.0 &  2.2            & 17.4 & 60-120     & \nl
   \tablevspace{2pt}
   Birkinshaw \etal\markcite{bgn78a} 1978a
     & BS+PS &  10.6 &  4.5            & 15.0 & \nodata    & $\ast$ \nl
   \tablevspace{2pt}
   Birkinshaw \etal\markcite{bgn78b} 1978b
     & BS+PS &  10.6 &  4.5            & 15.0 & \nodata    & $\ast$ \nl
   \tablevspace{2pt}
   Perrenod \& Lada\markcite{pl79} 1979
     & BS+PS &  31.4 &  3.5            &  8.0 & \nodata    & \nl
   \tablevspace{2pt}
   Schallwich\markcite{sch79} 1979 
     & BS+DS &  10.7 &  1.2            &  8.2 & 15         & $\ast$ \nl
   \tablevspace{2pt}
   Lake \& Partridge\markcite{lp80} 1980
     & BS+PS &  31.4 &  3.6            &  9.0 & \nodata    & \nl
   \tablevspace{2pt}
   Birkinshaw \etal\markcite{bgm81} 1981a
     & BS+PS &  10.7 &  3.3            & 14.4 & \nodata    & $\ast$ \nl
   \tablevspace{2pt}
   Birkinshaw \etal\markcite{bgn81} 1981b
     & BS+PS &  10.6 &  4.5            & 15.0 & \nodata    & \nl
   \tablevspace{2pt}
   Schallwich\markcite{sch82} 1982
     & BS+DS &  10.7 &  1.2            &  8.2 & 15         & \nl
   \tablevspace{2pt}
   Andernach \etal\markcite{and83} 1983
     & BS+DS &  10.7 &  1.2            &  3.2, 8.2  & 15   & \nl
   \tablevspace{2pt}
   Lasenby \& Davies\markcite{ld83} 1983
     & BS+PS &   5.0 & $8 \times 10$   & 30 & \nodata  & \nl
   \tablevspace{2pt}
   Birkinshaw \& Gull\markcite{bg84} 1984
     & BS+PS &  10.7 &  3.3            & 20.0 & \nodata        & \nl
    
     & BS+PS &  10.7 &  3.3            & 14.4 & \nodata        & \nl
   
     & BS+PS &  20.3 &  1.8            &  7.2 & \nodata        & \nl
   \tablevspace{2pt}
   Birkinshaw \etal\markcite{bgh84} 1984
     & BS+PS &  20.3 &  1.8            &  7.2 & \nodata        & \nl
   \tablevspace{2pt}
   Uson \& Wilkinson\markcite{uw84} 1984
     & BS+PS &  19.5 &  1.8            &  8.0 & \nodata        & $\ast$ \nl
   \tablevspace{2pt}
   Uson\markcite{uson85} 1985
     & BS+PS &  19.5 &  1.8            &  8.0 & \nodata        & $\ast$ \nl
   \tablevspace{2pt}
   Andernach \etal\markcite{and86} 1986
     & BS+DS &  10.7 &  1.18           & 3.2, 8.2 & 15         & \nl
   \tablevspace{2pt}
   Birkinshaw\markcite{b86} 1986
     & BS+PS &  10.7 &  1.78           & 7.15 & \nodata        & $\ast$ \nl
   \tablevspace{2pt}
   Birkinshaw \& Moffet\markcite{bm86} 1986
     & BS+PS &  10.7 &  1.78           & 7.15 & \nodata        & $\ast$ \nl
   \tablevspace{2pt}
   Radford \etal\markcite{rad86} 1986 
     & BS+DS &  90   &  1.3            & 4.0 & 10              & \nl

     & BS+PS &  90   &  1.2            & 4.3 & \nodata         & \nl
   
     & BS+PS & 105   &  1.7            & 19  & \nodata         & \nl
   \tablebreak
   Uson\markcite{uson86} 1986
     & BS+PS &  19.5 &  1.8            &  8.0 & \nodata        & $\ast$ \nl
   \tablevspace{2pt}
   Uson \& Wilkinson\markcite{uw88} 1988
     & BS+PS &  19.5 &  1.8            &  8.0 & \nodata        & \nl
   \tablevspace{2pt}
   Birkinshaw\markcite{b90} 1990
     & BS+PS &  20.3 &  1.78           &  7.15 & \nodata       & $\ast$ \nl
   \tablevspace{2pt}
   Klein \etal\markcite{kle91} 1991
     & BS+DS &  24.5 &  0.65           &  1.90 & 6.0           & \nl
   \tablevspace{2pt}
   Herbig \etal\ 1995 \markcite{herbig95}
     & BS+PS &  32.0 &  7.35           & 22.16 & \nodata       & \nl
   \tablevspace{2pt}
   Myers \etal\markcite{myers97} 1997
     & BS+PS &  32.0 &  7.35           & 22.16 & \nodata       & \nl
   \tablevspace{2pt}
   Uyaniker \etal\markcite{uyan97} 1997
     & DS    &  10.6 &  1.15           & \nodata & 10          & \nl
   \tablevspace{2pt}
   Birkinshaw \etal\markcite{bghm} 1998
     & BS+PS &  20.3 &  1.78           &  7.15 & \nodata       & \nl
   \tablevspace{2pt}
   Tsuboi \etal\markcite{tsu98} 1998
     & BS+PS &  36.0 &  0.82           & 6.0 & \nodata         & \nl
\enddata
\tablecomments{The technique codes are BS for beam-switching, PS for
position-switching, DS for drift- or driven-scanning. $\nu$ is the
central frequency of observation. $\theta_{\rm h}$ is the FWHM of the
telescope. $\theta_{\rm b}$ is the beam-switching angle (if 
beam-switching was used), and $\theta_{\rm s}$ is the scan length (for
drift or driven scans). $\ast$ in the final column indicates that the
paper contains data that are also included in a later paper in the
Table.}
\end{deluxetable}
\clearpage
%
%
\begin{deluxetable}{lcccccl}
\tablecolumns{7}
\tablecaption{\label{tab-bolom}
 Bolometric measurements of the \SZ\ effects}
\tablehead{
 \colhead{Paper} & \colhead{Technique} 
                 & \colhead{$\nu$} 
                 & \colhead{$\theta_{\rm h}$}
                 & \colhead{$\theta_{\rm b}$}
                 & \colhead{$\theta_{\rm s}$}
                 & \colhead{Notes} \\
 \colhead{}      & \colhead{}
                 & \colhead{$(\GHz)$}
                 & \colhead{$(\arcmin)$}
                 & \colhead{$(\arcmin)$}
                 & \colhead{$(\arcmin)$}
                 & \colhead{}
}
\startdata
 Meyer, Jeffries \& Weiss\markcite{mjw83} 1983
   & BS+PS & 90$-$300 & 5.0   & 5.0 & \nodata & \nl
 \tablevspace{2pt}
 Chase \etal\markcite{cjra87} 1987
   & BS+PS & 261      & 1.9   & 2.9  & \nodata                   & \nl
 \tablevspace{2pt}
 McKinnon \etal\markcite{moe90} 1990
   & BS+PS & 90       & 1.2   & 4.0  & \nodata              & Non-thermal \nl
 \tablevspace{2pt}
 Wilbanks \etal\markcite{wil94} 1994
   & BS+DS & 136      & 1.4   & 2.2, 4.3 & 26, 34 & $1 \times 3$ array \nl
 \tablevspace{2pt}
 Andreani \etal\markcite{and96} 1996
   & BS+PS & 150      & 0.73  & 2.3 & \nodata & \nl

   & BS+PS & 250      & 0.77  & 2.3 & \nodata & \nl
 \tablevspace{2pt}
 Holzapfel \etal\markcite{holz97a} 1997a
   & BS+DS & 143      & 1.7   & 2.3, 4.6 & 30 & $2 \times 3$~array \nl
 \tablevspace{2pt}
 Holzapfel \etal\markcite{holz97b} 1997b
   & BS+DS & 143      & 1.7   & 2.3, 4.6 & 30 & $2 \times 3$~array \nl

   & BS+DS & 214      & 1.7   & 2.3, 4.6 & 30 & $2 \times 3$~array \nl

   & BS+DS & 273      & 1.7   & 2.3, 4.6 & 30 & $2 \times 3$~array \nl
 \tablevspace{2pt}
 Silverberg \etal\markcite{silv97} 1997
   & PS+BS & 165      & 28    &  40   & 90 & Balloon \nl

   & PS+BS & 290      & 28    &  40   & 90 & Balloon \nl

   & PS+BS & 486      & 28    &  40   & 90 & Balloon \nl

   & PS+BS & 672      & 28    &  40   & 90 & Balloon \nl
\enddata
\tablecomments{The technique codes are BS for beam-switching, PS for
position-switching, DS for drift- or driven-scanning. $\nu$ is the
central frequency of observation: it is not possible to accurately
describe the four bands used by Meyer \etal\markcite{mjw83} (1983) in
this way, and 
only a range of frequencies is stated in this case. $\theta_{\rm h}$
is the FWHM of the telescope. $\theta_{\rm b}$ is the beam-switching
angle (if beam-switching was used), and $\theta_{\rm s}$ is the scan
length (for drift or driven scans).}
\end{deluxetable}
\clearpage
%
%
\begin{deluxetable}{lcccl}
\tablecolumns{5}
\tablecaption{\label{tab-interf} Interferometric measurements of the
 \SZ\ effects} 
\tablehead{
 \colhead{Paper} & \colhead{$\nu$}
                 & \colhead{$\theta_{\rm p}$}
                 & \colhead{$B$}
                 & \colhead{Telescope} \\
                 & \colhead{$(\GHz)$}
                 & \colhead{$(\arcmin)$}
                 & \colhead{$({\rm m})$}
                 & \colhead {}
}
\startdata
 Partridge \etal\markcite{part87} 1987
   & 4.9            & 9 & 35$-$1030 & VLA \nl
 Jones \etal\markcite{jon93} 1993
   & 15\phantom{.0} & 6 & 18$-$\phantom{1}108 & RT \nl
 Grainge \etal\markcite{gr93} 1993
   & 15\phantom{.0} & 6 & 18$-$\phantom{1}108 & RT \nl
 Jones\markcite{jon95} 1995
   & 15\phantom{.0} & 6 & 18$-$\phantom{1}108 & RT \nl
 Saunders\markcite{sau95} 1995 
   & 15\phantom{.0} & 6 & 18$-$\phantom{1}108 & RT \nl
 Liang\markcite{liang95} 1995
   &            8.8 & 5 & 31$-$\phantom{1}153 & ATCA \nl
 Carlstrom, Joy \& Grego\markcite{cjg96} 1996
   & 29\phantom{.0} & 4 & 20$-$\phantom{11}75 & OVMMA \nl
 Grainge\markcite{grthesis} 1996 
   & 15\phantom{.0} & 6 & 18$-$\phantom{1}288 & RT \nl
 Grainge \etal\markcite{gr96} 1996
   & 15\phantom{.0} & 6 & 18$-$\phantom{1}288 & RT \nl
 Jones \etal\markcite{jon97} 1997
   & 15\phantom{.0} & 6 & 18$-$\phantom{1}108 & RT \nl
 Matsuura \etal\markcite{mats96} 1996
   & 15\phantom{.0} & 6 & 18$-$\phantom{1}108 & RT \nl
\enddata
\tablecomments{$\nu$ is the central frequency of
observation. $\theta_{\rm p}$ is the corresponding FWHM of the beam
provided by the primary antennas of the interferometer. $B$ is the
range of baselines that were used in the work. 
The telescopes are the Very Large Array (VLA), Ryle
Telescope (RT), Australia Telescope Compact Array (ATCA), and Owens
Valley Radio Observatory Millimeter Array (OVMMA).}
\end{deluxetable}
\clearpage
%
%
\begin{deluxetable}{llccl}
\small
\tablecolumns{5}
\tablecaption{\label{tab-szdata} Final cluster center results}
\tablehead{
\colhead{Object} & \colhead{Redshift}
                 & \colhead{$\rm \DTRJ \ (mK)$}
                 & \colhead{O/C}
                 & \colhead{Reference}
}
\startdata
 Abell 71    & 0.0724 & $+0.29           \pm 0.54          $ & O & Birkinshaw \etal\markcite{bgn81} 1981b \nl
 \tablevspace{2pt}
 Abell 347   & 0.0187 & $+0.34           \pm 0.29          $ & O & Birkinshaw \etal\markcite{bgn81} 1981b \nl
 \tablevspace{2pt}
 Abell 370   & 0.373  & $                 > -0.20          $ & O & Liang\markcite{liang95} 1995 \nl
 \tablevspace{2pt}
 Abell 376   & 0.0489 & $+1.88           \pm 0.78          $ & O & Lake \& Partridge\markcite{lp80} 1980 \nl
             &        & $+1.22           \pm 0.35          $ & O & Birkinshaw \etal\markcite{bgn81} 1981b \nl
 \tablevspace{2pt}
 Abell 401   & 0.0748 & $-0.4\phantom{0} \pm 1.2\phantom{0}$ & O & Rudnick\markcite{rud78} 1978 \nl
             &        & $+0.78           \pm 0.62          $ & O & Birkinshaw \etal\markcite{bgn81} 1981b \nl
             &        & $-0.64           \pm 0.18          $ & O & Uson\markcite{uson86} 1986 \nl
 \tablevspace{2pt}
 Abell 426   & 0.0183 & $+3.67           \pm 1.12          $ & O & Lake \& Partridge\markcite{lp80} 1980 \nl
 \tablevspace{2pt}
 Abell 478   & 0.0900 & $-0.71           \pm 0.47          $ & O & Birkinshaw \etal\markcite{bgn81} 1981b \nl
             &        & $+0.44           \pm 0.32          $ & O & Birkinshaw \& Gull\markcite{bg84} 1984 \nl
             &        & $+2.0\phantom{0} \pm 3.2\phantom{0}$ & O & Radford \etal\markcite{rad86} 1986 \nl
             &        & $-0.2\phantom{0} \pm 1.0\phantom{0}$ & O & Chase \etal\markcite{cjra87} 1987 \nl
             &        & $-0.38           \pm 0.03          $ & O & Myers \etal\markcite{myers97} 1997 \nl
 \tablevspace{2pt}
 Abell 480   & [0.24] & $-2.08           \pm 1.49          $ & O & Birkinshaw \& Gull\markcite{bg84} 1984 \nl
 \tablevspace{2pt}
 Abell 506   & 0.1561 & $+0.63           \pm 0.76          $ & O & Perrenod \& Lada\markcite{pl79} 1979 \nl
 \tablevspace{2pt}
 Abell 508   & 0.1479 & $+1.62           \pm 1.27          $ & O & Birkinshaw \& Gull\markcite{bg84} 1984 \nl
 \tablevspace{2pt}
 Abell 518   & 0.1804 & $-1.56           \pm 0.83          $ & O & Perrenod \& Lada\markcite{pl79} 1979 \nl
 \tablevspace{2pt}
 Abell 545   & 0.1540 & $+1.68           \pm 0.45          $ & O & Lake \& Partridge\markcite{lp80} 1980 \nl
             &        & $+0.51           \pm 0.43          $ & O & Uson\markcite{uson85} 1985 \nl
 \tablevspace{2pt}
 Abell 576   & 0.0381 & $-1.27           \pm 0.28          $ & O & Lake \& Partridge\markcite{lp80} 1980 \nl
             &        & $-1.12           \pm 0.17          $ & O & Birkinshaw \etal\markcite{bgn81} 1981b \nl
             &        & $+1.10           \pm 0.44          $ & O & Lasenby \& Davies\markcite{ld83} 1983 \nl
             &        & $-0.14           \pm 0.29          $ & O & Birkinshaw \& Gull\markcite{bg84} 1984 \nl
             &        & $+0.50           \pm 0.29          $ & O & Radford \etal\markcite{rad86} 1986 \nl
 \tablevspace{2pt}
 Abell 586   & 0.1710 & $-0.09           \pm 0.38          $ & O & Birkinshaw \& Gull\markcite{bg84} 1984 \nl
 \tablebreak
 Abell 665   & 0.1816 & $-1.30           \pm 0.59          $ & O & Perrenod \& Lada\markcite{pl79} 1979 \nl
             &        & $-1.04           \pm 0.70          $ & O & Lake \& Partridge\markcite{lp80} 1980 \nl
             &        & $-0.53           \pm 0.22          $ & O & Birkinshaw \etal\markcite{bgn81} 1981b \nl
             &        & $+0.03           \pm 0.25          $ & O & Birkinshaw \& Gull\markcite{bg84} 1984 \nl
             &        & $-0.37           \pm 0.14          $ & O & Uson\markcite{uson86} 1986 \nl
             &        & $-0.24           \pm 0.04          $ & O & Grainge\markcite{grthesis} 1996 \nl
             &        & $-0.37           \pm 0.07          $ & O & Birkinshaw \etal\markcite{bghm} 1998 \nl
 \tablevspace{2pt}
 Abell 669   & [0.32] & $+0.38           \pm 0.24          $ & O & Birkinshaw \& Gull\markcite{bg84} 1984 \nl
 \tablevspace{2pt}
 Abell 697   & 0.282  & $-0.13           \pm 0.02          $ & O & Grainge\markcite{grthesis} 1996 \nl
 \tablevspace{2pt}
 Abell 773   & 0.1970 & $-0.18           \pm 0.04          $ & O & Grainge \etal\markcite{gr93} 1993 \nl
             &        & $-0.31           \pm 0.04          $ & O & Carlstrom \etal\markcite{cjg96} 1996 \nl
 \tablevspace{2pt}
 Abell 777   & 0.2240 & $-0.22           \pm 0.45          $ & O & Lake \& Partridge\markcite{lp80} 1980 \nl
 \tablevspace{2pt}
 Abell 910   & 0.2055 & $+0.22           \pm 0.56          $ & O & Lake \& Partridge\markcite{lp80} 1980 \nl
 \tablevspace{2pt}
 Abell 990   & 0.144  & $-0.13           \pm 0.03          $ & O & Grainge\markcite{grthesis} 1996 \nl
 \tablevspace{2pt}
 Abell 1204  & 0.1706 & $-0.10           \pm 0.36          $ & C & Matsuura \etal\markcite{mats96} 1996 \nl
 \tablevspace{2pt}
 Abell 1413  & 0.1427 & $-2.7\phantom{0} \pm 11.1          $ & O & Radford \etal\markcite{rad86} 1986 \nl
             &        & $+2.55           \pm 0.92          $ & O & Radford \etal\markcite{rad86} 1986 \nl
             &        & $+0.15           \pm 0.39          $ & O & Radford \etal\markcite{rad86} 1986 \nl
             &        & $-0.15           \pm 0.02          $ & O & Grainge \etal\markcite{gr96} 1996 \nl
 \tablevspace{2pt}
 Abell 1472  & [0.30] & $-1.26           \pm 1.02          $ & O & Perrenod \& Lada\markcite{pl79} 1979 \nl
 \tablevspace{2pt}
 Abell 1656  & 0.0232 & $-1.0\phantom{0} \pm 0.5\phantom{0}$ & O & Parijskij\markcite{par72} 1972 \nl
             &        & $+0.8\phantom{0} \pm 1.8\phantom{0}$ & O & Rudnick\markcite{rud78} 1978 \nl
             &        & $-0.20           \pm 0.22          $ & O & Lake \& Partridge\markcite{lp80} 1980 \nl
             &        & $+0.88           \pm 0.50          $ & O & Birkinshaw \etal\markcite{bgn81} 1981b \nl
             &        & $-0.27           \pm 0.03          $ & O & Herbig \etal\ 1995 \markcite{herbig95}\cr
             &        & $-0.31           \pm 0.40          $ & C & Silverberg \etal\markcite{silv97} 1997 \nl
 \tablevspace{2pt}
 Abell 1689  & 0.1810 & $-1.15           \pm 0.87          $ & O & Lake \& Partridge\markcite{lp80} 1980 \nl
             &        & $+0.24           \pm 0.38          $ & O & Birkinshaw \& Gull\markcite{bg84} 1984 \nl
             &        & $-1.87           \pm 0.32          $ & C & Holzapfel \etal\markcite{holz97b} 1997b \nl
 \tablevspace{2pt}
 Abell 1704  & 0.2200 & $                 > -0.12          $ & O & Carlstrom \etal\markcite{cjg96} 1996 \nl
 \tablebreak
 Abell 1763  & 0.1870 & $-0.36           \pm 0.25          $ & O & Uson\markcite{uson85} 1985 \nl
 \tablevspace{2pt}
 Abell 1795  & 0.0616 & $+0.2\phantom{0} \pm 0.9\phantom{0}$ & O & Meyer \etal\markcite{mjw83} 1983 \nl
 \tablevspace{2pt}
 Abell 1904  & 0.0708 & $+0.55           \pm 0.40          $ & O & Birkinshaw \etal\markcite{bgn81} 1981b \nl
 \tablevspace{2pt}
 Abell 1914  & 0.171  & $-0.15           \pm 0.04          $ & O & Grainge\markcite{grthesis} 1996 \nl
 \tablevspace{2pt}
 Abell 1995  & 0.318  & $-0.17           \pm 0.05          $ & O & Grainge\markcite{grthesis} 1996 \nl
 \tablevspace{2pt}
 Abell 2009  & 0.1530 & $-0.67           \pm 0.37          $ & O & Radford \etal\markcite{rad86} 1986 \nl
 \tablevspace{2pt}
 Abell 2079  & 0.0662 & $-0.05           \pm 0.25          $ & O & Lake \& Partridge\markcite{lp80} 1980 \nl
 \tablevspace{2pt}
 Abell 2125  & 0.2465 & $+0.73           \pm 0.45          $ & O & Lake \& Partridge\markcite{lp80} 1980 \nl
             &        & $-0.39           \pm 0.22          $ & O & Birkinshaw \etal\markcite{bgn81} 1981b \nl
             &        & $-0.31           \pm 0.39          $ & O & Birkinshaw \& Gull\markcite{bg84} 1984 \nl
 \tablevspace{2pt}
 Abell 2142  & 0.0899 & $-0.48           \pm 0.78          $ & O & Lake \& Partridge\markcite{lp80} 1980 \nl
             &        & $-1.4\phantom{0} \pm 1.0\phantom{0}$ & O & Birkinshaw \etal\markcite{bgn81} 1981b \nl
             &        & $-0.44           \pm 0.03          $ & O & Myers \etal\markcite{myers97} 1997 \nl
 \tablevspace{2pt}
 Abell 2163  & 0.201  & $-1.62           \pm 0.22          $ & C & Holzapfel \etal\markcite{holz97b} 1997b \nl
             &        & $                 > -0.19          $ & O & Liang\markcite{liang95} 1995 \nl
 \tablevspace{2pt}
 Abell 2199  & 0.0302 & $-2.2\phantom{0} \pm 1.2\phantom{0}$ & O & Rudnick\markcite{rud78} 1978 \nl
 \tablevspace{2pt}
 Abell 2218  & 0.1710 & $-1.04           \pm 0.48          $ & O & Perrenod \& Lada\markcite{pl79} 1979 \nl
             &        & $+0.81           \pm 0.39          $ & O & Lake \& Partridge\markcite{lp80} 1980 \nl
             &        & $-1.05           \pm 0.21          $ & O & Birkinshaw \etal\markcite{bgn81} 1981b \nl
             &        & $-1.84           \pm 0.33          $ & O & Schallwich\markcite{sch82} 1982 \nl
             &        & $+0.18           \pm 0.57          $ & O & Lasenby \& Davies\markcite{ld83} 1983 \nl
             &        & $-0.38           \pm 0.19          $ & O & Birkinshaw \& Gull\markcite{bg84} 1984 \nl
             &        & $-0.29           \pm 0.24          $ & O & Uson\markcite{uson85} 1985 \nl
             &        & $+3.5\phantom{0} \pm 2.4\phantom{0}$ & O & Radford \etal\markcite{rad86} 1986 \nl
             &        & $+0.10           \pm 0.27          $ & O & Radford \etal\markcite{rad86} 1986 \nl
             &        & $+0.26           \pm 0.20          $ & O & Radford \etal\markcite{rad86} 1986 \nl
             &        & $+0.4\phantom{0} \pm 0.7\phantom{0}$ & C & Partridge \etal\markcite{part87}1987 \nl
             &        & $-0.6\phantom{0} \pm 0.2\phantom{0}$ & O & Klein \etal\markcite{kle91} 1991 \nl
             &        & $-0.90           \pm 0.10          $ & C & Jones\markcite{jon95} 1995 \nl
             &        & $-0.68           \pm 0.20          $ & O & Uyaniker \etal\markcite{uyan97} 1997 \nl
             &        & $-0.40           \pm 0.05          $ & O & Birkinshaw \etal\markcite{bghm} 1998 \nl
             &        & $-0.52           \pm 0.15          $ & O & Tsuboi \etal\markcite{tsu98} 1998 \nl
 \tablebreak
 Abell 2255  & 0.0800 & $+1.5\phantom{0} \pm 3.0\phantom{0}$ & O & Rudnick\markcite{rud78} 1978 \nl
 \tablevspace{2pt}
 Abell 2256  & 0.0601 & $-0.24           \pm 0.03          $ & O & Myers \etal\markcite{myers97} 1997 \nl
 \tablevspace{2pt}
 Abell 2319  & 0.0564 & $+1.0\phantom{0} \pm 3.0\phantom{0}$ & O & Rudnick\markcite{rud78} 1978 \nl
             &        & $+1.37           \pm 0.94          $ & O & Perrenod \& Lada\markcite{pl79} 1979 \nl
             &        & $-0.14           \pm 0.20          $ & O & Lake \& Partridge\markcite{lp80} 1980 \nl
             &        & $-0.40           \pm 0.29          $ & O & Birkinshaw \etal\markcite{bgn81} 1981b \nl
             &        & $+0.82           \pm 0.60          $ & O & Birkinshaw \& Gull\markcite{bg84} 1984 \nl
 \tablevspace{2pt}
 Abell 2507  & 0.1960 & $+16.9           \pm 1.1           $ & O & Birkinshaw \& Gull\markcite{bg84} 1984 \nl
 \tablevspace{2pt}
 Abell 2645  & 0.2510 & $+2.35           \pm 0.70          $ & O & Lake \& Partridge\markcite{lp80} 1980 \nl
 \tablevspace{2pt}
 Abell 2666  & 0.0265 & $+0.62           \pm 0.31          $ & O & Lake \& Partridge\markcite{lp80} 1980 \nl
             &        & $+0.34           \pm 0.29          $ & O & Birkinshaw \etal\markcite{bgn81} 1981b \nl
 \tablevspace{2pt}
 Abell 2744  & 0.308  & $-2.1\phantom{0} \pm 0.7\phantom{0}$ & O & Andreani \etal\markcite{and96} 1996 \nl
 \tablevspace{2pt}
 Abell 3444  & 0.254  & $                 > -0.19          $ & O & Liang\markcite{liang95} 1995 \nl
 \tablevspace{2pt}
 CL~0016+16  & 0.5455 & $-0.9\phantom{0} \pm 0.9\phantom{0}$ & O & Andernach \etal\markcite{and83} 1983 \nl
             &        & $-0.72           \pm 0.18          $ & O & Birkinshaw \& Gull\markcite{bg84} 1984 \nl
             &        & $-0.50           \pm 0.59          $ & O & Radford \etal\markcite{rad86} 1986 \nl
             &        & $-0.48           \pm 0.12          $ & O & Uson\markcite{uson86} 1986 \nl
             &        & $-1.6\phantom{0} \pm 1.0\phantom{0}$ & O & Chase \etal\markcite{cjra87} 1987 \nl
             &        & $-0.43           \pm 0.03          $ & O & Carlstrom \etal\markcite{cjg96} 1996 \nl
             &        & $-0.33           \pm 0.03          $ & O & Grainge\markcite{grthesis} 1996 \nl
             &        & $-0.62           \pm 0.09          $ & O & Birkinshaw \etal\markcite{bghm} 1998 \nl
 \tablevspace{2pt}
 S 295       & 0.299  & $                 > -2.9\phantom{0}$ & O & Andreani \etal\markcite{and96} 1996 \nl
 \tablevspace{2pt}
 S 1077      & 0.312  & $-2.9\phantom{0} \pm 1.0\phantom{0}$ & O & Andreani \etal\markcite{and96} 1996 \nl
 \tablevspace{2pt}
 J 1780.5BL  & 0.49   & $                 > -0.39          $ & O & Liang\markcite{liang95} 1995 \nl
 \tablevspace{2pt}
 CL 1305+29  & 0.241  & $-0.28           \pm 0.22          $ & O & Birkinshaw \& Gull\markcite{bg84} 1984 \nl
 \tablevspace{2pt}
 Zw 1370     & 0.216  & $                 > -0.25          $ & O & Grainge\markcite{grthesis} 1996 \nl
 \tablevspace{2pt}
 MS 2137-23  & 0.313  & $                 > -0.11          $ & O & Liang\markcite{liang95} 1995 \nl
 \tablebreak
 PHL 957     & 2.3128 & $-0.60           \pm 0.15          $ & O & Andernach \etal\markcite{and86} 1986 \nl
 \tablevspace{2pt}
 MS00365     & 1.25   & $                 > -0.20          $ & O & Jones \etal\markcite{jon97} 1997 \nl
 \tablevspace{2pt}
 PG 0117+213 & 1.493  & $                 > -0.20          $ & O & Jones \etal\markcite{jon97} 1997 \nl
 \tablevspace{2pt}
 PC 1643+4631& 3.83   & $-0.13           \pm 0.04          $ & O & Jones \etal\markcite{jon97} 1997 \nl
\enddata
\tablecomments{The redshifts are shown in square brackets, as $[0.30]$,
when they are uncertain. Column~4 indicates whether the value
for $\DTRJ$ in the table is as observed (code~O), or a
calculated central decrement (code~C).}
\end{deluxetable}
\clearpage
%
%
\begin{deluxetable}{cll}
\tablecolumns{3}
\tablecaption{\label{tab-a2218consist1} Abell~2218 internal consistency}
\tablehead{
 \colhead{$\rm \DTRJ \ (mK)$} & \colhead{Reference} 
                              & \colhead{Telescope; frequency}
}
\startdata
 $-1.94 \pm 0.54$ & Gull \& Northover 1976\markcite{gn76}    & Chilbolton 25-m; 10.6~GHz \nl
 $-1.09 \pm 0.28$ & Birkinshaw \etal\markcite{bgn78a} 1978a  & Chilbolton 25-m; 10.6~GHz \nl
 $-1.49 \pm 0.23$ & Birkinshaw \etal\markcite{bgn78b} 1978b  & Chilbolton 25-m; 10.6~GHz \nl
 $-1.05 \pm 0.21$ & Birkinshaw \etal\markcite{bgn81} 1981b   & Chilbolton 25-m; 10.6~GHz \nl
 \tablevspace{2pt}
 $-0.38 \pm 0.19$ & Birkinshaw \& Gull\markcite{bg84} 1984   & OVRO 40-m; 10.7~GHz \nl
 \tablevspace{2pt}
 $-0.34 \pm 0.05$ & Birkinshaw \etal\markcite{bgh84} 1984    & OVRO 40-m; 20.3~GHz \nl
 $-0.31 \pm 0.13$ & Birkinshaw \& Gull\markcite{bg84} 1984   & OVRO 40-m; 20.3~GHz \nl
 $-0.39 \pm 0.03$ & Birkinshaw \& Moffet\markcite{bm86} 1986 & OVRO 40-m; 20.3~GHz \nl
 $-0.36 \pm 0.10$ & Birkinshaw\markcite{b86} 1986            & OVRO 40-m; 20.3~GHz \nl
 $-0.35 \pm 0.09$ & Birkinshaw\markcite{b90} 1990            & OVRO 40-m; 20.3~GHz \nl
 $-0.40 \pm 0.05$ & Birkinshaw \etal\markcite{bghm} 1998     & OVRO 40-m; 20.3~GHz \nl
 \tablevspace{2pt}
\enddata
\end{deluxetable}
\clearpage
%
%
\begin{deluxetable}{ll}
\tablewidth{250pt}
\tablecolumns{2}
\tablecaption{\label{tab-a2218consist2} Abell~2218 external consistency}
\tablehead{
 \colhead{$\rm \DTRJzero \ (mK)$} & \colhead{Reference}
}
\startdata
 $-2.6\phantom{0} \pm 1.2\phantom{0}$ & Perrenod \& Lada\markcite{pl79} 1979   \nl
 $+2.2\phantom{0} \pm 1.1\phantom{0}$ & Lake \& Partridge\markcite{lp80} 1980  \nl
 $-3.04           \pm 0.61          $ & Birkinshaw \etal\markcite{bgn81} 1981b \nl
 $-4.49           \pm 0.80          $ & Schallwich\markcite{sch82} 1982        \nl
 $+0.8\phantom{0} \pm 2.4\phantom{0}$ & Lasenby \& Davies\markcite{ld83} 1983  \nl
 $-0.77           \pm 0.38          $ & Birkinshaw \& Gull\markcite{bg84} 1984 \nl
 $-0.48           \pm 0.39          $ & Uson\markcite{uson85} 1985             \nl
 $+7.8\phantom{0} \pm 5.3\phantom{0}$ & Radford \etal\markcite{rad86} 1986     \nl
 $+0.21           \pm 0.57          $ & Radford \etal\markcite{rad86} 1986     \nl
 $+0.46           \pm 0.36          $ & Radford \etal\markcite{rad86} 1986     \nl
 $+0.40           \pm 0.70          $ & Partridge \etal\markcite{part87} 1987  \nl
 $-3.2\phantom{0} \pm 1.1\phantom{0}$ & Klein \etal\markcite{kle91} 1991       \nl
 $-0.90           \pm 0.10          $ & Jones\markcite{jon95} 1995             \nl
 $-0.88           \pm 0.26          $ & Uyaniker \etal\markcite{uyan97} 1997   \nl
 $-0.67           \pm 0.08          $ & Birkinshaw \etal\markcite{bghm} 1998   \nl
 $-0.68           \pm 0.19          $ & Tsuboi \etal\markcite{tsu98} 1998      \nl
\enddata
\end{deluxetable}
\clearpage
%
%
\begin{deluxetable}{lll}
\tablewidth{400pt}
\tablecolumns{3}
\tablecaption{\label{tab-reliableSZ}Clusters with reliable \SZ\ effects \hfil}
\tablehead{
 \colhead{Cluster} & \colhead{Recent measurement}
                   & \colhead{Independent confirmation}
}
\startdata
 Abell~478  & Myers \etal\markcite{myers97} 1997
            & \nodata \nl
 Abell~665  & Birkinshaw \etal\markcite{bghm} 1998
            & Grainge\markcite{grthesis} 1996 \nl
 Abell~697  & Grainge\markcite{grthesis} 1996
            & \nodata \nl
 Abell~773  & Carlstrom \etal\markcite{cjg96} 1996
            & Grainge \etal\markcite{gr93} 1993 \nl
 Abell~990  & Grainge \etal\markcite{gr96} 1996
            & \nodata \nl
 Abell~1413 & Grainge \etal\markcite{gr96} 1996
            & \nodata \nl
 Abell~1656 & Herbig \etal\markcite{herbig95} 1995
            & \nodata \nl
 Abell~1689 & Holzapfel \etal\markcite{holz97b} 1997b
            & \nodata \nl
 Abell~2142 & Myers \etal\markcite{myers97} 1997
            & \nodata \nl
 Abell~2163 & Holzapfel \etal\markcite{holz97b} 1997b
            & \nodata \nl
 Abell~2218 & Birkinshaw \etal\markcite{bghm} 1998
            & Jones\markcite{jon95} 1995 \nl
 Abell~2256 & Myers \etal\markcite{myers97} 1997
            & \nodata \nl
 CL~0016+16 & Carlstrom \etal\markcite{cjg96} 1996
            & Birkinshaw \etal\markcite{bghm} 1998 \nl
\enddata
\end{deluxetable}
\clearpage
%
%
\begin{deluxetable}{llcl}
\tablewidth{400pt}
\tablecolumns{4}
\tablecaption{\label{tab-nonthermalSZ}Non-thermal \SZ\ effect results}
\tablehead{
 \colhead{Object} & \colhead{Redshift}
                  & \colhead{$\rm \DTRJ \ (mK)$}
                  & \colhead{Reference}
}
\startdata
 0742+318 & 0.462 & $+0.25 \pm 0.59$ & McKinnon \etal\markcite{moe90} 1990 \nl
 \tablevspace{2pt}
 1721+343 & 0.206 & $+0.05 \pm 1.08$ & McKinnon \etal\markcite{moe90} 1990 \nl
 \tablevspace{2pt}
 2221-02  & 0.057 & $-1.74 \pm 0.76$ & McKinnon \etal\markcite{moe90} 1990 \nl
 \tablevspace{2pt}
 2349+32  & 0.671 & $+1.18 \pm 0.91$ & McKinnon \etal\markcite{moe90} 1990 \nl
\enddata
\end{deluxetable}
\clearpage
%
%
\end{document}